\documentclass[review]{elsarticle}
\usepackage[body={16.cm,23.cm}]{geometry}

\usepackage{lineno,hyperref}
\modulolinenumbers[5]

\journal{Annals of Physics}

\usepackage{fouriernc}

\renewcommand{\d}{\mathrm{d}}
\renewcommand{\v}[1]{\mathbf{#1}}

\renewcommand{\rm}[1]{\mathrm{#1}}

\usepackage{titlesec}
\renewcommand{\thesection}{\arabic{section}}
\titleformat{\section}{\sc\large}{\thesection.}{.5em}{}
\titleformat{\subsection}{\sc}{\thesubsection.}{.5em}{}

\usepackage{amsmath}
\usepackage{amsmath,amssymb}

\usepackage{mathrsfs} 
\usepackage{amscd}

\begin{document}

\begin{frontmatter}

\title{\sc Equation of State of Neutron-rich Matter in $d$-Dimensions}
%

\author{Bao-Jun Cai\fnref{fn1}}
\address{Quantum Machine Learning Laboratory, Shadow Creator Inc., Shanghai 201208, China}
\author{Bao-An Li\fnref{fn2}}
\address{Department of Physics and Astronomy, Texas A$\&$M University-Commerce, Commerce, TX 75429-3011, USA}

\fntext[fn1]{bjcai87@gmail.com}
\fntext[fn2]{Bao-An.Li@tamuc.edu}

%


%

\begin{abstract}
Nuclear systems under constraints, with high degrees of symmetries and/or collectivities may be considered as moving effectively in spaces with reduced spatial dimensions. We first derive analytical expressions for the nucleon specific energy $E_0(\rho)$,  pressure $P_0(\rho)$, incompressibility coefficient $K_0(\rho)$ and skewness coefficient $J_0(\rho)$ of symmetric nucleonic matter (SNM), the quadratic symmetry energy $E_{\rm{sym}}(\rho)$, its slope parameter $L(\rho)$ and curvature coefficient $K_{\rm{sym}}(\rho)$ as well as the fourth-order symmetry energy $E_{\rm{sym,4}}(\rho)$ of neutron-rich matter 
in general $d$ spatial dimensions (abbreviated as ``$d$D'') in terms of the isoscalar and isovector parts of the isospin-dependent single-nucleon potential according to the generalized Hugenholtz-Van Hove (HVH) theorem. 
The equation of state (EOS) of nuclear matter in $d$D can be linked to that in the conventional 3-dimensional (3D) space by the $\epsilon$-expansion which is a perturbative approch successfully used previously in treating second-order phase transitions and related critical phenomena in solid state physics and more recently in studying the EOS of cold atoms. 
The $\epsilon$-expansion of nuclear EOS in $d$D based on a reference dimension $d_{\rm{f}}=d-\epsilon$ is shown to be effective with $-1\lesssim\epsilon\lesssim1$ starting from $1\lesssim d_{\rm{f}}\lesssim3$ in comparison with the exact expressions derived using the HVH theorem. Moreover, the EOS of SNM (with/without considering its potential part) is found to be reduced (enhanced) in lower (higher) dimensions, indicating in particular that the many-nucleon system tends to be more bounded but saturate at higher densities in spaces with lower dimensions. The symmetry energy perturbed from its counterpart in 3D is found to strongly depend on the momentum-dependence of the nucleon isovector potential. Moreover, the specific structure of the fourth-order symmetry energy in $d$D is also analyzed generally, and it is found to be naturally small, confirming the parabolic approximation for the EOS of neutron-rich matter from an even wider viewpoint. The links between the EOSs in 3D and $d$D spaces from the $\epsilon$-expansion provide new perspectives to the EOS of neutron-rich matter. Further studies and potential applications of these links in nuclear physics and/or astrophysics are discussed.

\end{abstract}


\end{frontmatter}

\newpage

\tableofcontents

\section{Introduction}
Dimensionality reflects distinct natures or independent degrees of freedom of a single particle, a many-body physical system or an object.  The concept of dimensionality is very general and wide and could even not be limited to conventional physical systems,  e.g.,  the dimensionality of (animal) behavior could be defined as the minimum number of features of the past needed to make the maximally informative predictions about the future\,\cite{Bialek2022}. Fundamentally, the spatial dimension of a physical system is normally defined as the minimum number of coordinates needed to specify any point within it.
Since ordinary lifes and conventional physical systems live in three dimensions, their spatial degree of freedom are often frozen to three. It does not, however, mean that the problem in other dimensions is unimportant or irrelevant. In fact, it really helped in history to develop modern physical methodologies or improve the understanding on certain types of important theoretical issues, by considering the spatial dimension $d$ as a continuous or a discrete variable. For example, the dimension regularization technique in quantum field theories\,\cite{Collins1984} writing a Feynman integral as an integral depending on the space-time dimension $d+1$ instead of four solved for the first time the divergence problem of the gauge fields\,\cite{Hooft1972}.
On the other hand, the exact solution of the two-dimensional squared Ising model by L. Onsager\,\cite{Onsager1944} not only provided deep insight into the phase transition problems\,\cite{Stanley1971} but also largely prompted the development of modern statistical physics, such as the low- and high- temperature expansions, the critical phenomena and the renormalization group, see, e.g., Ref.\,\cite{Kardar2007} for a modern introduction on these issues.
The last three to four decades actually have witnessed many exciting novel features as well as important breakthroughs in systems with reduced dimensions, such as the 2-dimensional electron gas\,\cite{Ando1982}, the graphene\,\cite{Sarma2011}, and the topological insulator\,\cite{Hasan2010,Qi2011}.
One of the crucial characteristics of these 2-dimensional systems is the constant density of state which is independent of energy or momentum\,\cite{Ashcroft1976}, and it is this feature (partially) makes the 2D systems novel and peculiar.

Thanks to the developments of new technologies in the past years such as the Feshbach resonance\,\cite{Chin2010}, the laser cooling\,\cite{Schreck2021}, the optical lattices/boxes\,\cite{Vale2021,Navon2021} and the artificial gauge potentials\,\cite{Dalibard2011}, many exciting and fundamental experimental findings in various dimensions are emerging. To name a few,
we mention (1) the realization of spin-orbit interaction in degenerate quantum Fermi gas\,\cite{Zhai2015}, (2) the observation of the quantum-limited spin transport\,\cite{Luciuk2017,Sommer2011} as well as the sound propagation and damping in a 2D Fermi gases\,\cite{Bohlen2020} and a 2D Bose gas\,\cite{Ville2018}, the universal (first-) sound diffusion in a strongly cylindrical optical box\,\cite{Patel2020}, the second-sound attenuation near the quantum criticality\,\cite{LiX2022,Christodoulou2021}, (3) the observation of a non-Hermitian phase transition in an optical quantum gas\,\cite{Ozturk2021}, (4) the observation of superfluidity in a strongly correlated 2D Fermi gas\,\cite{Sobirey2021} together with the quantum scale anomaly and the spatial coherence of the system\,\cite{Murthy2019}, the synthetic dissipation and the turbulent flow in low dimensions\,\cite{Navon2016,Navon2019,Gauthier2019,Johnstone2019}, (5) the experimental verification of the generalized hydrodynamics in a strongly interacting 1D Bose gas\,\cite{Malvania2021}, the spin-charge separation in a 1D Fermi gas\,\cite{Senaratne2022} and the direct measurements of the compressibility and the EOS of a 2D photon gas inside a box potential\,\cite{Busley2022}, see Refs.\,\cite{Bloch2008,Hadzibabic2011,Bottcher2021,Sowinski2019,
Aidelsburger2018,Tannoudji2011,Pit2016,Gardiner2017,Zhai2021} for related discussions on these topics.
On the theoretical side, a two-species Fermi gas with mixed
dimensions\,\cite{Nishida2011}, a 3D resonant Bose-Fermi mixture at
zero temperature\,\cite{Bertaina2013} as well as the 2D Fermi-Bose dimers with a stable p-wave resonant interaction\,\cite{Bazak2018} were studied.
It was also demonstrated that a multiple-body problem could be effectively mapped to a two-body problem in
a higher dimension\,\cite{Guo2018}, and the scattering hypervolume for Fermions in 2D was recently investigated\,\cite{Wang2022}.
In addition, as an important but complicated problem for a strongly-interacting Fermi gas\,\cite{Bloch2008,Giorgini2008}, namely the determination of its EOS at unitary limit (i.e., the system with an infinite large scattering length), Ref.\,\cite{Nishida2006} successfully adopted the $\epsilon$-expansion method originated from the theory of second-order phase transition\,\cite{Wilson1974,MaSK1976,Wallace1976}, based on the observation that the unitary Fermi gas in 4D is actually an ideal Bose gas, thus the expansion via $\epsilon=d-4$ could relatively be easier to develop.
These studies vividly show that explorations of many-body systems in spaces with different dimensions may provide important new insights into certain interesting physical problems. In fact, the dimensionality is not only important for physical problems but also plays very fundamental roles in solving scientific issues for other fields, such as in the massive data subject in computational sciences\,\cite{Blum2020,FanJQ2020,Vershynin2018,Wainwright2018}, the technique of principal component analysis\,\cite{Jolliffe2002} is often applied to find the effective low-dimensional manifold (representation) of the high-dimensional data.

On the other hand, either the finite nucleus or the infinite nucleonic matter is an outstanding many-body system\,\cite{Fetter2003}. Their properties originate from different important aspects such as the complicated nucleon-nucleon interactions with momentum-, density-, spin- and isospin-dependence\,\cite{LiBA2008}, the (ultra-) relativistic effects\,\cite{Walecka1974,Chin1977,Serot1986} produced in heavy-ion reactions as well as from extremely astrophysical environments like those in neutron stars and/or around black holes\,\cite{Shapiro1983}, the finite-size effects due to the non-ideal parameters coming into the problem (e.g., the finite scattering length or the large but finite nucleon numbers in the nucleus), and also the dimensionality.
Almost all the existing treatments up to today for solving the nuclear many-body problems are done in 3 dimensions. However, there exist low-dimensional or quasi low-dimensional problems. For instance, a long time ago J. Wheeler suggested that nuclei in hydrodynamical equilibrium may obtain toroidal and/or spherical bubble topologies. The possiblilites of such shapes have been extensively studied both theoretically and experimentally, see, e.g., Refs.\,\cite{Siemens1967,Wong1972,Wong1973,Cao2019} and references therein. The high degree of symmetries in such nuclei may reduce effectively the dimensionality of the nucleons contained in them. 
Moreover, the production of particle jets from the participant region in a specific direction may effectively create an approximately 1D sub-system and in the meanwhile the collective
flow in the reaction plane may create approximately a 2D subsystem in intermediate-relativistic energy heavy-ion reactions\,\cite{LiBA2008}. Similarly, neutron star mergers may create $\gamma$-ray bursts
or other particles preferentially in certain directions\,\cite{Baiotti2017,Radice2020,Burns2020}. Moreover, it is well known that the dimensionality plays an important role in simulating neutrino-induced supernova explosions\,\cite{Burrows1990,Burrows2013,Burrows2021,Woosley1986,Janka2012}.
Consider for example the crust in neutron stars, its thickness could be roughly estimated to be about $t\approx1\,\rm{km}$. Compared with the typical radius of a neutron star with $R\approx10\,\rm{km}$, one sees that $t/4\pi R^2\approx10^{-3}\ll1$. In this case, one may ask if the crust can effectively be treated as a quasi 2D object. Furthermore, given an EOS the core structure of a non-rotating neutron star is determined uniquely by the well-known Tolman-Oppenheimer-Volkoff (TOV) equation depending on the radial coordinate only. Can the core of the neutron star be described effectively as a 1D system? 
If we can, then the relevant EOSs for the crust and core should be those with reduced dimensions (relative to the conventional 3D forms). The EOS of asymmetric nucleonic matter (ANM) is among the most important quantities for understanding all properties of neutron stars, such as their thermodynamic features and the transport characteristics\,\cite{Shapiro1983,Haensel2007,Oertel2017}.
The questions mentioned above naturally call for investigating the nuclear physical quantities (including the EOS of ANM) in a general dimension $d$ and to explore their possibly new features with respect to the conventional 3D space.

To introduce the relevant characteristics of ANM EOS, lets first recall here a few definitions and terminologies.
The EOS of ANM at zero temperature can be described by the energy per nucleon (specific energy) $E(\rho,\delta)$
with $\rho=\rho_{\rm{n}}+\rho_{\rm{p}}$ being the nucleon density and $\delta=(\rho_{\rm{n}}-\rho_{\rm{p}})/\rho$ the isospin asymmetry of the system, here $\rho_{\rm{n}}$ and $\rho_{\rm{p}}$ are the neutron and proton densities, respectively. Preserving the neutron-proton exchange symmetry, it is usually expanded in even powers of $\delta$ as\,\cite{LiBA2008}
\begin{equation}\label{eos}
E(\rho,\delta)\approx E_0(\rho)+E_{\rm{sym}}(\rho)\delta^2+E_{\rm{sym,4}}(\rho)\delta^4+\cdots,
\end{equation}
in terms of the energy per nucleon $E_0(\rho)\equiv E(\rho,0)$ in symmetric nuclear matter (SNM), the (quadratic) symmetry energy $E_{\rm{sym}}(\rho)\equiv 2^{-1}\partial^2E(\rho,\delta)/\partial\delta^2|_{\delta=0}$, the fourth-order (quartic) symmetry energy $E_{\rm{sym,4}}(\rho)\equiv {24}^{-1}\partial^4E(\rho,\delta)/\partial\delta^4|_{\delta=0}$, etc.
If we truncate the above expansion to order $\delta^2$, the resulting EOS is often called the parabolic approximation\,\cite{Bombaci1991}.
Moreover, the EOS of SNM, the symmetry energy as well as the  fourth-order symmetry energy could be further expanded around the saturation density $\rho_0$ as 
\begin{align}
E_0(\rho)\approx& E_0(\rho_0)+\frac{1}{2}K_0\chi^2+\frac{1}{6}J_0\chi^3+\cdots,~~\\
E_{\rm{sym}}(\rho)\approx &E_{\rm{sym}}(\rho_0)+L\chi+\frac{1}{2}K_{\rm{sym}}\chi^2+\frac{1}{6}J_{\rm{sym}}\chi^3+\cdots,\\
E_{\rm{sym,4}}(\rho)\approx &E_{\rm{sym,4}}(\rho_0)+L_{\rm{sym,4}}\chi+\frac{1}{2}K_{\rm{sym,4}}\chi^2+\frac{1}{6}J_{\rm{sym,4}}\chi^3+\cdots,
\end{align}
defining the incompressibility $K_0\equiv 9\rho_0^2\d^2E_0(\rho)/\d\rho^2|_{\rho=\rho_0}$ and skewness $J_0\equiv 27\rho_0^3\d^3E_0(\rho)/\d\rho^3|_{\rho=\rho_0}$ of the SNM, the slope coefficient $L\equiv3\rho_0\d E_{\rm{sym}}(\rho)/\d\rho|_{\rho=\rho_0}$, the curvature coefficient $K_{\rm{sym}}\equiv 9\rho_0^2\d^2E_{\rm{sym}}(\rho)/\d\rho^2|_{\rho=\rho_0}$ and the skewness coefficient $J_{\rm{sym}}\equiv 27\rho_0^3\d^3E_{\rm{sym}}(\rho)/\d\rho^3|_{\rho=\rho_0}$ of the symmetry energy, etc., here the dimensionless quantity $\chi$ is defined as $\chi=(\rho-\rho_0)/3\rho_0$.
The EOS of ANM provides an important and basic input for various applications in both nuclear physics and astrophysics\,\cite{LiBA2008,Steiner2005,Lattimer2007,Oertel2017,Burgio2021PPNP,Baiotti2019PPNP,
Maza2018PPNP,Kyutoku2021,Drischler2021}.

In this work, we explore the EOS of ANM in a space of general dimension $d$. The establishment of the general formalism is the first step to investigate possibly some novel features itself and to explore more interesting problems for the nuclear many-body system in dimensions different from three. The organization and the main conclusions of the work are outlined as follows: 

In section \ref{SEC_GF}, we establish the general formalism for the EOS of ANM including the characteristics $K_0,J_0$ and $L$, etc., via the Hugenholtz-Van Hove (HVH) theorem\,\cite{HVH1958}.
The HVH theorem is very useful for establishing the connection between the EOS of the system (which is a thermodynamic quantity) and the single particle energy (including the potential as well as the kinetic parts)\,\cite{XuC2010,XuC2011,ChenR2012,CaiBJ2012PLB,LiXH2013PLB,LiXH2015PLB}.
The results obtained from the HVH theorem are essentially model independent. Interesting features of the EOS of ANM may naturally emerge as we shift the focus from $d=3$ to certain other dimensions. 

In section \ref{SEC_2D}, we investigate the EOS of ANM in space with $d=2$ and discuss its qualitative properties. Here we find that the parabolic approximation for the kinetic EOS of ANM is exact, meaning all the higher-order kinetic symmetry energies including the quatic term are exactly zero.
The parabolic approximation in spaces with a general dimension $d$ is also investigated here and it is found to be naturally good considering the specific structure of the fourth-order symmetry energy.
Moreover, all the characteristic coefficients of the kinetic symmetry energy beyond the slope parameter $L$ are also exactly zero, i.e., the curvature $K_{\rm{sym}}$ as well as the skewness $J_{\rm{sym}}$ vanish.
The origin of these features could be traced back to the linear relation between the Fermi momentum $k_{\rm{F}}$ and the nucleon density $\rho$.

Section \ref{SEC_EXP} is denoted to the study on the $\epsilon$-expansion of the EOS of ANM based on the conventional dimension $d=3$. While we have no direct information about the EOS in low dimensions (e.g. 2D), the $\epsilon$-expansion gives us some indications about the qualitative characteristics of the 2D EOS. In particular, we find that the EOS of SNM will probably be reduced if $d$ is downwardly perturbed, showing that the EOS of SNM is more bounded in 2D. The symmetry energy in 2D, on other hand, strongly depends on the momentum dependence of the symmetry (isovector) potential. Its qualitative tendency in 2D needs more accurate determination of the aforementioned momentum dependence of the nucleon potential and/or higher-order contributions from the $\epsilon$ expansion.
In addition, the effectiveness of the $\epsilon$-expansion is demonstrated by comparing the exact kinetic EOS (which is obtained in section \ref{SEC_GF}) and the perturbative EOS.
Discussions in section \ref{SEC_EXP} show that the 3D EOS actually encapsulates relevant information on the EOS in $d_{\rm{f}}$D with $d_{\rm{f}}$ being near 3.

In section \ref{SEC_TOY} we further investigate the effects generated by varying the dimension $d$, adopting a toy model for the single-nucleon potential. Here the model nucleon potential is designed to effectively re-produce certain empirical knowledge on the 3D EOS\,\cite{LiBA2008}, such as the saturation density $\rho_0$, the binding energy $E_0(\rho_0)$ at the saturation density, the incompressibility coefficient $K_0$, the skewness $J_0$ of the SNM, the magnitude $E_{\rm{sym}}(\rho_0)$ and the slope coefficient $L$ of the symmetry energy at the saturation density $\rho_0$,  as well as the single-nucleon potential extracted from optical model fittings of nucleon-nucleus scattering data\,\cite{Hama1990}. Moreover, the pressure in SNM at densities around about $(1.5-4.5)\rho_0$ is also required to be consistent with the constraint from analyzing collective flow in relativisitc heavy-ion reactions. Then we find that the EOS of SNM as well as the symmetry energy are generally enhanced (reduced) as the dimension $d$ increases (decreases),  from the viewpoint of the $\epsilon$-expansion approach developed in section \ref{SEC_EXP}.
These results may find relevant applications in both astrophysics and nuclear physics.

In section \ref{SEC_HMT}, we introduce the short-range-correlations (SRC) induced high-momentum-tail (HMT) in the single-nucleon momentum distribution function\,\cite{Hen2014,Duer2018,Schmookler2019,Schmidt2020} into the kinetic EOS of ANM in dimension $d$. The SRC/HMT is known to be due to mostly the repulsive core in the central force and the tensor force in the neutron-proton isosinglet channel. Although we can only modify the kinetic part at this point, the SRC-induced HMT is obviously an effective representation of the nucleon-nucleon interactions, thus useful information may be obtained by investigating how the kinetic EOS may change as the $d$ varies with and without considering the SRC/HMT. Interestingly, we find that as $d$ decreases, the kinetic EOS with or without the SRC-induced HMT becomes close, indicating that the EOS in low dimensions can be treated as nearly free. The calculations of the relative nucleon momentum fluctuation also confirm this conclusion. This connection may find potential applications in the future to explore the 3D EOS of ANM from the low-dimensional counterpart, providing a possible alternative.

In section \ref{SEC_Esym2D-O2}, we perform the $\epsilon$-expansion
for the symmetry energy to order $\epsilon^2$, and analyze its main features. We find that the second-order contribution is generally negative. However, the prediction from the linear-order calculation may not change qualitatively, indicating that the perturbation is stable even when an $\epsilon=-1$ is applied.
Section \ref{SEC_SUM} is a summary of the present work, where we also suggest a few questions to stimulate further studies about the EOS of neutron-rich matter in $d$ dimensions.

\setcounter{equation}{0}
\section{General formalism}\label{SEC_GF}

\subsection{Review of the generalized HVH theorem in 3D and the EOS of neutron-rich matter}
In this section, we derive the general formalism for the EOS of ANM in $d$D space via the generalized HVH theorem\,\cite{HVH1958}.
We first briefly review the derivations in 3D and give the necessary ingredients\,\cite{XuC2010,XuC2011,ChenR2012,CaiBJ2012PLB}.
The single-nucleon potential $U_J(\rho,\delta,|\v{k}|)$ (where $J=\rm{n,p}$) could be expanded around $\delta=0$ as,
\begin{equation}\label{cck-1}
U_J(\rho,\delta,|\v{k}|)\approx U_0(\rho,|\v{k}|)+U_{\rm{sym}}(\rho,|\v{k}|)\tau_3^J\delta
+U_{\rm{sym,2}}(\rho,|\v{k}|)\delta^2+\cdots,
\end{equation}
where $\tau_3^{\rm{n}}=+1$ and $\tau_3^{\rm{p}}=-1$, and the symmetry potentials are defined as,
\begin{equation}\label{sa_1}
U_{\rm{sym},\ell}(\rho,|\v{k}|)=\left.
\frac{1}{\ell!}\frac{\d^\ell}{\d\delta^\ell}\frac{U_{\rm{n}}(\rho,\delta,|\v{k}|)+(-1)^{\ell}U_{\rm{p}}(\rho,\delta,|\v{k}|)}{2}\right|_{\delta=0}
.\end{equation} 
With $l$ being odd (even), one obtains the isovector (isoscalar) single-nucleon potentials. 
If only the first two terms on the right hand side (RHS) of (\ref{cck-1}) are kept, the corresponding approximation reduces to the well-known Lane's potential\,\cite{Lane1962}.

As mentioned in the {\sc Introduction}, the HVH theorem is independent of the nucleon-nucleon interactions adopted and should be satisfied by all nuclear many-body theories. It expresses the exact relation between the single-particle energy and the thermodynamic quantities in a many-body system.
The HVH theorem in the non-relativistic case for the SNM simply reads
$
P_0+\varepsilon_0=\mu_0\rho$,
where $P_0$ is the pressure of the SNM and $\varepsilon_0$ is the energy density, both are thermodynamic quantities, $\mu_0
$ is the nucleon chemical potential (single particle quantity). For SNM at saturation density $\rho_0$ where $P_0=0$, the HVH theorem is reduced to the familar relation $\mu_0=\varepsilon_0/\rho_0=E_0(\rho_0)$.
The power of the HVH theorem is established via the well-known relation between the Fermi momentum and the nucleon density, i.e., $
k_{\rm{F}}=(3\pi^2\rho/2)^{1/3}$ (in 3D).
According to the thermodynamic definition of the pressure in SNM at an arbitrary density $\rho$, i.e., $
P_0=\rho^2\partial
E_0/\partial\rho$,
one could obtain the following nucleon chemical potential
\begin{equation}\label{mu0}
 \mu_0=\rho\frac{\partial
E_0}{\partial\rho}+\frac{\varepsilon_0}{\rho}=\rho\frac{\partial(\varepsilon_0/\rho-M)}{\partial\rho}
+\frac{\varepsilon_0}{\rho}=\frac{\partial\varepsilon_0}{\partial\rho}.
\end{equation}
In fact, it is just the definition of the nucleon chemical potential in SNM.

For nucleons in ANM,  the non-relativistic nucleon chemical potential $\mu_J$ with $J$ denoting protons or neutrons is written as,
\begin{equation}
\mu_J(\rho,\delta,k_{\rm{F}}^J)={T}_J(k_{\rm{F}}^J)+{U_J}(\rho,\delta,k_{\rm{F}}^J),
\end{equation}
where $T_J(k_{\rm{F}}^J)$ is the nucleon kinetic energy and $k_{\rm{F}}^J=k_{\rm{F}}(1+\tau_3^J\delta)^{1/3}$ is the nucleon Fermi momentum.
Generalizing the relation (\ref{mu0}) to neutrons and protons leads to,
\begin{align}
{T}_{\rm{n}}(k_{\rm{F}}^{\rm{n}})+U_{\rm{n}}(\rho,\delta,k_{\rm{F}}^{\rm{n}})=\frac{\partial[\rho
E(\rho,\delta)]}{\partial\rho_{\rm{n}}}\label{4qHVHn},\\
{T}_{\rm{p}}(k_{\rm{F}}^{\rm{p}})+U_{\rm{p}}(\rho,\delta,k_{\rm{F}}^{\rm{p}})=\frac{\partial[\rho
E(\rho,\delta)]}{\partial\rho_{\rm{p}}}\label{4qHVHp}.
\end{align}
Subtracting Eq.\,(\ref{4qHVHn}) and Eq.\,(\ref{4qHVHp}) gives,
\begin{equation}\label{eq27}
{
[{T}_{\rm{n}}(k_{\rm{F}}^{\rm{n}})-{T}_{\rm{p}}(k_{\rm{F}}^{\rm{p}})]
+[U_{\rm{n}}(\rho,\delta,k_{\rm{F}}^{\rm{n}})-U_{\rm{p}}(\rho,\delta,k_{\rm{F}}^{\rm{p}})]
=\frac{\partial[\rho
E(\rho,\delta)]}{\partial\rho_{\rm{n}}}-\frac{\partial[\rho
E(\rho,\delta)]}{\partial\rho_{\rm{p}}},}
\end{equation}
 both sides could be expanded according to the isospin asymmetry $\delta$.
Since
\begin{equation}
\frac{\partial}{\partial \rho_J}=\frac{\partial \rho}{\partial
\rho_J}\frac{\partial}{\partial \rho}
+\frac{\partial\delta}{\partial
\rho_J}\frac{\partial}{\partial\delta}=\frac{\partial}{\partial
\rho}+\frac{2\tau_3^J\rho_{\overline{J}}}{\rho^2}\frac{\partial}{\partial\delta}
,\end{equation}
where $\rho_{\overline{\rm{p}}}=\rho_{\rm{n}}$, etc., one obtains
\begin{equation}
\frac{\partial(\rho E)}{\partial\rho_{\rm{n}}}-\frac{\partial(\rho
E)}{\partial\rho_\rm{p}}=\frac{2}{\rho}\frac{\partial(\rho
E)}{\partial\delta}=2\frac{\partial
E}{\partial\delta}.
\end{equation}
On the other hand, if one writes the EOS of ANM as $
E(\rho,\delta)=E_0(\rho)+E_{\rm{sym}}(\rho)\delta^2
+E_{\rm{sym,4}}(\rho)\delta^4+\cdots=E_0(\rho)+\sum_{i=1}E_{\rm{sym},2i}(\rho)\delta^{2i}$, then we obtain the following series in $\delta$,
\begin{equation}
\frac{\partial(\rho E)}{\partial\rho_{\rm{n}}}-\frac{\partial(\rho
E)}{\partial\rho_{\rm{p}}}=\sum_{i=1}4iE_{\rm{sym},2i}(\rho)\delta^{2i-1}\approx4E_{\rm{sym}}(\rho)\delta
+8E_{\rm{sym},4}(\rho)\delta^3+\cdots.
\end{equation}

The Eq. (\ref{eq27}) can then be rewritten as
\begin{equation}\label{4qDetEsym}
\sum_{J=\rm{n,p}}\tau_3^J\left[{T}_J(k_{\rm{F}}^J)+{U_J}(\rho,\delta,k_{\rm{F}}^J)\right]=\sum_{i=1}4iE_{\rm{sym},2i}(\rho)\delta^{2i-1}
,\end{equation} 
where $E_{\rm{sym},2}(\rho)\equiv
E_{\rm{sym}}(\rho)$ is the conventional (quadratic) symmetry energy.
It is obvious that in order to obtain the symmetry energy one needs to expand the single-nucleon potential and the kinetic energy as power series in $\delta$. They should be expanded to order $\delta^3$ if one wants to obtain the fourth-order symmetry energy, etc. In addition, if one subtracts the Eq.\,(\ref{4qHVHn}) and Eq.\,(\ref{4qHVHp}), one obtains the following equation determining the slope parameter of the symmetry energy at an arbitrary density $\rho$
\begin{equation}\label{4qDetLsym}
\sum_{J=\rm{n,p}}\left[{T}_J(k_{\rm{F}}^J)+{U_J}(\rho,\delta,k_{\rm{F}}^J)\right]=2\frac{\partial}{\partial\rho}\left[\rho
E_0(\rho) +\sum_{i=1}\rho E_{\rm{sym},2i}(\rho)\delta^{2i}\right].
\end{equation}
For example, in order to obtain the slope parameter of the conventional symmetry energy, $L(\rho)\equiv
L_{\rm{sym},2}(\rho)$, we need to expand the single-nucleon energy to order $\delta^2$. Moreover, the determination of the slope parameter of the fourth-order symmetry energy, i.e., $L_{\rm{sym},4}(\rho)
$, needs the expansion of the single-nucleon energy to order $\delta^4$, etc.
The zeroth-order of the Eq.\,(\ref{4qDetLsym}) gives the formula for determining the EOS of SNM.
Then, according to the terms proportional to $\delta$ in Eq.\,(\ref{4qDetEsym}) and the terms proportional to $\delta^2$ in Eq.\,(\ref{4qDetLsym}), one obtains the formulae determining the symmetry energy as well as its slope parameter as,
\begin{align}
E_{\rm{sym}}(\rho)=&\frac{1}{4}\times\mbox{coefficient of $\delta$ terms in expanding}\sum_{J=\rm{n,p}}\tau_3^J\left[{T}_J(k_{\rm{F}}^J)+{U_J}(\rho,\delta,k_{\rm{F}}^J)\right]\label{4qEsymF},\\
L(\rho)=&\frac{3}{2}\times\mbox{coefficient of $\delta^2$ terms in expanding}
\sum_{J=\rm{n,p}}\left[{T}_J(k_{\rm{F}}^J)+{U_J}(\rho,\delta,k_{\rm{F}}^J)\right]+3E_{\rm{sym}}(\rho)\label{4qLF}
,\end{align}
where the following relation is used when deriving the second relation,
\begin{align}
2\frac{\partial}{\partial\rho}[\rho
E_{\rm{sym}}(\rho)\delta^2]
&=2\left[E_{\rm{sym}}(\rho)\delta^2+\rho\frac{\partial
E_{\rm{sym}}(\rho)}{\partial\rho}\delta^2+2\rho
E_{\rm{sym}}(\rho)\delta\frac{\partial\delta}{\partial\rho} \right]
=\left[-2E_{\rm{sym}}(\rho)+\frac{2}{3}L(\rho)\right]\delta^2,\label{fffff_a}
\end{align}
similarly the fourth-order symmetry energy is given by
\begin{equation}
E_{\rm{sym},4}(\rho)=\frac{1}{8}\times\mbox{coefficient of $\delta^3$ terms in expanding}\sum_{J=\rm{n,p}}\tau_3^J\left[{T}_J(k_{\rm{F}}^J)+{U_J}(\rho,\delta,k_{\rm{F}}^J)\right].\label{4qEsym4F}
\end{equation}

\subsection{Generalization to $d$D space}

Now consider the EOS of ANM in $d$D space, the total nucleon number $A=N+Z$ could be written as,
\begin{equation}
A=2\cdot2\cdot\left(\frac{R}{2\pi}\right)^d\int_0^{k_{\rm{F}}}
\d^d\v{k}=4\left(\frac{R}{2\pi}\right)^d\int_0^{k_{\rm{F}}}\frac{d\pi^{d/2}k^{d-1}\d
k}{\Gamma(d/2+1)}=\frac{2^{2-d}dR^d\pi^{-d/2}k_{\rm{F}}^d}{\Gamma(d/2+1)},\end{equation} 
here the pre-factor $4=2\cdot2$ is taken for considering the isospin and spin degeneracy.
By introducing the volume $
V_d=R^d$, one obtains the relation between the nucleon density $\rho$ and the Fermi momentum $k_{\rm{F}}$ as,
\begin{equation}\label{def_kF_d}
k_{\rm{F}}=\left[\rho 2^{d-2}\pi^{d/2}\Gamma\left({d}/{2}+1\right)\right]^{1/d}\sim\rho^{1/d},
\end{equation}
here $[\rho]\sim[\rm{fm}^{-d}]$.
We use the notation $\rho$ as the density in $d$D in situations with no confusion, and when we discuss the perturbative expansion over the dimension $d$ as in section \ref{SEC_EXP},  the notation $\rho_d$ is adopted to denote the density in $d$D and there $\rho$ is the conventional 3D density (it should be clear from the context).
The expression of $k_{\rm{F}}$ provides the mathematical foundations of the effectiveness of the perturbative calculation based on the dimension $d$.
Very similarly, the Fermi momenta of neutrons and protons are given by
$
k_{\rm{F}}^{\rm{n}}=k_{\rm{F}}(1+\delta)^{1/d}$ and $
k_{\rm{F}}^{\rm{p}}=k_{\rm{F}}(1-\delta)^{1/d}$,
which could be written in a unified form,
\begin{equation}
k_{\rm{F}}^{J}=k_{\rm{F}}(1+\tau_3^J\delta)^{1/d}=\left[
\rho2^{d-2}\pi^{d/2}\Gamma\left(\frac{d}{2}+1\right)\left(1+\tau_3^J\delta\right)\right]^{1/d}
.\end{equation}
Mathematically if the dimension $d$ is very large and the Fermi momentum $k_{\rm{F}}$ is fixed at certain value, then the nucleon density $\rho=k_{\rm{F}}^d/2^{d-2}\pi^{d/2}\Gamma(d/2+1)$ decreases as $d$ increases and approaches zero if $d\to\infty$.
Moreover, most of the density $\rho$ is contained in an annulus of width $k_{\rm{F}}(1-1/d)$ near the Fermi surface.

For the convenience of later calculations, here we list a few relevant perturbative quantities and relations. For the expansion based on the isospin asymmetry $\delta$, we have
\begin{align}
(1+\tau_3^J\delta)^{m/d}\approx&1+\frac{m\tau_3^J\delta}{d}
+\frac{m}{2d}\left(\frac{m}{d}-1\right)\delta^2
+\frac{m\tau_3^J}{6d}\left(\frac{m}{d}-1\right)\left(\frac{m}{d}-2\right)\delta^3\notag\\
&+\frac{m}{24d}\left(\frac{m}{d}-1\right)\left(\frac{m}{d}-2\right)\left(\frac{m}{d}-3\right)\delta^4+\mathcal{O}(\delta^5).
\end{align}
The situation where $m=2$ is relevant for computing the kinetic EOS.
This expansion is very intuitive since all even-order terms in $\delta$ starting from $\delta^4$ contain the factor $(m/d-1)(m/d-2)$ which is identically zero for $d=1$ or $d=2$ (if $m=2$ is adopted).
This means the kinetic fourth-order symmetry energy and terms beyond in low dimensions are exactly zero.
Similarly for the perturbative expansion around the reference density $\rho_{\rm{f}}$, by introducing the small quantity,
\begin{equation}\label{sjk_1}
\theta_{\rm{f}}=\frac{\rho-\rho_{\rm{f}}}{3\rho_{\rm{f}}},\end{equation}
one obtains $\theta_{\rm{f}}=\theta_0\equiv\chi$ with $\chi$ being the conventional dimensionless quantity $(\rho-\rho_0)/3\rho_0$ for $\rho_{\rm{f}}=\rho_0$.
The Fermi momentum at density $\rho$ can be expanded in terms of that at the reference density ${\rho_{\rm{f}}}$ as
\begin{align}
k_{\rm{F}}(\rho)=&k_{\rm{F}}\left(\rho_{\rm{f}}+3\rho_{\rm{f}}\theta_{\rm{f}}\right)
\approx k_{\rm{F}}(\rho_{\rm{f}})+\left.\frac{\d
k_{\rm{F}}}{\d
\rho}\right|_{\rho=\rho_{\rm{f}}}\cdot3\rho_{\rm{f}}\theta_{\rm{f}}+\left.\frac{1}{2}\frac{\d^2k_{\rm{F}}}{
\d \rho^2}\right|_{\rho=\rho_{\rm{f}}}\cdot9\rho_{\rm{f}}^2\theta_{\rm{f}}^2+\mathcal{O}(\theta_{\rm{f}}^3)\notag\\
=&k_{\rm{F}}(\rho_{\rm{f}})+\frac{a}{d}\rho_{\rm{f}}^{1/d-1}\cdot3{\rho_{\rm{f}}}\theta_{\rm{f}}
+\frac{1}{2}\frac{a}{d}\left(\frac{1}{d}-1\right)\rho_{\rm{f}}^{1/d-2}\cdot9\rho_{\rm{f}}^2\theta_{\rm{f}}^2+\mathcal{O}(\theta_{\rm{f}}^3)\notag\\
=&k_{\rm{F}}(\rho_{\rm{f}})\times\left[1+\frac{3}{d}\theta_{\rm{f}}
+\frac{9}{2d}\left(\frac{1}{d}-1\right)\theta_{\rm{f}}^2\right]
+\mathcal{O}(\theta_{\rm{f}}^3),
\end{align}
where,
\begin{equation}\label{def_ad}
a=a(d)=\left[2^{d-2}\pi^{d/2}\Gamma\left(\frac{d}{2}+1\right)\right]^{1/d}
.\end{equation} 
Moreover, one has for the square of the Fermi momentum as,
\begin{equation}
k_{\rm{F}}^2(\rho)\approx
k_{\rm{F}}^2({\rho_{\rm{f}}})\times\left[1+\frac{6}{d}\theta_{\rm{f}}
+\frac{9}{d}\left(\frac{2}{d}-1\right)\theta_{\rm{f}}^2\right]
+\mathcal{O}(\theta_{\rm{f}}^3) .\end{equation}

In the remaining part of this subsection, we give the general expressions for the EOS of SNM $E_0(\rho)$, the symmetry energy $E_{\textmd{sym}}(\rho)$, the fourth-order symmetry energy $E_{\textmd{sym,4}}(\rho)$, the slope parameter $L(\rho)$ of the symmetry energy, the incompressibility coefficient $K_0(\rho)$ as well as the skewness $J_0(\rho)$ of the EOS of SNM as functions of density $\rho$ in dimensions $d$, in terms of the single-nucleon potential together with its isospin expansions.
The single-nucleon energy in $d$D still reads
$
e_J(\rho,\delta,|\v{k}|)={\v{k}^2}/{2M}+{U_J}(\rho,\delta,|\v{k}|)$,
with the first term the kinetic energy and the second term the potential energy. Considering the free Fermi gas (FFG) model, we have for the kinetic part of the EOS as\,\cite{CaiBJ2022},
\begin{align}
E^{\rm{kin}}(\rho,\delta)=\frac{d}{d+2}\frac{k_{\rm{F}}^2}{2M}\frac{1}{2}\left[(1+\delta)^{1+2/d}+(1-\delta)^{1+2/d}\right]
=\frac{d}{d+2}{E}_{\rm{F}}F_{1+2/d}(\delta),
\end{align}
where ${E}_{\rm{F}}=k_{\rm{F}}^2/2M$ is the nucleon Fermi energy, and in addition,
\begin{align}
F_{1+2/d}(\delta)=&\frac{1}{2}\left[(1+\delta)^{1+2/d}+(1-\delta)^{1+2/d}\right]
\approx1+\frac{d+2}{d^2}\delta^2
+\frac{(d^2-4)(d-1)}{6d^4}\delta^4+\mathcal{O}(\delta^6).
\end{align}
By expanding the $E^{\rm{kin}}(\rho,\delta)$ around $\delta=0$, we obtain
\begin{align}
E^{\rm{kin}}(\rho,\delta)\approx&\frac{d{E}_{\rm{F}}}{d+2}
\left[1+\frac{d+2}{d^2}\delta^2
+\frac{(d^2-4)(d-1)}{6d^4}\delta^4\right]
=\frac{d{E}_{\rm{F}}}{d+2}+\frac{{E}_{\rm{F}}}{d}\delta^2+\frac{d^2-3d+2}{6d^3}{E}_{\rm{F}}\delta^4
+\mathcal{O}(\delta^6) ,\end{align} and consequently
\begin{align}
E^{\rm{kin}}_0(\rho)=&\frac{d}{d+2}{E}_{\rm{F}},~~
E^{\rm{kin}}_{\rm{sym}}(\rho)=\frac{1}{d}{E}_{\rm{F}},~~
E^{\rm{kin}}_{\rm{sym,4}}(\rho)=\frac{d^2-3d+2}{6d^3}{E}_{\rm{F}}.
\end{align}

The contribution from the single-nucleon potential to the EOS of SNM is similarly given as
\begin{equation}
E^{\rm{pot}}_0(\rho)=\frac{1}{\rho}\int_0^{\rho}U_0\left(f,k_{\rm{F}}^f\right)\d
f,~~k_{\rm{F}}^f=\left[f2^{d-2}\pi^{d/2}\Gamma\left(\frac{d}{2}+1\right)\right]^{1/d},
\end{equation}
and by combining the kinetic part and the potential contribution, one has the general expression for the EOS of SNM,
\begin{align}\label{kk_E0}
E_0(\rho)=\frac{d{E}_{\rm{F}}}{d+2}+\frac{1}{\rho}\int_0^{\rho}U_0\left(f,k_{\rm{F}}^f\right)\d
f=\frac{d}{d+2}\frac{1}{2M}\left[\rho2^{d-2}\pi^{d/2}\Gamma\left(\frac{d}{2}+1\right)\right]^{
2/d}+\frac{1}{\rho}\int_0^{\rho}U_0\left(f,k_{\rm{F}}^f\right)\d f.
\end{align}

Next, the perturbative expansions for the kinetic energy and the single-nucleon potential ${U_J}(\rho,\delta,|\v{k}|)$ could be given.
Since our aim is to give the analytic expressions for quantities including the fourth-order symmetry energy, the relevant expansions on $\delta$ need to be kept at order $\delta^3$.
For the nucleon kinetic energy, one then has
\begin{align}
T_J(k_{\rm{F}}^J)\approx\frac{k_{\rm{F}}^2}{2M}+\frac{1}{d}\frac{k_{\rm{F}}^2}{M}\tau_3^J\delta
+\frac{1}{d}\left(\frac{2}{d}-1\right)\frac{k_{\rm{F}}^2}{2M}\delta^2
+\frac{1}{d}\left(\frac{2}{d}-1\right)\left(\frac{2}{d}-2\right)
\frac{k_{\rm{F}}^2}{6M}\tau_3^J\delta^3+\mathcal{O}(\delta^4)
.\end{align} The single-nucleon potential also needs to be expanded to order $\delta^3$, and when it is taken at the Fermi momentum $k_{\rm{F}}$, we obtain,
\begin{align}
U_0(\rho,k_{\rm{F}}^J)\approx& U_0(\rho,k_{\rm{F}})
+k_{\rm{F}}\left[\frac{\tau_3^J\delta}{d}+\frac{1}{2d}\left(\frac{1}{d}-1
\right)\delta^2+\frac{1}{6d}\left(\frac{1}{d}-1\right)
\left(\frac{1}{d}-2\right)\tau_3^J\delta^3\right]\left.\frac{\partial
U_0}{\partial|\v{k}|}\right|_{|\v{k}|=k_{\rm{F}}}\notag\\
&+\frac{1}{2}k_{\rm{F}}^2\left[\frac{\tau_3^J\delta}{d}+\frac{1}{2d}\left(\frac{1}{d}-1
\right)\delta^2\right]^2\left.\frac{\partial^2
U_0}{\partial|\v{k}|^2}\right|_{|\v{k}|=k_{\rm{F}}} +
\frac{1}{6}k_{\rm{F}}^3\left(\frac{\tau_3^J\delta}{d}\right)^3\left.\frac{\partial^3
U_0}{\partial|\v{k}|^3}\right|_{|\v{k}|=k_{\rm{F}}},
\end{align}
and similarly,
\begin{align} U_{\rm{sym}}(\rho,k_{\rm{F}}^J)\approx&
U_{\rm{sym}}(\rho,k_{\rm{F}})
+k_{\rm{F}}\left[\frac{\tau_3^J\delta}{d}+\frac{1}{2d}\left(\frac{1}{d}-1
\right)\delta^2\right]\left.\frac{\partial
U_{\rm{sym}}}{\partial|\v{k}|}\right|_{|\v{k}|=k_{\rm{F}}}
+\frac{1}{2}k_{\rm{F}}^2\left(\frac{\tau_3^J\delta}{d}\right)^2\left.\frac{\partial^2
U_{\rm{sym}}}{\partial|\v{k}|^2}\right|_{|\v{k}|=k_{\rm{F}}},\\
U_{\rm{sym},2}(\rho,k_{\rm{F}}^J)\approx&
U_{\rm{sym},2}(\rho,k_{\rm{F}})+\left.\frac{k_{\rm{F}}}{d}\frac{\partial
U_{\rm{sym},2}}{\partial|\v{k}|}\right|_{|\v{k}|=k_{\rm{F}}}\tau_3^J\delta,~~
U_{\rm{sym},3}(\rho,k_{\rm{F}}^J)\approx
U_{\rm{sym},3}(\rho,k_{\rm{F}})+\mathcal{O}(\delta^4).
\end{align}
Combining these contributions, one obtains the nucleon potential at the Fermi surface as
\begin{align}
{U_J}(\rho,\delta,k_{\rm{F}}^J)\approx&
U_0(\rho,k_{\rm{F}})+\tau_3^J\delta\left[\frac{k_{\rm{F}}}{d}\frac{\partial
U_0}{\partial|\v{k}|}+U_{\rm{sym}}(\rho,|\v{k}|)\right]_{|\v{k}|=k_{\rm{F}}}\notag\\
&+\delta^2\left[\frac{k_{\rm{F}}^2}{2d^2}\frac{\partial^2U_{0}}{\partial|\v{k}|^2}
+\frac{k_{\rm{F}}}{2d}\left(\frac{1}{d}-1\right)\frac{\partial
U_{0}}{\partial|\v{k}|} +\frac{k_{\rm{F}}}{d}\frac{\partial
U_{\rm{sym}}}{\partial|\v{k}|}+U_{\rm{sym},2}(\rho,|\v{k}|)\right]_{|\v{k}|=k_{\rm{F}}}\notag\\
&+\tau_3^J\delta^3\Bigg[\frac{k_{\rm{F}}^3}{6d^3}\frac{\partial^3U_{0}}{\partial|\v{k}|^3}
+\frac{k_{\rm{F}}^2}{2d^2}\left(\frac{1}{d}-1\right)\frac{\partial^2U_{0}}{\partial|\v{k}|^2}+\frac{k_{\rm{F}}}{6d}\left(\frac{1}{d}-1\right)
\left(\frac{1}{d}-2\right)\frac{\partial U_{0}}{\partial|\v{k}|}
+\frac{k_{\rm{F}}^2}{2d^2}\frac{\partial^2U_{\rm{sym}}}{\partial|\v{k}|^2}\notag\\
&\hspace*{2cm}-\frac{k_{\rm{F}}}{2d}\left(\frac{1}{d}-1\right)\frac{\partial
U_{\rm{sym}}}{\partial|\v{k}|}+\frac{k_{\rm{F}}}{d}\frac{\partial
U_{\rm{sym},2}}{\partial|\v{k}|}+U_{\rm{sym},3}(\rho,|\v{k}|)\Bigg]_{|\v{k}|=k_{\rm{F}}}.
\end{align}

According to the formulas derived earlier from the HVH theorem, i.e., relations (\ref{4qEsymF}), (\ref{4qLF}) and (\ref{4qEsym4F}), we then obtain the expressions for $E_{\textmd{sym}}(\rho),L(\rho)$, and $E_{\textmd{sym,4}}(\rho)$ as,
\begin{align}
E_{\rm{sym}}(\rho)=&\frac{1}{2d}\frac{k_{\rm{F}}^2}{M}+\left.\frac{k_{\rm{F}}}{2d}\frac{\partial
U_0}{\partial
|\v{k}|}\right|_{|\v{k}|=k_{\rm{F}}}+\frac{1}{2}U_{\rm{sym}}(\rho,k_{\rm{F}})\label{4qEsymFF_1},\\
L(\rho)=&\frac{3}{d^2}\frac{k_{\rm{F}}^2}{M}+\left[\frac{3k_{\rm{F}}^2}{2d^2}\frac{\partial^2
U_0}{\partial|\v{k}|^2}+\frac{3k_{\rm{F}}}{2d^2}\frac{\partial
U_0}{\partial|\v{k}|}\right]_{|\v{k}|=k_{\rm{F}}}+\left.\frac{3k_{\rm{F}}}{d}\frac{\partial
U_{\rm{sym}}}{\partial|\v{k}|}\right|_{|\v{k}|=k_{\rm{F}}}\notag\\
&+\frac{3}{2}U_{\rm{sym}}(\rho,k_{\rm{F}})+3U_{\rm{sym},2}(\rho,k_{\rm{F}})\label{4qLFF_1},\\
E_{\rm{sym},4}(\rho)=&\frac{1}{24d}\left(\frac{2}{d}-1\right)
\left(\frac{2}{d}-2\right)\frac{k_{\rm{F}}^2}{M}\notag\\
&+\left[\frac{k_{\rm{F}}^3}{24d^3}\frac{\partial^3U_{0}}{\partial|\v{k}|^3}
+\frac{k_{\rm{F}}^2}{8d^2}\left(\frac{1}{d}-1\right)\frac{\partial^2U_{0}}{\partial|\v{k}|^2}
+\frac{k_{\rm{F}}}{24d}\left(\frac{1}{d}-1\right)
\left(\frac{1}{d}-2\right)\frac{\partial
U_{0}}{\partial|\v{k}|}\right]_{|\v{k}|=k_{\rm{F}}}\notag\\
&+\left[\frac{k_{\rm{F}}^2}{8d^2}\frac{\partial^2U_{\rm{sym}}}{\partial|\v{k}|^2}
-\frac{k_{\rm{F}}}{8d}\left(\frac{1}{d}-1\right)\frac{\partial
U_{\rm{sym}}}{\partial|\v{k}|}\right]_{|\v{k}|=k_{\rm{F}}}
+\left.\frac{k_{\rm{F}}}{4d}\frac{\partial
U_{\rm{sym},2}}{\partial|\v{k}|}\right|_{|\v{k}|=k_{\rm{F}}}
+\frac{1}{4}U_{\rm{sym},3}(\rho,k_{\rm{F}})\label{4qEsym4FF_1}.
\end{align}
The symmetry energy could be cast into the following simple form,
\begin{equation}
E_{\rm{sym}}(\rho)=\frac{k_{\rm{F}}^2}{2dM_{\rm{s}}^{\ast}(\rho,k_{\rm{F}})}
+\frac{1}{2}U_{\rm{sym}}(\rho,k_{\rm{F}}),
\end{equation}
by introducing the scalar nucleon Landau effective mass $M_{\rm{s}}^{\ast}$ defined as\,\cite{LiBA2018PPNP},
\begin{equation}\label{f_landau}
M_{\rm{s}}^{\ast}(\rho,|\v{k}|)=M\left(1+\frac{M}{|\v{k}|}\frac{\partial
U_0}{\partial|\v{k}|}\right)^{-1}.
\end{equation}
Taking the momentum at the Fermi surface gives the effective mass as a function of density.

In order to derive the expressions for $K_0(\rho)$ and $J_0(\rho)$,
we need to use the zeroth-order equation of the HVH theorem, i.e., 
$P_0(\rho)=\rho[\mu_0(\rho)-E_0(\rho)]$, where
$\mu_0(\rho)={k_{\rm{F}}^2}/{2M}+U_0(\rho,k_{\rm{F}})$
is the nucleon chemical potential. Moreover, the $U_0$ appeared here depends only on the density $\rho$ (instead of also on the momentum), the zeroth-order equation could further be written in the following form by recasting the HVH theorem using
$P_0(\rho)=3^{-1}\rho L_0(\rho)$
\begin{equation}\label{fff}
\frac{1}{3}L_0(\rho)=\frac{k_{\rm{F}}^2}{2M}+U_0(\rho)-E_0(\rho)
,\end{equation} 
where $L_0(\rho)=3\rho\d E_0(\rho)/\d\rho$ which is identically zero at $\rho_0$ (definition of the saturation density $\rho_0$).
A few related relations are listed here for convenience,
\begin{align}
E_0(\rho)&=E_0({\rho_{\rm{f}}})+L_0({\rho_{\rm{f}}})\theta_{\rm{f}}+\frac{1}{2}K_0(\rho_{\rm{f}})\theta_{\rm{f}}^2,\\
L_0(\rho)&=L_0({\rho_{\rm{f}}})+\left.3{\rho_{\rm{f}}}\frac{\d
L_0(\rho)}{\d
\rho}\right|_{{\rho=\rho_{\rm{f}}}}\theta_{\rm{f}}+\left.\frac{9{\rho^2_{\rm{f}}}}{2}\frac{\d^2
L_0(\rho)}{\d \rho^2}\right|_{{\rho=\rho_{\rm{f}}}}\theta_{\rm{f}}^2,\\
\frac{\d L_0(\rho)}{\d \rho}&=3\frac{\d
E_0(\rho)}{\d \rho}+3\rho\frac{\d^2E_0(\rho)}{\d
\rho^2},~~ \frac{\d^2L_0(\rho)}{\d
\rho^2}=6\frac{\d^2E_0(\rho)}{\d
\rho^2}+3\rho\frac{\d^3E_0(\rho)}{\d \rho^3}.
\end{align}
By expanding the equation (\ref{fff}) around the reference density ${\rho_{\rm{f}}}$, one has
\begin{align}
\rm{LHS}\approx&\frac{1}{3}L_0({\rho_{\rm{f}}})+\theta_{\rm{f}}\left[L_0({\rho_{\rm{f}}})+\frac{1}{3}K_0({\rho_{\rm{f}}})\right]
+\theta_{\rm{f}}^2\left[K_0({\rho_{\rm{f}}})+\frac{1}{6}J_0({\rho_{\rm{f}}})\right],\\
\rm{RHS}\approx&\frac{k_{\rm{F}}^2({\rho_{\rm{f}}})}{2M}+U_0({\rho_{\rm{f}}})-E_0({\rho_{\rm{f}}})
+\theta_{\rm{f}}\left[\frac{3k_{\rm{F}}^2({\rho_{\rm{f}}})}{Md}+\left.3{\rho_{\rm{f}}}\frac{\d
U_0}{\d \rho}\right|_{{\rho=\rho_{\rm{f}}}}-L_0({\rho_{\rm{f}}})\right]\notag\\
&+\theta_{\rm{f}}^2\left[\frac{9k_{\rm{F}}^2({\rho_{\rm{f}}})}{2M}\frac{1}{d}\left(\frac{2}{d}-1\right)
+\left.\frac{9{\rho_{\rm{f}}^2}}{2}\frac{\d^2U_0}{\d
\rho^2}\right|_{{\rho=\rho_{\rm{f}}}}-\frac{1}{2}K_0({\rho_{\rm{f}}})\right],
\end{align}
where the total-derivative of the $U_0$ with respect to the density $\rho$ is understood here as $\d/\d\rho=\partial/\partial\rho+(\d k_{\rm{F}}/\d\rho)\cdot\partial/\partial k_{\rm{F}}$, by comparing terms at the same order and removing the subscript ``f'', we have
\begin{align}
K_0(\rho)=&\frac{9k_{\rm{F}}^2}{Md}+9\rho\frac{\d U_0}{\d
\rho}-6L_0(\rho),\\
J_0(\rho)=&\frac{27k_{\rm{F}}^2}{M}\frac{1}{d}\left(\frac{2}{d}-1\right)
+27\rho^2\frac{\d^2U_0}{\d \rho^2}-9K_0(\rho)\notag\\
=&\frac{54k_{\rm{F}}^2}{M}\frac{1}{d}\left(\frac{1}{d}-2\right)
+27\rho^2\frac{\d^2U_0}{\d \rho^2}-81\rho\frac{\d U_0}{\d
\rho}+54L_0(\rho),
\end{align}
where
\begin{align}
L_0(\rho)=&3\rho\frac{\d E_0(\rho)}{\d
\rho}=\frac{3k_{\rm{F}}^2}{(d+2)M}+3\rho\frac{\d
E_0^{\rm{pot}}(\rho)}{\d \rho}\notag\\
=&\frac{3k_{\rm{F}}^2}{(d+2)M}-\frac{3}{\rho}\int_0^\rho
U_0\left(f,k_{\rm{F}}^f\right)\d
f+3U_0(\rho,k_{\rm{F}})\notag\\
=&\frac{3k_{\rm{F}}^2}{(d+2)M}-\frac{3}{\rho}\int_0^\rho
U_0\left(f,k_{\rm{F}}^f\right)\d
f+3U_0(\rho).\label{deff_L0}
\end{align}
Thus the $K_0$ and $J_0$ could also be rewritten in the following forms as (where $U(\rho)=U(\rho,k_{\rm{F}})$),
\begin{align}
K_0(\rho)=&-\frac{1}{d}\frac{d-2}{d+2}\frac{9k_{\rm{F}}^2}{M}
+9\rho\frac{\d U_0}{\d \rho}+\frac{18}{\rho}\int_0^\rho
U_0\left(f,k_{\rm{F}}^f\right)\d
f-18U_0(\rho),\label{kk_K0}\\
J_0(\rho)=&\frac{(d-1)(d-2)}{d^2(d+2)}\frac{54k_{\rm{F}}^2}{M}
+27\rho^2\frac{\d^2U_0}{\d \rho^2}-81\rho\frac{\d U_0}{\d \rho}
-\frac{162}{\rho}\int_0^\rho U_0\left(f,k_{\rm{F}}^f\right)\d
f+162U_0(\rho).\label{kk_J0}
\end{align}

\subsection{A relationship between the isovector $U_{\rm{sym}}(\rho,|\v{k}|)$ and isoscalar $U_{\rm{sym},2}(\rho,|\v{k}|)$ potentials}

Using the HVH theorem as shown above, we have derived separate expressions for the symmetry energy $E_{\rm{sym}}(\rho)$ and its slope parameter $L(\rho)$ in terms of the isoscalar and isovector components of the single-nucleon potential. On the other hand, the basic definition of $L(\rho)$ as the density derivative of $E_{\rm{sym}}(\rho)$ also gives an expression for $L(\rho)$ once the $E_{\rm{sym}}(\rho)$ is obtained from the HVH theorem. Thus, when equating these two expressions for $L(\rho)$, an identity follows.
This identity connects the strength of the (isovector) symmetry potential $U_{\rm{sym}}$ with that of the (isoscalar) second-order potential $U_{\rm{sym},2}$ in $d$D as,
\begin{equation}
U_{\rm{sym},2}(\rho,k_{\rm{F}})=\frac{1}{2}\left(\rho\frac{\partial
U_{\rm{sym}}(\rho,|\v{k}|)}{\partial
\rho}-U_{\rm{sym}}(\rho,|\v{k}|)\right)_{|\v{k}|=k_{\rm{F}}}
-\frac{k_{\rm{F}}}{2d}\left(\frac{\partial
U_{\rm{sym}}(\rho,|\v{k}|)}{\partial|{\v{k}}|}-\rho\frac{\partial^2U_0(\rho,|\v{k}|)}{\partial
\rho\partial
|{\v{k}}|}\right)_{|{\v{k}}|=k_{\rm{F}}}.
\label{U2-U1}
\end{equation}
This relation (\ref{U2-U1}) holds for any single-nucleon potential at any density, i.e., it is model-independent, which can be used to check the consistency of a given $U_J$ or act as a guidance in constructing phenomenological single-nucleon potentials.
Using the relation between the density $\rho$ and the Fermi momentum $k_{\rm{F}}$ in $d$D, i.e., $k_{\rm{F}}\sim\rho^{1/d}$ or $\rho\d k_{\rm{F}}/\d\rho=k_{\rm{F}}/d$, we can rewrite (\ref{U2-U1}) in the following form,
\begin{equation}\label{U2-U1-kk}
U_{\rm{sym},2}(\rho,k_{\rm{F}})
=\frac{k_{\rm{F}}}{2d}\left[
\left(\frac{\partial }{\partial k_{\rm{F}}}
-\frac{\partial }{\partial |\v{k}|}-\frac{d}{k_{\rm{F}}}\right)U_{\rm{sym}}(\rho,|\v{k}|)+\frac{k_{\rm{F}}}{d}\frac{\partial^2U_0(\rho,|\v{k}|)}{\partial k_{\rm{F}}\partial|\v{k}|}
\right]_{|\v{k}|=k_{\rm{F}}}.
\end{equation}
If the single nucleon potential $U_J$ has no momentum dependence, then (\ref{U2-U1}) is reduced to 
\begin{equation}
U_{\rm{sym},2}(\rho)=\frac{1}{2}\left(\rho\frac{\partial
U_{\rm{sym}}(\rho)}{\partial
\rho}-U_{\rm{sym}}(\rho)\right).
\label{U2-U1-nomom}
\end{equation}
Interestingly, it is independent of the spacial dimension $d$.

As a basic application of the relation (\ref{U2-U1}), we now check if the Skyrme pseudo-potential\,\cite{WangR2018} fulfills the constraint given above. For example, the $D^{[4]}$-term in the single-nucleon potential\,\cite{WangR2018} reads as 
\begin{equation}
\frac{D^{[4]}}{16}\left[\frac{k_{\rm{F}}^3|\v{k}|^4}{3\pi^2}\cdot(1+\tau_3^J\delta)
+\frac{2k_{\rm{F}}^5|\v{k}|^2}{3\pi^2}\cdot(1+\tau_3^J\delta)^{5/3}+\frac{k_{\rm{F}}^7}{7\pi^2}\cdot(1+\tau_3^J\delta)^{7/3}
\right],
\end{equation}
which depends on both the density $k_{\rm{F}}$ and the momentum $|\v{k}|$.
The symmetry potential and the second-order potential could be obtained straightforwardly by expanding the above expression around $\delta=0$. Putting them into the relation (\ref{U2-U1}) with $d=3$, both sides readily give the result $D^{[4]}k_{\rm{F}}^7/27\pi^2$, i.e., $U_{\rm{sym,2}}(\rho,k_{\rm{F}})=D^{[4]}k_{\rm{F}}^7/27\pi^2$.
Other momentum- and density-dependent terms of the Skyrme-like potential of Ref.\,\cite{WangR2018} could be checked similarly. Another potential application of (\ref{U2-U1}) is to construct consistent single-nucleon potentials.
For instance, consider a simple nucleon potential depending on the density alone as $U_J(\rho)=[\alpha+\alpha_{\rm{IS}}f( \Gamma_{\vartheta}\tau_3^J\delta)]\widetilde{\rho}^{\vartheta}$ where $\widetilde{\rho}=\rho/\rho_0$ is the reduced density and $\Gamma_{\vartheta}$ is a factor needed to fulfill the identity (\ref{U2-U1-nomom}).
When using the relation (\ref{U2-U1-nomom}) the isospin-independent term $\alpha$ plays no role and only the isospin-dependent term represented by the $\alpha_{\rm{IS}}$ is relevant.
Consequently, we have $U_{\rm{sym}}(\rho)=\alpha_{\rm{IS}}\Gamma_{\vartheta}f'(0)\widetilde{\rho}^{\vartheta}$ where the derivative of $f$ is taken with respect to the isospin asymmetry $\delta$, and similarly $U_{\rm{sym},2}(\rho)=2^{-1}\alpha_{\rm{IS}}\Gamma_{\vartheta}^2f''(0)\widetilde{\rho}^{\vartheta}$.
Putting the $U_{\rm{sym}}(\rho)$ and $U_{\rm{sym,2}}(\rho)$ into (\ref{U2-U1-nomom}) gives the relation,
\begin{equation}\label{Gammay}
\Gamma_{\vartheta}=\frac{f'(0)}{f''(0)}\cdot({\vartheta}-1).
\end{equation}
This formula excludes the linear function of $\Gamma_{\vartheta}\tau_3^J\delta$ since the second-order derivative in the denominator is required to be non-zero. For instance, if $f(\Gamma_{\vartheta}\tau_3^J\delta)=\exp(\Gamma_{\vartheta}\tau_3^J\delta)$, then the $\Gamma_{\vartheta}$ is given as $\Gamma_{\vartheta}={\vartheta}-1$, i.e., the single-nucleon potential having the form $[\alpha+\alpha_{\rm{IS}}e^{(\vartheta-1)\tau_3^J\delta}]\widetilde{\rho}^{\vartheta}$ fulfills the constraint from the HVH theorem.
Moreover if $f(x)=ax+bx^2$ with $a,b>0$ and $x=\Gamma_{\vartheta}\tau_3^J\delta$, then the $\Gamma_{\vartheta}$ factor is given by $\Gamma_{\vartheta}=a({\vartheta}-1)/2b$, simply using the relation (\ref{Gammay}). Similarly, the potential $\sim\widetilde{\rho}^{\vartheta}[1-(\vartheta-1)\tau_3^J\delta/2]^{-1}$ fulfills the requirement of Eq. (\ref{U2-U1-nomom}).
There is no doubt that if the single-nucleon potential is momentum-dependent, the corresponding requirement by the HVH theorem will become more complicated. In these situations, one needs to adopt the more general relation (\ref{U2-U1}).

\setcounter{equation}{0}
\section{The two-dimensional nucleonic EOS}\label{SEC_2D}
In this section, we discuss the EOS of ANM in 2D.
In 2D, the Fermi momentum $k_{\rm{F}}$ is determined via the density $\rho$ as $
k_{\rm{F}}=(\pi \rho)^{1/2}$, see the general formula (\ref{def_kF_d}), thus the dimensions of density and Fermi momentum scale as $[\rho]=[\rm{fm}]^{-2}=[\rm{MeV}]^2$. The corresponding formulas for the EOS of ANM are given as
\begin{align}
E_0(\rho)=&\frac{\pi\rho}{4M}+\frac{1}{\rho}\int_0^{\rho}U_0\left(f,k_{\rm{F}}^f\right)\d
f=\frac{\pi \rho}{4M}+\frac{1}{\rho}\int_0^\rho
U_0\left(f,k_{\rm{F}}^f\right)\d
f,~~k_{\rm{F}}^f=(\pi f)^{1/2},\label{dd_E0_2}\\
E_{\rm{sym}}(\rho)=&\frac{\pi \rho}{4M}+\left.\frac{\sqrt{\pi
\rho}}{4}\frac{\partial U_0}{\partial
|\v{k}|}\right|_{|\v{k}|=k_{\rm{F}}}+\frac{1}{2}U_{\rm{sym}}(\rho,k_{\rm{F}})\label{dd_Esym_2},\\
E_{\rm{sym},4}(\rho)=&\frac{k_{\rm{F}}}{64}\left(
\frac{\partial
U_{0}}{\partial|\v{k}|}-k_{\rm{F}}\frac{\partial^2U_{0}}{\partial|\v{k}|^2}
+\frac{k_{\rm{F}}^2}{3}\frac{\partial^3U_{0}}{\partial|\v{k}|^3}
\right)_{|\v{k}|=k_{\rm{F}}}+\frac{k_{\rm{F}}}{32}\left(\frac{\partial
U_{\rm{sym}}}{\partial|\v{k}|}+k_{\rm{F}}\frac{\partial^2U_{\rm{sym}}}{\partial|\v{k}|^2}\right)_{|\v{k}|=k_{\rm{F}}}\notag\\
&+\left.\frac{k_{\rm{F}}}{8}\frac{\partial
U_{\rm{sym},2}}{\partial|\v{k}|}\right|_{|\v{k}|=k_{\rm{F}}}
+\frac{1}{4}U_{\rm{sym},3}(\rho,k_{\rm{F}})\label{dd_Esym_4},\\
P_0(\rho)=&\frac{3\pi \rho^2}{4M}-\int_0^\rho
U_0\left(f,k_{\rm{F}}^f\right)\d f+\rho
U_0(\rho,k_{\rm{F}}),\label{dd_p0_2}\\
K_0(\rho)=&9\rho\frac{\d U_0}{\d \rho}+\frac{18}{\rho}\int_0^\rho
U_0\left(f,k_{\rm{F}}^f\right)\d
f-18U_0(\rho,k_{\rm{F}}),\label{dd_K0}\\
J_0(\rho)=&27\rho^2\frac{\d^2U_0}{\d \rho^2}-81\rho\frac{\d U_0}{\d
\rho} -\frac{162}{\rho}\int_0^\rho U_0\left(f,k_{\rm{F}}^f\right)\d
f
+162U_0(\rho,k_{\rm{F}}),\label{dd_J0}\\
L(\rho)=&\frac{3\pi \rho}{4M}+\frac{3\pi
\rho}{8}\left[\frac{\partial^2
U_0}{\partial|\v{k}|^2}+\frac{1}{\sqrt{\pi \rho}}\frac{\partial
U_0}{\partial|\v{k}|}\right]_{|\v{k}|=k_{\rm{F}}}+\left.\sqrt{\pi
\rho}\frac{\partial
U_{\rm{sym}}}{\partial|\v{k}|}\right|_{|\v{k}|=k_{\rm{F}}}\notag\\
&+\frac{3}{2}U_{\rm{sym}}(\rho,k_{\rm{F}})+3U_{\rm{sym},2}(\rho,k_{\rm{F}})\label{dd_L},\\
K^{\rm{SNM}}(\rho)=&9\frac{\partial P_0(\rho)}{\partial\rho}=\frac{9\pi \rho}{2M}+9\rho\frac{\d U_0}{\d
\rho}.\label{dd_KSNM}
\end{align}
The coefficient $K^{\rm{SNM}}(\rho)$ defined above is generally different from $K_0(\rho)=9\rho^2\d^2E_0(\rho)/\d\rho^2$, and the relation between them is $K^{\rm{SNM}}(\rho)=K_0(\rho)+6L_0(\rho)$. Thus only at the saturation density $\rho_0$ these two coefficients are equal $K^{\rm{SNM}}(\rho_0)=K_0(\rho_0)\equiv K_0$ due to the vanishing of the pressure $P_0(\rho_0)=L_0(\rho_0)\rho_0/3$.

All the kinetic parts of the characteristic coefficients beyond the linear terms $L_0(\rho)$ and $L(\rho)$ are zero (including the coefficients $K_0(\rho),J_0(\rho)$ and $K_{\rm{sym}}(\rho)$), i.e., the EOS are totally represented by the effective potentials. Furthermore, the higher-order kinetic symmetry energies including the $E_{\rm{sym,4}}(\rho)$ are also zero (the sixth-order term is given by $E_{\rm{sym},6}^{\rm{kin}}(\rho)=(3d-2)(d-1)(d-2)(2d-1)k_{\rm{F}}^2/180d^5M$) which vanishes with $d=2$, indicating that the conventional parabolic approximation of the kinetic EOS of ANM is exact in 2D.
Since the potential contribution to the EOS is essentially smaller than its kinetic part (especially at low densities), the absence of the kinetic part in the $E_{\rm{sym,4}}(\rho)$ already indicates the parabolic approximation should be good in 2D.

Since the Fermi momentum $k_{\rm{F}}$ in dimensions 2 scales with the density as $\rho^{1/2}$, the kinetic symmetry energy is linear in density. 
If the single-nucleon potential $U_0$ depending on the momentum $k$ via the manner $U_0\approx a+bk+ck^2+\cdots$ and $U_{\rm{sym}}(\rho)\approx a'+b'k_{\rm{F}}+c'k_{\rm{F}}^2$, then $\partial U_0/\partial k\approx b+2ck+\cdots$ (and similarly for the symmetry potential), thus the density dependence of the $E_0(\rho)$ or the symmetry energy is given as
\begin{equation}
\sim f_1\sqrt{\rho}+f_2\rho+f_3\rho^{3/2}+\cdots,
\end{equation}
i.e., the effective expansion of the $E_0(\rho)$ or the symmetry energy on density is based on $\rho^{1/2}$, which is different from the situation in dimensions 3 where the expansion element is $k_{\rm{F}}\sim\rho^{1/3}$.
As an example, assume that the $U_0$ is a function of density $\rho$ alone (without momentum dependence), in the polynomial form as,
\begin{equation}
U_0(\rho)=\sum_{i=1}c_ik_{\rm{F}}^i,~~k_{\rm{F}}=\sqrt{\pi\rho},
\end{equation}
consequently,
\begin{align}
\frac{\d U_0}{\d\rho}=&\frac{\d U_0}{\d k_{\rm{F}}}\frac{\d k_{\rm{F}}}{\d\rho}=\frac{1}{2}\sqrt{\frac{\pi}{\rho}}\frac{\d U_0}{\d k_{\rm{F}}}
=\frac{1}{2}\sqrt{\frac{\pi}{\rho}}\frac{1}{k_{\rm{F}}}\sum_{i=1}c_iik_{\rm{F}}^i=\frac{1}{2\rho}\sum_{i=1}ic_ik_{\rm{F}}^i,\\
\frac{1}{\rho}\int_0^{\rho}U_0(f)\d f=&\frac{1}{\rho}
\sum_{i=1}c_i\pi^{i/2}\int_0^{\rho}f^{i/2}\d f=
\frac{1}{\rho}\sum_{i=1}c_i\pi^{i/2}\frac{1}{i/2+1}\rho^{i/2+1}=2\sum_{i=1}\frac{c_i}{i+2}k_{\rm{F}}^i.
\end{align}
The EOS of SNM is then expressed as,
\begin{equation}
E_0(\rho)=\frac{\pi\rho}{4M}+2\sum_{i=1}\frac{c_i}{i+2}k_{\rm{F}}^i,
\end{equation}
while the pressure and incompressibility are respectively given by
\begin{equation}
P_0(\rho)=\rho\left(\frac{3\pi\rho}{4M}+\sum_{i=1}\frac{i}{i+2}c_ik_{\rm{F}}^i\right),~~K_0(\rho)
=9\sum_{i=1}\frac{i}{2}\frac{i-2}{i+2} c_ik_{\rm{F}}^i.
\end{equation}
If the density expansion contains only two effective terms, i.e., $c_1,c_2$, we then have
\begin{equation}
E_0(\rho)=\frac{2}{3}c_1\sqrt{\pi\rho}+\left(\frac{1}{4M}+\frac{c_2}{2}\right)\pi\rho,~~P_0(\rho)/\rho
=\frac{1}{3}c_1\sqrt{\pi\rho}+\left(\frac{3}{4M}+\frac{c_2}{2}\right)\pi\rho,~~K_0(\rho)=-\frac{3}{2}c_1\sqrt{\pi\rho}.
\end{equation}

\begin{figure}[h!]
\centering
\includegraphics[width=6.5cm]{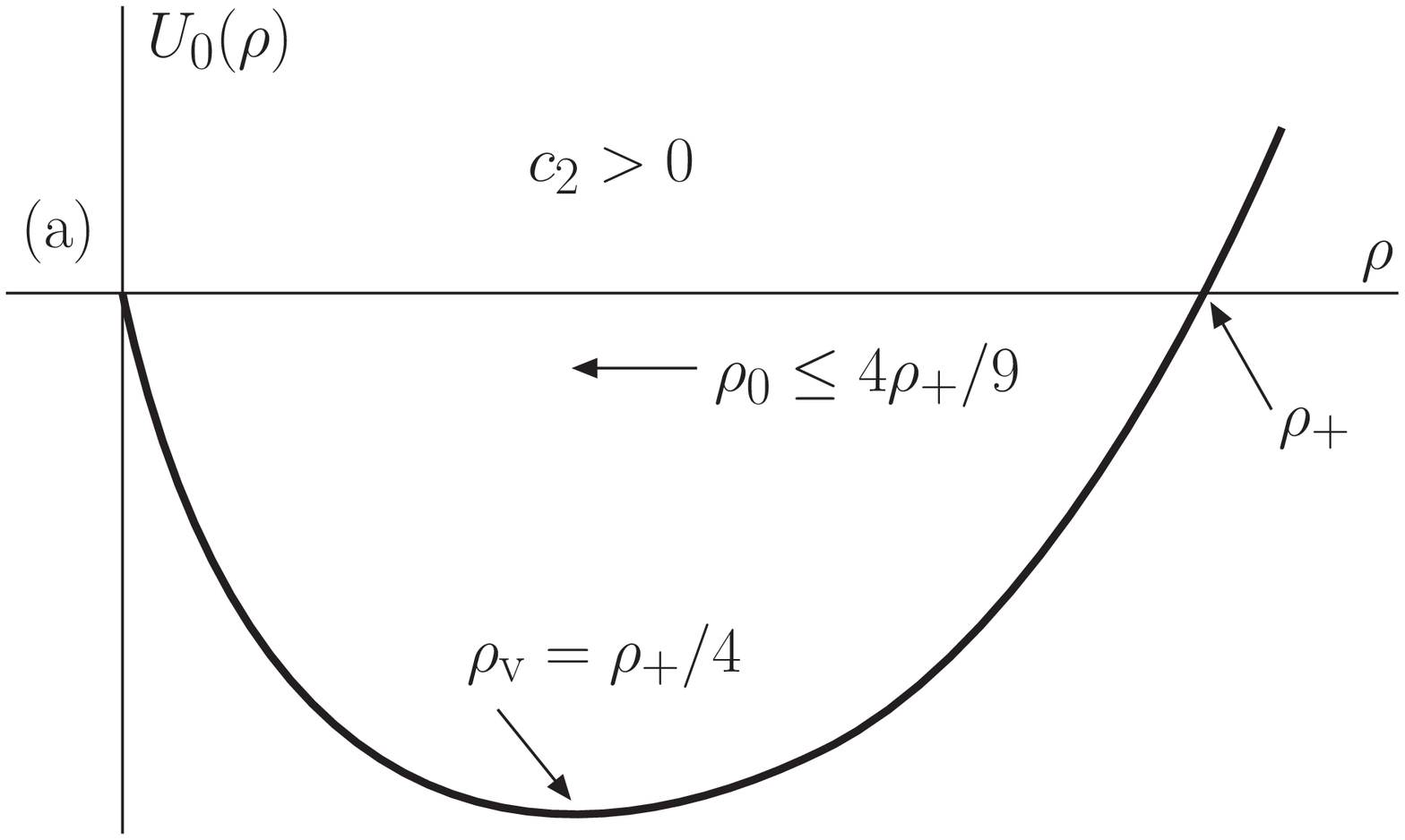}\qquad
\includegraphics[width=6.5cm]{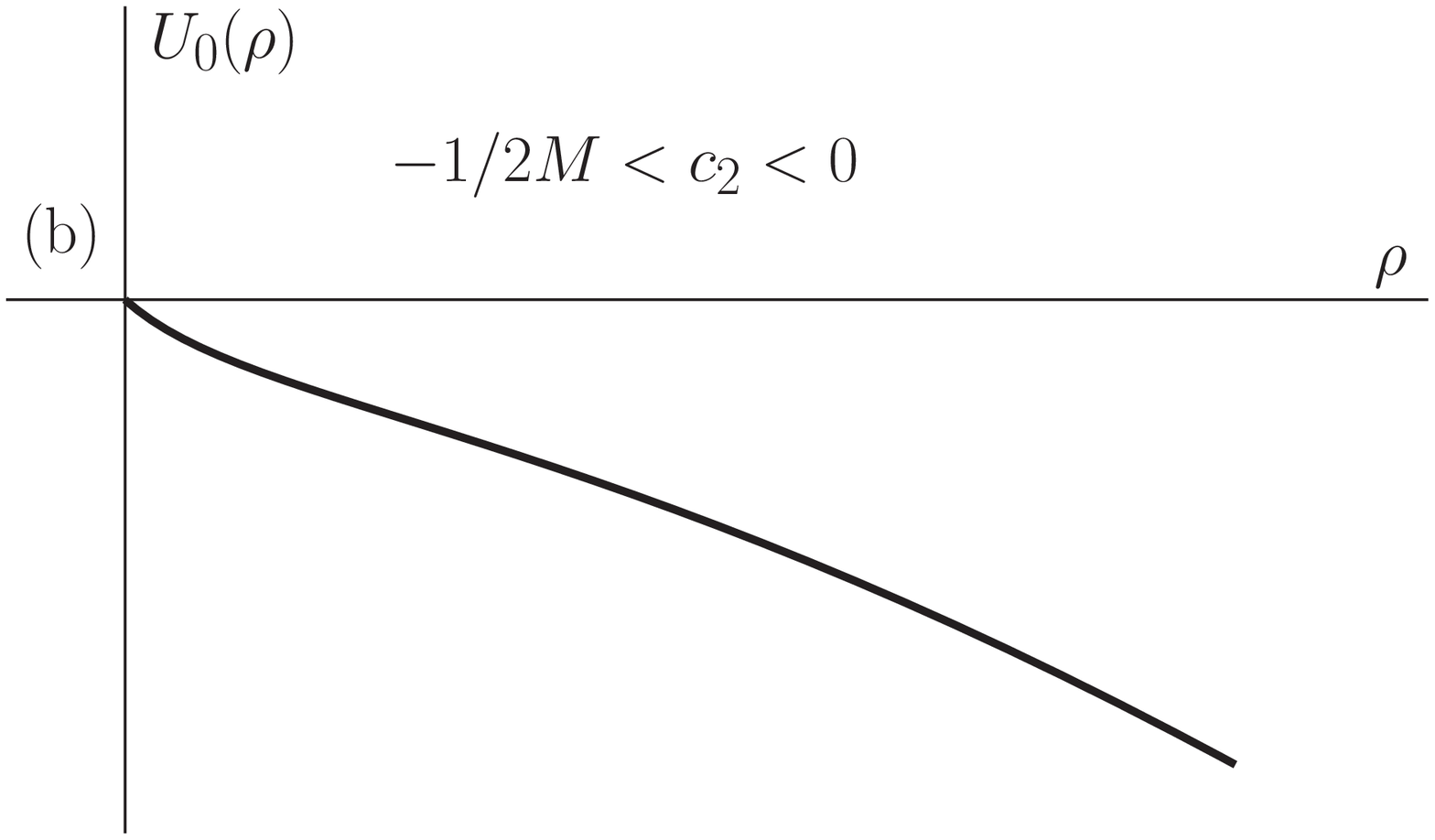}
\caption{Sketch of the single-nucleon potential $U_0(\rho)$ as a function of $\rho$ in 2D.}\label{fig_U0rho_p}
\end{figure}

Assuming the nuclear system in dimensions 2 behaves similarly as the one in dimensions 3 (at least qualitatively), i.e., the $E_0(\rho)$ is negative at low densities and eventually increases to be positive at densities larger than the saturation density $\rho_0$ (corresponding to the vanishing of the pressure), we than have the condition that 
\begin{equation}
c_1<0,
\end{equation}
 and similarly $1/4M+c_2/2>0$ as well as $3/4M+c_2/2>0$,  or 
$
 c_2>-{1}/{2M}$. The first condition namely $c_1<0$ could also be obtained from the expression for $K_0(\rho)$, since there would be a positive value for the $K_0$ at the saturation density $\rho_0$.
Moreover, a $\sqrt{\rho}$-dependence of the $K_0(\rho)$ is also predicted under the above assumptions. 
Furthermore, if one assumes that the potential $U_0$ is not always negative as density $\rho$ increases, the second-order coefficient $c_2$ should be greater than zero, i.e.,
$
c_2>0$.
Under this assumption, we can define three densities: The saturation density $\rho_0$ is defined as the point where the pressure vanishes, i.e., $P_0(\rho_0)=0$, the density $\rho_{\rm{v}}$ corresponding to the minimum of the potential $U_0$ is defined via $\d U_0/\d\rho|_{\rho=\rho_{\rm{v}}}=0$, and finally the crossing density $\rho_+$ above which the single-nucleon potential $U_0$ turns positive, $U_0(\rho_+)=0$. We have the expressions for these three densities as,
\begin{equation}
\sqrt{\pi\rho_0}=-\frac{c_1}{3}\cdot\frac{1}{3/4M+c_2/2},~~\sqrt{\pi\rho_{\rm{v}}}=-\frac{c_1}{2c_2},~~\sqrt{\pi\rho_+}
=-\frac{c_1}{c_2}.
\end{equation}
Consequently, we obtain,
\begin{equation}
\rho_+/\rho_{\rm{v}}=4,~~\rho_+/\rho_0=\frac{9}{4}\left(1+\frac{3}{2Mc_2}\right)^2\geq\frac{9}{4}.
\end{equation}
These relations show that at low densities, the potential $U_0$ is attractive, as the $\rho$ increases the potential eventually decrease to reach its minimum at $\rho_{\rm{v}}$ which could either be larger than $\rho_0$ or smaller than $\rho_0$ depending on the coefficient $c_2$ since $\rho_{\rm{v}}/\rho_0=(9/16)\cdot(1+3/2Mc_2)^2$, as the density increases even further above $\rho_{\rm{v}}$, the potential starts to increases and finally becomes repulsive when crossing $\rho_+$.
See the left panel of Fig.\,\ref{fig_U0rho_p} for a sketch of the density dependence of $U_0$.
On the other hand, if one assumes that $-1/2M<c_2<0$, the potential $U_0$ always decreases as $\rho$ increases, i.e., the attractive interaction the nucleons feel becomes deeper and deeper, see the right panel of Fig.\,\ref{fig_U0rho_p}.
The density behavior of the skewness $J_0$ could be analyzed in a similar manner. Further justifications on the $U_0$ as well as possible constraints on the coefficient $c_i$ need a detailed model of the nucleon potential.
For example, when the third-order term $c_3k_{\rm{F}}^3$ is included in the potential $U_0$, the qualitative analysis should be correspondingly modified.

\renewcommand*\figurename{\small Fig.}

We can also analyze the strength of the symmetry energy and the fourth-order symmetry energy, starting from the formulas (\ref{dd_Esym_2}) and (\ref{dd_Esym_4}). Here we assume that the single-nucleon potential has a sizable first-order symmetry potential $U_{\rm{sym}}$ but all the terms beyond the linear order are small, i.e., $U_{\rm{sym,2}}\approx U_{\rm{sym,3}}\approx\cdots\approx0$. In addition, the momentum-dependence of the isospin-dependent part of the potential is also assumed to be weak, then we can approximate (\ref{dd_Esym_2}) and (\ref{dd_Esym_4}) by,
\begin{equation}
E_{\rm{sym}}(\rho)\approx\frac{k_{\rm{F}}^2}{4M}+\left.\frac{k_{\rm{F}}}{4}\frac{\partial U_0}{\partial|\v{k}|}\right|_{|\v{k}|=k_{\rm{F}}}
+\frac{1}{2}U_{\rm{sym}},~~E_{\rm{sym,4}}(\rho)\approx\left.\frac{k_{\rm{F}}}{64}\frac{\partial U_0}{\partial|\v{k}|}\right|_{|\v{k}|=k_{\rm{F}}}.
\end{equation}
In 3D, the scalar nucleon effective mass is about $M_{\rm{s}}^{\ast}(\rho_0)\gtrsim0.8M$\,\cite{LiBA2018PPNP} and the symmetry potential (decreasing with increasing nucleon kinetic energy) has a maximum value of about $U_{\rm{sym}}\approx36.7\,\rm{MeV}$ at zero nucleon kinetic energy\,\cite{LiXH2013PLB,LiXH2015PLB}, one can estimate the ratio between $E_{\rm{sym},4}(\rho)$ and $E_{\rm{sym}}(\rho)$ if these two constraints holds approximately in 2D (since $k_{\rm{F}}^2/4M\approx18.4\,\rm{MeV}\approx2^{-1}U_{\rm{sym}}$ and $M/M_{\rm{s}}^{\ast}\approx5/4$),
\begin{equation}
\frac{E_{\rm{sym,4}}(\rho_0)}{E_{\rm{sym}}(\rho_0)}
\approx\left.\frac{k_{\rm{F}}^2}{64}\left(\frac{1}{M_{\rm{s}}^{\ast}}-\frac{1}{M}\right)\right/\left(\frac{k_{\rm{F}}^2}{4M}+\frac{1}{2}U_{\rm{sym}}\right)
\lesssim\frac{1}{32}\left(\frac{M}{M_{\rm{s}}^{\ast}}-1\right)\lesssim\frac{1}{128},
\end{equation}
indicating the parabolic approximation for the EOS of ANM including the potential part may still be reasonable (or even better) in 2D. Of course, as we will discuss in the next section, the quantities in 3D can be perturbed to give estimates of their correspondences in 2D. Indeed, we find qualitatively similar results.

Similarly, for the 1D situation, the symmetry energy and the fourth-order symmetry energy are given via (\ref{4qEsymFF_1}) and (\ref{4qEsym4FF_1}) as,
\begin{align}
E_{\rm{sym}}(\rho)
=&\frac{k_{\rm{F}}^2}{2M}+\left.\frac{k_{\rm{F}}^2}{2}\frac{\partial U_0}{\partial |\v{k}|}\right|_{|\v{k}|=k_{\rm{F}}}+\frac{1}{2}U_{\rm{sym}},\\
E_{\rm{sym,4}}(\rho)
=&\left.\frac{k_{\rm{F}}^3}{24}\frac{\partial^3U_0}{\partial |\v{k}|^3}\right|_{|\v{k}|=k_{\rm{F}}}+\left.\frac{k_{\rm{F}}^2}{8}\frac{\partial^2 U_{\rm{sym}}}{\partial |\v{k}|^2}\right|_{|\v{k}|=k_{\rm{F}}}
+\left.\frac{k_{\rm{F}}}{4}\frac{\partial U_{\rm{sym,2}}}{\partial |\v{k}|}\right|_{|\v{k}|=k_{\rm{F}}}+\frac{1}{4}U_{\rm{sym,3}},
\end{align}
where $k_{\rm{F}}=\pi\rho/4$.
Under the similar assumption that higher order symmetry potentials as well as their momentum dependence are weak, now the fourth-order symmetry energy could even be approximated as zero, i.e., the parabolic approximation for the EOS of ANM in 1D behaves even better.

Finally, in the large-$d$ limit, due to the pre-factors like $1/24d,1/8d^2$ and $1/24d^3$, etc., as shown from (\ref{4qEsym4FF_1}), the fourth-order symmetry energy is still expected to be small, and if the term $2^{-1}U_{\rm{sym}}$ in the symmetry energy is assumed to be larger than the corresponding term $4^{-1}U_{\rm{sym,3}}$ in the fourth-order symmetry energy\,\cite{LiXH2013PLB}, i.e., comparing the first-order term with the third-order term of the expansion $U_J$ over $\delta$, the ratio ${E_{\rm{sym,4}}(\rho_0)}/{E_{\rm{sym}}(\rho_0)}$ stays small, indicating that the parabolic approximation holds again in the large-$d$ limit. In the infinite-$d$ limit, we have then ${E_{\rm{sym}}(\rho_0)}/{E_{\rm{sym,4}}(\rho_0)}=U_{\rm{sym,3}}/2U_{\rm{sym}}\ll1$.
Writing out the EOS of ANM in a general dimension $d$ thus gives us extra insight into the strength of different expansion terms, and from this viewpoint, the goodness of the conventional parabolic approximation in 3D has no magic origin\,\cite{CaiBJ2022}.

\setcounter{equation}{0}
\section{The $\epsilon$-expansion of the EOS in $d\rm{D}$ space based on the EOS with $d_{\rm{f}}=3$}\label{SEC_EXP}

We have in this work encountered up to now two types of perturbative calculations, one is based on the isospin asymmetry $\delta$ and the other is based on the expansion around some reference density $\rho_{\rm{f}}$ through the quantity $\theta_{\rm{f}}$ (see (\ref{sjk_1})).
These two expansions both give us useful/informative information on the EOS of ANM, e.g., as we discussed in the last section, the $\delta^2$-expansion generally leads to the parabolic approximation, i.e., the fourth-order symmetry energy and beyond are smaller than the quadratic symmetry energy.
In addition, via the HVH theorem, a novel expansion related to the $\delta$ was developed, leading to the useful expressions for the symmetry energy as well as its slope parameter in terms of the single-nucleon potentials, with the latter being extractable from experiments\,\cite{LiXH2013PLB,LiXH2015PLB}.
If one treats the dimension $d$ as a continuous variable, one can also develop corresponding perturbation theories based on certain (fixed) dimension $d_{\rm{f}}$ by introducing the (apparently) small quantity $\epsilon=d-d_{\rm{f}}$. This approach is often called the $\epsilon$-expansion theory. Some problems are well defined and easier to analyze in certain dimensions, while they are difficult or even impossible to study in other dimensions.
The $\epsilon$-expansion was invented in the 1970's to deal with the second-order phase transition problems and related critical phenomena with impressive successes\,\cite{Wilson1974,MaSK1976,Wallace1976}. A recent successful example of the $\epsilon$-expansion is in studying the EOS of a unitary Fermi gas. The latter could simply be written as $\xi\cdot(3E_{\rm{F}}/5)$ where $E_{\rm{F}}=k_{\rm{F}}^2/2M$ is the Fermi energy with $M$ being the mass of the atoms.
Here the universal dimensionless quantity $\xi$ characterizes the interacting nature of the unitary gas, and it is often called the Bertsch's parameter\,\cite{Randeria2012,Bulgac2005}.
Due to the infinite scattering length $a_{\rm{s}}$ near the unitary limit, there are currently no self-consistent theories on the BEC-BCS crossover physics\,\cite{Bloch2008,Giorgini2008,Strinati2018} and thus lacks direct theories on the Bertsch's parameter.
The effective theories based on the $\epsilon$-expansion to order $\epsilon^{5/2}\ln\epsilon$\,\cite{Nishida2006} and later to higher orders in $\epsilon$ successfully predicted the value of $\xi\approx0.367$\,\cite{Arnold2007}, which was found to be very close to the experimental result of about 0.376\,\cite{Ku2012}.

In this section, we expand the EOS of ANM near the conventional 3D situation by introducing the perturbative dimension $\epsilon=d-3$ and investigate the general features when a nonzero $\epsilon$ is applied.
Then we shall show that the $\epsilon$-expansion is effective (e.g., to 2D) at least for the kinetic EOS of ANM by comparing the perturbed kinetic EOS with the exact one given in section \ref{SEC_2D}.
In the next section, we shall apply the $\epsilon$-expansion method to calculate the EOS of ANM adopting a toy-model nucleon potential and explore how the EOS in $d$D space modifies as the $d$ varies from 3 to other values.
In section \ref{SEC_HMT}, we adopt the $\epsilon$-expansion to calculate the kinetic EOS of ANM when the SRC-induced HMT in the single-nucleon momentum distribution is considered. In section \ref{SEC_Esym2D-O2}, we then perform the expansion to order $\epsilon^2$ for the symmetry energy, and investigate the corresponding features from the high order corrections.
The discussions of this section and the relevant parts of section \ref{SEC_TOY} and section \ref{SEC_HMT} on the $\epsilon$-expansion are mostly intuitive and probably preliminary. Our main motivation here is to demonstrate the interesting features of the $\epsilon$-expansion method and to stimulate further studies on this topic.

In order to give the EOS of ANM in dimension $d=d_{\rm{f}}+\epsilon=3+\epsilon$ with $\epsilon$ being the perturbative dimension, we need to expand the relevant quantities assuming $\epsilon$ is small.
The expansion of the function $a(d)$ (defined in (\ref{def_ad})) around $\epsilon=0$ is 
\begin{align}
a(d)=&a(3+\epsilon)=\left[2^{1+\epsilon}\pi^{3/2+\epsilon/2}\Gamma\left(\frac{\epsilon}{2}+\frac{5}{2}\right)\right]^{{1}/({3+\epsilon})}
\approx\left(\frac{3\pi^2}{2}\right)^{1/3}\left(1+\sigma
\epsilon\right)+\mathcal{O}(\epsilon^2),\end{align} where,
\begin{equation}\label{defff_sig}
\sigma=\frac{4}{9}-\frac{\gamma_{\rm{E}}}{6}+\frac{1}{18}\ln\left(\frac{4}{9\pi}\right)
\approx0.2396,\end{equation}
where $\gamma_{\rm{E}}\approx0.5773$ is Euler's constant.
Here we do not make perturbative expansion on the density $\rho$, and the only perturbative contribution comes from the function $a(d)$ by writing the $d$ as $3+\epsilon$.
Thus the Fermi momentum $k_{\rm{F}}$ in $3+\epsilon$ dimensions to linear order of $\epsilon$ is \begin{equation}
k_{\rm{F}}=\rho^{1/3}a(d)\approx \overline{k}_{\rm{F}}(1+\sigma
\epsilon)+\mathcal{O}(\epsilon^2),
\end{equation}
where $
\overline{k}_{\rm{F}}=({3\pi^2\rho}/{2})^{1/3}$ is the conventional 3D Fermi momentum (indicated by the ``$\bar{~~}$'' over the quantity).

The expansion of $U_0$ is
\begin{equation}
U_0(\rho,k_{\rm{F}})=U_0\left(\rho,\overline{k}_{\rm{F}}+\overline{k}_{\rm{F}}\sigma
\epsilon\right)\approx
U_0\left(\rho,\overline{k}_{\rm{F}}\right)+\left.\frac{\partial
U_0}{\partial|\v{k}|}\right|_{|\v{k}|=\overline{k}_{\rm{F}}}\cdot\overline{k}_{\rm{F}}\sigma
\epsilon,\end{equation}
to linear order in $\epsilon$, thus
\begin{align}
\frac{1}{\rho}\int_0^\rho U_0\left(f,k_{\rm{F}}^f\right)\d f \approx&
\frac{1}{\rho}\int_0^\rho
U_0\left(f,\overline{k}_{\rm{F}}^f\right)\d f +\epsilon
\left[\frac{\sigma}{\rho}\int_0^{\rho}\left( \left.\frac{\partial
U_0}{\partial|\v{k}|}\right|_{|\v{k}|=\overline{k}^f_{\rm{F}}}\cdot\overline{k}^f_{\rm{F}}\right)\d
f\right],\end{align} 
where $
\overline{k}_{\rm{F}}^f=\left({3\pi^2f}/{2}\right)^{1/3}$. Similarly,
\begin{align}
U_{\rm{sym}}(\rho,k_{\rm{F}})&\approx
U_{\rm{sym}}\left(\rho,\overline{k}_{\rm{F}}\right)+\left.\frac{\partial
U_{\rm{sym}}}{\partial|\v{k}|}\right|_{|\v{k}|=\overline{k}_{\rm{F}}}\cdot\overline{k}_{\rm{F}}\sigma
\epsilon,\\
U_{\rm{sym,2}}(\rho,k_{\rm{F}})&\approx
U_{\rm{sym,2}}\left(\rho,\overline{k}_{\rm{F}}\right)+\left.\frac{\partial
U_{\rm{sym,2}}}{\partial|\v{k}|}\right|_{|\v{k}|=\overline{k}_{\rm{F}}}\cdot\overline{k}_{\rm{F}}\sigma
\epsilon.
\end{align}
The first-order derivatives of the $U_0$ and $U_{\rm{sym}}$ with respect to $|\v{k}|$ could be obtained as
\begin{align}
\left.\frac{\partial
U_0}{\partial|\v{k}|}\right|_{|\v{k}|=k_{\rm{F}}}&\approx
\left(\frac{\partial
U_0}{\partial|\v{k}|}
+\frac{\partial^2
U_0}{\partial|\v{k}|^2}\cdot\overline{k}_{\rm{F}}\sigma
\epsilon\right)_{|\v{k}|=\overline{k}_{\rm{F}}},\\
\left.\frac{\partial
U_{\rm{sym}}}{\partial|\v{k}|}\right|_{|\v{k}|=k_{\rm{F}}}
&\approx\left(\frac{\partial
U_{\rm{sym}}}{\partial|\v{k}|}
+\frac{\partial^2
U_{\rm{sym}}}{\partial|\v{k}|^2}\cdot\overline{k}_{\rm{F}}\sigma
\epsilon\right)_{|\v{k}|=\overline{k}_{\rm{F}}}.
\end{align}
Very similarly, the second-order derivative of the $U_0$ is given by,
\begin{equation}
\left.\frac{\partial^2
U_0}{\partial|\v{k}|^2}\right|_{|\v{k}|=k_{\rm{F}}}\approx
\left(\frac{\partial^2
U_0}{\partial|\v{k}|^2}+\frac{\partial^3
U_0}{\partial|\v{k}|^3}\cdot\overline{k}_{\rm{F}}\sigma
\epsilon\right)_{|\v{k}|=\overline{k}_{\rm{F}}}.
\end{equation}

The total derivative of $U_0$ with respect to the density is thus obtained as,
\begin{align}
\frac{\d U_0}{\d \rho}=&\frac{\partial U_0}{\partial \rho} +
\left.\frac{\partial
U_0}{\partial|\v{k}|}\right|_{|\v{k}|=k_{\rm{F}}}\cdot\frac{\partial
k_{\rm{F}}}{\partial \rho}
\approx\frac{\partial U_0}{\partial \rho} +(1+\sigma
\epsilon)\cdot\left(\frac{\partial
U_0}{\partial|\v{k}|}+\frac{\partial^2
U_0}{\partial|\v{k}|^2}\cdot\overline{k}_{\rm{F}}\sigma
\epsilon\right)_{|\v{k}|=\overline{k}_{\rm{F}}}\cdot\frac{\partial \overline{k}_{\rm{F}}}{\partial
\rho}\notag\\
=&\frac{\partial U_0}{\partial \rho}+
\left.\frac{\overline{k}_{\rm{F}}}{3\rho}\frac{\partial
U_0}{\partial|\v{k}|}\right|_{|\v{k}|=\overline{k}_{\rm{F}}}
\times\left[1+\sigma
\epsilon\left(1+\overline{k}_{\rm{F}}\cdot\left.\frac{\partial
U_0^2}{\partial|\v{k}|^2}\right/\frac{\partial
U_0}{\partial|\v{k}|}\right)_{|\v{k}|=\overline{k}_{\rm{F}}}\right]\notag\\
=&\frac{\partial U_0}{\partial \rho}+
\left.\frac{\overline{k}_{\rm{F}}}{3\rho} \frac{\partial
U_0}{\partial|\v{k}|}\right|_{|\v{k}|=\overline{k}_{\rm{F}}}
+\epsilon\sigma\frac{\overline{k}_{\rm{F}}}{3\rho}\left[\frac{\partial
U_0}{\partial|\v{k}|}+\frac{\partial^2
U_0}{\partial|\v{k}|^2}\cdot\overline{k}_{\rm{F}}\right]_{|\v{k}|=\overline{k}_{\rm{F}}},
\end{align}
and the total second-order derivative with respect to $\rho$ is,
\begin{align}
\frac{\d^2U_0}{\d \rho^2}=&\frac{\d}{\d \rho}\left[\frac{\partial
U_0}{\partial \rho} + \left.\frac{\partial
U_0}{\partial|\v{k}|}\right|_{|\v{k}|=k_{\rm{F}}}\cdot\frac{\partial
k_{\rm{F}}}{\partial \rho}\right] =\frac{\d }{\d \rho}\frac{\partial
U_0}{\partial \rho} +\frac{\d}{\d \rho}\left[\left.\frac{\partial
U_0}{\partial|\v{k}|}\right|_{|\v{k}|=k_{\rm{F}}}\cdot\frac{\partial
k_{\rm{F}}}{\partial \rho}\right]\notag\\
=&\frac{\partial^2U_0}{\partial \rho^2}
+\left.\frac{\partial}{\partial|\v{k}|}\frac{\partial U_0}{\partial
\rho}\right|_{|\v{k}|=k_{\rm{F}}}\cdot\frac{\partial
k_{\rm{F}}}{\partial \rho} +\frac{\partial }{\partial
\rho}\left[\left.\frac{\partial
U_0}{\partial|\v{k}|}\right|_{|\v{k}|=k_{\rm{F}}}\cdot\frac{\partial
k_{\rm{F}}}{\partial \rho}\right]
+\frac{\partial}{\partial|\v{k}|}\left[\left.\frac{\partial
U_0}{\partial|\v{k}|}\right|_{|\v{k}|=k_{\rm{F}}}\cdot\frac{\partial
k_{\rm{F}}}{\partial
\rho}\right]_{|\v{k}|=k_{\rm{F}}}\cdot\frac{\partial
k_{\rm{F}}}{\partial \rho}\notag\\
=&\frac{\partial^2U_0}{\partial
\rho^2}+\left.2\frac{\partial^2U_0}{\partial \rho\partial
|\v{k}|}\right|_{|\v{k}|=k_{\rm{F}}}\cdot\frac{\partial
k_{\rm{F}}}{\partial \rho} +\left.\frac{\partial U_0}{\partial
|\v{k}|}\right|_{|\v{k}|=k_{\rm{F}}}\cdot\frac{\partial^2
k_{\rm{F}}}{\partial \rho^2} +\left.\frac{\partial^2U_0}{\partial
|\v{k}|^2}\right|_{|\v{k}|=k_{\rm{F}}}\cdot\left(\frac{\partial
k_{\rm{F}}}{\partial \rho}\right)^2.
\end{align}
By expanding the terms around $\overline{k}_{\rm{F}}$ and using the relations \begin{equation}
{\partial\overline{k}_{\rm{F}}}/{\partial
\rho}={\overline{k}_{\rm{F}}}/{3\rho},~~{\partial^2
\overline{k}_{\rm{F}}}/{\partial
\rho^2}=-{2\overline{k}_{\rm{F}}}/{9\rho^2},
\end{equation}
 one has
\begin{align}
\frac{\d^2U_0}{\d \rho^2}=&\frac{\partial^2U_0}{\partial
\rho^2}+\left.\frac{2\overline{k}_{\rm{F}}}{3\rho}\frac{\partial^2U_0}{\partial
\rho\partial
|\v{k}|}\right|_{|\v{k}|=\overline{k}_{\rm{F}}}-\left.\frac{2\overline{k}_{\rm{F}}}{9\rho^2}\frac{\partial
U_0}{\partial |\v{k}|}\right|_{|\v{k}|=\overline{k}_{\rm{F}}}
+\left.\frac{\overline{k}_{\rm{F}}^2}{9\rho^2}\frac{\partial^2U_0}{\partial
|\v{k}|^2}\right|_{|\v{k}|=\overline{k}_{\rm{F}}}\notag\\
&+\epsilon\sigma\left\{\frac{2\overline{k}_{\rm{F}}}{3\rho}\frac{\partial^2U_0}{\partial
\rho\partial
|\v{k}|}-\frac{2\overline{k}_{\rm{F}}}{9\rho^2}\frac{\partial
U_0}{\partial |\v{k}|}
+\frac{2\overline{k}_{\rm{F}}^2}{9\rho^2}\frac{\partial^2U_0}{\partial
|\v{k}|^2}+\overline{k}_{\rm{F}}\left[
\frac{2\overline{k}_{\rm{F}}}{3\rho}\frac{\partial^3U_0}{\partial
\rho\partial
|\v{k}|^2}-\frac{2\overline{k}_{\rm{F}}}{9\rho^2}\frac{\partial^2
U_0}{\partial |\v{k}|^2}
+\frac{\overline{k}_{\rm{F}}^2}{9\rho^2}\frac{\partial^3U_0}{\partial
|\v{k}|^3}\right]\right\}_{|\v{k}|=\overline{k}_{\rm{F}}},\notag\\
=&\frac{\partial^2U_0}{\partial
\rho^2}+\left.\frac{2\overline{k}_{\rm{F}}}{3\rho}\frac{\partial^2U_0}{\partial
\rho\partial
|\v{k}|}\right|_{|\v{k}|=\overline{k}_{\rm{F}}}-\left.\frac{2\overline{k}_{\rm{F}}}{9\rho^2}\frac{\partial
U_0}{\partial |\v{k}|}\right|_{|\v{k}|=\overline{k}_{\rm{F}}}
+\left.\frac{\overline{k}_{\rm{F}}^2}{9\rho^2}\frac{\partial^2U_0}{\partial
|\v{k}|^2}\right|_{|\v{k}|=\overline{k}_{\rm{F}}}\notag\\
&+\epsilon\sigma\left[\frac{2\overline{k}_{\rm{F}}}{3\rho}\frac{\partial^2U_0}{\partial
\rho\partial
|\v{k}|}-\frac{2\overline{k}_{\rm{F}}}{9\rho^2}\frac{\partial
U_0}{\partial |\v{k}|}+
\frac{2\overline{k}_{\rm{F}}^2}{3\rho}\frac{\partial^3U_0}{\partial
\rho\partial
|\v{k}|^2}+\frac{\overline{k}_{\rm{F}}^3}{9\rho^2}\frac{\partial^3U_0}{\partial
|\v{k}|^3}\right]_{|\v{k}|=\overline{k}_{\rm{F}}}.
\end{align}

We now have the relevant expansions to finally derive the EOS of ANM.
In particular, based on the formula (\ref{kk_E0}) for the EOS of SNM $E_0(\rho)$, (\ref{4qEsymFF_1}) for the symmetry energy $E_{\rm{sym}}(\rho)$, (\ref{deff_L0}) for $L_0(\rho)$ and consequently the pressure $P_0(\rho)=L_0(\rho)\rho/3$, (\ref{kk_K0}) for the incompressibility coefficient ($K_0(\rho)$ and $K^{\rm{SNM}}(\rho)=K_0(\rho)+6L_0(\rho)$), (\ref{kk_J0}) for the skewness $J_0(\rho)$ of the SNM, and (\ref{4qLFF_1}) for the slope parameter $L(\rho)$ of the symmetry energy, we write out the EOS of ANM and the related quantities in dimension $3+\epsilon$ to linear order of the perturbative dimension $\epsilon$ as,
\begin{align}
{E}_0(\rho)=&\frac{3\overline{k}_{\rm{F}}^2}{10M}+
\frac{1}{\rho}\int_0^\rho
U_0\left(f,\overline{k}_{\rm{F}}^f\right)\d
f+\epsilon
\left[\frac{3\overline{k}_{\rm{F}}^2}{5M}\left(\sigma+\frac{1}{15}\right)+\frac{\sigma}{\rho}\int_0^{\rho}\left(
\left.\frac{\partial
U_0}{\partial|\v{k}|}\right|_{|\v{k}|=\overline{k}^f_{\rm{F}}}\cdot\overline{k}^f_{\rm{F}}\right)\d
f\right],\label{ddef_E0}\\
E_{\rm{sym}}(\rho)=&\frac{\overline{k}_{\rm{F}}^2}{6M}+\left.\frac{\overline{k}_{\rm{F}}}{6}
\frac{\partial U_0}{\partial
|\v{k}|}\right|_{|\v{k}|=\overline{k}_{\rm{F}}}+\frac{1}{2}U_{\rm{sym}}(\rho,\overline{k}_{\rm{F}})\notag\\
&+\epsilon\left[\frac{\overline{k}_{\rm{F}}^2}{6M}\left(2\sigma-\frac{1}{3}\right)
+\frac{\overline{k}_{\rm{F}}}{6}\left({\displaystyle\sigma\overline{k}_{\rm{F}}\frac{\partial^2
U_0}{\partial
|\v{k}|^2} }+\left(\sigma-\frac{1}{3}\right)\frac{\partial
U_0}{\partial
|\v{k}|}\right)
+\frac{\sigma}{2}\frac{\partial U_{\rm{sym}}}{\partial
|\v{k}|}\cdot\overline{k}_{\rm{F}}\right]_{|\v{k}|=\overline{k}_{\rm{F}}},\label{ddef_Esym}\\
P_0(\rho)=&\frac{\overline{k}_{\rm{F}}^2\rho}{5M}+
\int_0^\rho U_0\left(f,\overline{k}_{\rm{F}}^f\right)\d
f+\rho U_0(\rho,\overline{k}_{\rm{F}})\notag\\
&+\epsilon\left[\frac{\rho\overline{k}_{\rm{F}}^2}{10M}\left(2\sigma-\frac{1}{5}\right)
-\sigma\int_0^{\rho}\left( \left.\frac{\partial
U_0}{\partial|\v{k}|}\right|_{|\v{k}|=\overline{k}^f_{\rm{F}}}\cdot\overline{k}^f_{\rm{F}}\right)\d
f+\rho\sigma\left.\frac{\partial U_0}{\partial
|\v{k}|}\right|_{|\v{k}|=\overline{k}_{\rm{F}}}\cdot\overline{k}_{\rm{F}}\right],\label{ddef_P0}\\
K_0(\rho)=&-\frac{3\overline{k}_{\rm{F}}^2}{5M}
+9\rho\left(\frac{\partial U_0}{\partial
\rho}+\left.\frac{\overline{k}_{\rm{F}}}{3\rho}\frac{\partial
U_0}{\partial|\v{k}|}\right|_{|\v{k}|=\overline{k}_{\rm{F}}}\right)
+\frac{18}{\rho}\int_0^\rho
U_0\left(f,\overline{k}_{\rm{F}}^f\right)\d
f-18U_0(\rho,\overline{k}_{\rm{F}})\notag\\
&+\epsilon\left[3\sigma\frac{\partial^2
U_0}{\partial|\v{k}|^2}
\cdot\overline{k}_{\rm{F}}^2-\frac{3\overline{k}_{\rm{F}}^2}{5M}\left(\frac{7}{15}+2\sigma\right)+\frac{18\sigma}{\rho}\int_0^{\rho}\left(
\left.\frac{\partial
U_0}{\partial|\v{k}|}\right|_{|\v{k}|=\overline{k}^f_{\rm{F}}}\cdot\overline{k}^f_{\rm{F}}\right)\d
f-15\sigma\frac{\partial
U_0}{\partial|\v{k}|}\cdot\overline{k}_{\rm{F}}
\right]_{|\v{k}|=\overline{k}_{\rm{F}}},\label{ddef_K0}\\
J_0(\rho)=&\frac{12\overline{k}_{\rm{F}}^2}{5M}+27\rho^2\left(\frac{\partial^2U_0}{\partial
\rho^2}+\frac{2\overline{k}_{\rm{F}}}{3\rho}\frac{\partial^2U_0}{\partial
\rho\partial
|\v{k}|}-\frac{2\overline{k}_{\rm{F}}}{9\rho^2}\frac{\partial
U_0}{\partial |\v{k}|}
+\frac{\overline{k}_{\rm{F}}^2}{9\rho^2}\frac{\partial^2U_0}{\partial
|\v{k}|^2}\right)_{|\v{k}|=\overline{k}_{\rm{F}}}\notag\\
&-81\rho\left(\frac{\partial U_0}{\partial \rho}+
\left.\frac{\overline{k}_{\rm{F}}}{3\rho}\frac{\partial
U_0}{\partial|\v{k}|}\right|_{|\v{k}|=\overline{k}_{\rm{F}}}\right)
-\frac{162}{\rho}\int_0^\rho
U_0\left(f,\overline{k}_{\rm{F}}^f\right)\d f
+162U_0(\rho,\overline{k}_{\rm{F}})\notag\\
&+\epsilon\left[
\frac{12\overline{k}_{\rm{F}}^2}{5M}\left(\frac{19}{30}+2\sigma\right)
+\sigma\left(18\rho\overline{k}_{\rm{F}}\frac{\partial^2U_0}{\partial
\rho\partial |\v{k}|}+
18\rho\overline{k}_{\rm{F}}^2\frac{\partial^3U_0}{\partial
\rho\partial
|\v{k}|^2}+3\overline{k}_{\rm{F}}^3\frac{\partial^3U_0}{\partial
|\v{k}|^3}\right)\right.\notag\\
&\hspace*{1.5cm}-\left.27\sigma\frac{\partial^2
U_0}{\partial|\v{k}|^2}\cdot\overline{k}_{\rm{F}}^2
-\frac{162\sigma}{\rho}\int_0^{\rho}\left( \left.\frac{\partial
U_0}{\partial|\v{k}|}\right|_{|\v{k}|=\overline{k}^f_{\rm{F}}}\cdot\overline{k}^f_{\rm{F}}\right)\d
f+129\sigma\frac{\partial
U_0}{\partial|\v{k}|}\cdot\overline{k}_{\rm{F}}
\right]_{|\v{k}|=\overline{k}_{\rm{F}}},\label{ddef_J0}
\end{align}
and
\begin{align}
L(\rho)=&\frac{\overline{k}_{\rm{F}}^2}{3M}+\left(\frac{\overline{k}_{\rm{F}}^2}{6}\frac{\partial^2
U_0}{\partial|\v{k}|^2}+\frac{\overline{k}_{\rm{F}}}{6}\frac{\partial
U_0}{\partial|\v{k}|}\right)_{|\v{k}|=\overline{k}_{\rm{F}}}+\left.\overline{k}_{\rm{F}}\frac{\partial
U_{\rm{sym}}}{\partial|\v{k}|}\right|_{|\v{k}|=\overline{k}_{\rm{F}}}
+\frac{3}{2}U_{\rm{sym}}(\rho,\overline{k}_{\rm{F}})
+3U_{\rm{sym},2}(\rho,\overline{k}_{\rm{F}})\notag\\
&+\epsilon\left[\left(2\sigma-\frac{2}{3}\right)\left(\frac{\overline{k}_{\rm{F}}^2}{3M}+
\frac{\overline{k}_{\rm{F}}^2}{6}\frac{\partial^2
U_0}{\partial|\v{k}|^2}\right)+\frac{\sigma}{6}\frac{\partial^3
U_0}{\partial|\v{k}|^3}\cdot\overline{k}_{\rm{F}}^3+
\frac{\sigma}{6}\frac{\partial^2
U_0}{\partial|\v{k}|^2}\cdot\overline{k}_{\rm{F}}^2+
\frac{1}{6}\left(\sigma-\frac{2}{3}\right)
\frac{\partial
U_0}{\partial|\v{k}|}\cdot \overline{k}_{\rm{F}}\right.
\notag\\
&\hspace*{1.5cm}\left.+\left(\frac{5}{2}\sigma-\frac{1}{3}\right)\frac{\partial
U_{\rm{sym}}}{\partial|\v{k}|}\cdot\overline{k}_{\rm{F}}+
\sigma\frac{\partial^2
U_{\rm{sym}}}{\partial|\v{k}|^2}\cdot\overline{k}_{\rm{F}}^2
+
3\sigma\frac{\partial
U_{\rm{sym,2}}}{\partial|\v{k}|}\cdot\overline{k}_{\rm{F}}\right]_{|\v{k}|=\overline{k}_{\rm{F}}},\label{ddef_L}\\
K^{\rm{SNM}}(\rho)=&\frac{3\overline{k}_{\rm{F}}^2}{M}+9\rho\left(\frac{\partial
U_0}{\partial \rho}+\left.\frac{\overline{k}_{\rm{F}}}{3\rho}
\frac{\partial
U_0}{\partial|\v{k}|}\right|_{|\v{k}|=\overline{k}_{\rm{F}}}\right)
+\epsilon\left[\frac{3\overline{k}_{\rm{F}}^2}{M}\left(\sigma-\frac{1}{3}\right)
+3\sigma\overline{k}_{\rm{F}}\left(\frac{\partial
U_0}{\partial|\v{k}|}
+\frac{\partial^2
U_0}{\partial|\v{k}|^2}\cdot\overline{k}_{\rm{F}}\right)\right]_{|\v{k}|=\overline{k}_{\rm{F}}}\label{ddef_KSNM}.
\end{align}

These expressions are physically intuitive and useful since all the quantities involved are those in 3D.  These quantities are well known and most of them are tightly constrained either theoretically or experimentally. Let's now discuss a few main features if a nonzero $\epsilon$ is applied. Besides the kinetic contribution, a non-trivial contribution to the EOS of SNM comes from the integration over the momentum-dependence of the potential $U_0$, i.e., $(\sigma/\rho)\int_0^{\rho}[\overline{k}_{\rm{F}}^f\cdot\partial U_0/\partial|\v{k}|]_{|\v{k}|=\overline{k}^f_{\rm{F}}}\d f$. One knows from nuclear optical model fitting to nucleon-nucleus scattering data that the dependence of the nucleon isoscalar potential $U_0$ on the nucleon momentum is positive\,\cite{LiXH2013PLB,LiXH2015PLB,Hama1990}, thus the integration is also positive since $\sigma>0$. Consequently the linear $\epsilon$-term contributes a positive term to the unperturbed EOS of SNM, as demonstrated in formula (\ref{ddef_E0}).
If we extrapolate the 3D EOS of SNM to lower dimensions (higher dimensions), it will be reduced (enhanced). Specifically, the 2D SNM is much more bounded than its 3D counterpart.
For the symmetry energy, we first neglect the momentum-dependence of the symmetry potential as well as the second derivative $\partial^2U_0/\partial|\v{k}|^2$, then the $\epsilon$-term of (\ref{ddef_Esym}) becomes
\begin{equation}\label{ckj-1}
\epsilon\left[\frac{\overline{k}_{\rm{F}}^2}{6M}\left(2\sigma-\frac{1}{3}\right)
+\frac{\overline{k}_{\rm{F}}}{6}\frac{\partial
U_0}{\partial
|\v{k}|}\left(\sigma-\frac{1}{3}\right)_{|\v{k}|=\overline{k}_{\rm{F}}}
\right]=\frac{\overline{k}_{\rm{F}}^2}{6M}\left[\frac{M}{M_{\rm{s}}^{\ast}}\left(\sigma-\frac{1}{3}\right)+\sigma\right]\epsilon\approx
\frac{\overline{k}_{\rm{F}}^2}{6M}\left(0.24-0.09\frac{M}{M_{\rm{s}}^{\ast}}\right)\epsilon,
\end{equation}
where $M_{\rm{s}}^{\ast}\gtrsim0.8M$\,\cite{LiBA2018PPNP} is the scalar Landau effective mass. The factor in the bracket here is positive with high probability and thus the linear $\epsilon$-correction is negative if $\epsilon<0$, e.g., the symmetry energy in 2D will also be reduced (like the EOS of SNM).
However, as there exist large uncertainties (at least qualitatively) about the symmetry potential $U_{\rm{sym}}$ especially its momentum dependence\,\cite{LiXH2013PLB,LiXH2015PLB,LiBA2018PPNP,Holt2016}, the final complete results and implications of Eq. (\ref{ckj-1}) and the term $(\sigma/2)[\overline{k}_{\rm{F}}\cdot\partial U_{\rm{sym}}/\partial|\v{k}|]_{|\v{k}|=\overline{k}_{\rm{F}}}$ are not quite definite and certainly need more analyses. 
For example, by writing the linear $\epsilon$-term in (\ref{ddef_Esym}) as 
\begin{equation}\label{edf}
\Pi=\sigma\left(\frac{\overline{k}_{\rm{F}}^2}{3M}+\left.\frac{\overline{k}_{\rm{F}}}{6}\frac{\partial U_0}{\partial|\v{k}|}\right|_{|\v{k}|=\overline{k}_{\rm{F}}}
+\left.\frac{\overline{k}_{\rm{F}}^2}{6}\frac{\partial^2U_0}{\partial|\v{k}|^2}\right|_{|\v{k}|=\overline{k}_{\rm{F}}}
+\left.\frac{\overline{k}_{\rm{F}}}{2}\frac{\partial U_{\rm{sym}}}{\partial|\v{k}|}\right|_{|\v{k}|=\overline{k}_{\rm{F}}}
\right)-\frac{\overline{k}_{\rm{F}}^2}{18M^{\ast}_{\rm{s}}},
\end{equation}
and adopting the nucleon optical model fitting values for the relevant terms given in Ref.\,\cite{LiXH2013PLB}, one obtains ${\overline{k}_{\rm{F}}^2}/{3M}+6^{-1}[{\overline{k}_{\rm{F}}}\cdot{\partial U_0}/{\partial|\v{k}|}]_{|\v{k}|=\overline{k}_{\rm{F}}}\approx37.7\,\rm{MeV}$, $6^{-1}[{\overline{k}_{\rm{F}}^2}\cdot{\partial^2U_0}/{\partial|\v{k}|^2}]_{|\v{k}|=\overline{k}_{\rm{F}}}\approx-2.3\,\rm{MeV}$
and $[\overline{k}_{\rm{F}}\cdot\partial U_{\rm{sym}}/\partial|\v{k}|]_{|\v{k}|=\overline{k}_{\rm{F}}}\approx-46.0\,\rm{MeV}$, thus $\Pi\approx(12.4\sigma-5.1)\epsilon\,\rm{MeV}\approx-2.1\epsilon\,\rm{MeV}$, i.e., the symmetry energy is now enhanced if the dimension is downward perturbed.
However, a slight change on $[\overline{k}_{\rm{F}}\cdot\partial U_{\rm{sym}}/\partial|\v{k}|]_{|\v{k}|=\overline{k}_{\rm{F}}}$ may essentially change the sign of $\Pi$.
Further accurate information on $U_{\rm{sym}}$ and/or higher-order calculations on $\epsilon$ are essential for this type of analyses. We will give more numerical results in section \ref{SEC_TOY}, see the relevant discussion given after Eq.\,(\ref{djk-2}), Tab.\,\ref{Tab_2DEsym} and Tab.\,\ref{Tab_2DL}.
In section \ref{SEC_Esym2D-O2}, we will extend the above linear-$\epsilon$ perturbation to second order of $\epsilon^2$.
The $\epsilon$-induced effects on the other quantities (\ref{ddef_P0})-(\ref{ddef_KSNM}) could be analyzed similarly, but would be omitted in the current work.

The perturbative expression (\ref{ddef_Esym}) can be cast into the following form,
\begin{align}
E_{\rm{sym}}(\rho)=&\frac{\overline{k}_{\rm{F}}^2}{6M}\left[1+\epsilon\left(2\sigma-\frac{1}{3}\right)\right]+\left.\frac{\overline{k}_{\rm{F}}}{6}
\left[1+\epsilon\left(2\sigma-\frac{1}{3}\right)\right]\frac{\partial U_0}{\partial
|\v{k}|}\right|_{|\v{k}|=\overline{k}_{\rm{F}}}\notag\\
&+\frac{1}{2}U_{\rm{sym}}(\rho,\overline{k}_{\rm{F}})+\epsilon\sigma\left(
\frac{\overline{k}_{\rm{F}}^2}{6}\frac{\partial^2U_0}{\partial|\v{k}|^2}+\frac{\overline{k}_{\rm{F}}}{2}\frac{\partial U_{\rm{sym}}}{\partial|\v{k}|}-\frac{\overline{k}_{\rm{F}}}{6}\frac{\partial U_0}{\partial|\v{k}|}\right)_{|\v{k}|=\overline{k}_{\rm{F}}}.
\end{align}
We introduce the effective mass $M_{\rm{eff}}$ (with respect to the bare nucleon mass), the effective isoscalar potential $U_0^{\rm{eff}}$, and the effective symmetry potential $U_{\rm{sym}}^{\rm{eff}}$ as,
\begin{align}
M_{\rm{eff}}=&M\left[1+\epsilon\left(2\sigma-\frac{1}{3}\right)\right]^{-1},\\
U_0^{\rm{eff}}(\rho,|\v{k}|)=&\left[1+\epsilon\left(2\sigma-\frac{1}{3}\right)\right] U_0,\\
U_{\rm{sym}}^{\rm{eff}}(\rho,|\v{k}|)=&U_{\rm{sym}}(\rho,|\v{k}|)+2\epsilon\sigma\left(
\frac{\overline{k}_{\rm{F}}^2}{6}\frac{\partial^2U_0}{\partial|\v{k}|^2}+\frac{\overline{k}_{\rm{F}}}{2}\frac{\partial U_{\rm{sym}}}{\partial|\v{k}|}
-\frac{\overline{k}_{\rm{F}}}{6}\frac{\partial U_0}{\partial|\v{k}|}\right)_{|\v{k}|=\overline{k}_{\rm{F}}}.
\end{align}
They are all dependent on the perturbative dimension $\epsilon$. We can then rewrite the symmetry energy as
\begin{equation}\label{ddef_Esym_eff}
E_{\rm{sym}}(\rho)=\frac{\overline{k}_{\rm{F}}^2}{6M_{\rm{eff}}}
\left.+\frac{\overline{k}_{\rm{F}}}{6}\frac{\partial U_0^{\rm{eff}}}{\partial |\v{k}|}\right|_{|\v{k}|=\overline{k}_{\rm{F}}}+\frac{1}{2}U^{\rm{eff}}_{\rm{sym}}(\rho,\overline{k}_{\rm{F}})
=\frac{\overline{k}_{\rm{F}}^2}{6M_{\rm{s},\rm{eff}}^{\ast}}+\frac{1}{2}U^{\rm{eff}}_{\rm{sym}}(\rho,\overline{k}_{\rm{F}}),
\end{equation}
where $M_{\rm{s},\rm{eff}}^{\ast}/M_{\rm{eff}}=[1+(M_{\rm{eff}}/|\v{k}|)\partial U_0^{\rm{eff}}/\partial|\v{k}|]^{-1}_{|\v{k}|=k_{\rm{F}}}$ is the scalar Landau effective mass based on $M_{\rm{eff}}$.
Interestingly, (\ref{ddef_Esym_eff}) has the same form as the unperturbed symmetry energy in 3D.
For the 2D situation ($\epsilon=-1$), we have $M_{\rm{eff}}>M$ and similarly $0<\partial U_0^{\rm{eff}}/\partial|\v{k}|<\partial U_0/\partial|\v{k}|$ since $-1<\epsilon(2\sigma-1/3)<0$, the expression (\ref{ddef_Esym_eff}) shows that the nucleon moves in $d$D space with a heavier mass in a weaker-momentum-dependent potential (compared with the 3D case). The effective form of the symmetry energy (\ref{ddef_Esym_eff}) in 2D provides a convenient starting point for the relevant investigations on the $E_{\rm{sym}}(\rho)$ in lower dimensions.

The expressions from (\ref{ddef_E0})-(\ref{ddef_KSNM}) are general and model-independent.
We have discussed the non-trivial contributions caused by the momentum-dependence of the nucleon potential to the EOS of SNM and to the symmetry energy.
On the other hand, if one artificially assumes that the $U_J$ is momentum independent, then the above $\epsilon$-expansion expressions are reduced to
\begin{align}
{E}_0(\rho)=&\frac{3\overline{k}_{\rm{F}}^2}{10M}+
\frac{1}{\rho}\int_0^\rho
U_0\left(f\right)\d
f+\frac{3\overline{k}_{\rm{F}}^2}{5M}\left(\sigma+\frac{1}{15}\right)\epsilon,\label{ddef_E0-nonm}\\
E_{\rm{sym}}(\rho)=&\frac{\overline{k}_{\rm{F}}^2}{6M}+\frac{1}{2}U_{\rm{sym}}(\rho)+\frac{\overline{k}_{\rm{F}}^2}{6M}\left(2\sigma-\frac{1}{3}\right)\epsilon,
\end{align}
as well as
\begin{align}
P_0(\rho)=&\frac{\overline{k}_{\rm{F}}^2\rho}{5M}+
\int_0^\rho U_0\left(f\right)\d
f+\rho U_0(\rho)+\frac{\rho\overline{k}_{\rm{F}}^2}{10M}\left(2\sigma-\frac{1}{5}\right)\epsilon,\\
K_0(\rho)=&-\frac{3\overline{k}_{\rm{F}}^2}{5M}
+9\rho\frac{\partial U_0}{\partial
\rho}+\frac{18}{\rho}\int_0^\rho
U_0\left(f\right)\d
f-18U_0(\rho)-\frac{3\overline{k}_{\rm{F}}^2}{5M}\left(\frac{7}{15}+2\sigma\right)\epsilon,\\
J_0(\rho)=&\frac{12\overline{k}_{\rm{F}}^2}{5M}+27\rho^2\frac{\partial^2U_0}{\partial
\rho^2}
-81\rho\frac{\partial U_0}{\partial \rho}
-\frac{162}{\rho}\int_0^\rho
U_0\left(f\right)\d f
+162U_0(\rho)+
\frac{12\overline{k}_{\rm{F}}^2}{5M}\left(\frac{19}{30}+2\sigma\right)\epsilon
,\\
L(\rho)=&\frac{\overline{k}_{\rm{F}}^2}{3M}+\frac{3}{2}U_{\rm{sym}}(\rho)
+3U_{\rm{sym},2}(\rho)+
\frac{\overline{k}_{\rm{F}}^2}{3M}\left(2\sigma-\frac{2}{3}\right)\epsilon,\label{ddef_L_nonm}\\
K^{\rm{SNM}}(\rho)=&\frac{3\overline{k}_{\rm{F}}^2}{M}+9\rho\frac{\partial
U_0}{\partial \rho}+\frac{3\overline{k}_{\rm{F}}^2}{M}\left(\sigma-\frac{1}{3}\right)\epsilon,
\end{align}
here the partial and total differentials involved are the same, i.e, $\partial \equiv\d$.
It is obvious now that  all the $\epsilon$-expansion terms come from the kinetic EOS.
Under such conditions, the linear-$\epsilon$ contributions both to the EOS of SNM and the symmetry energy are negative deterministically if $\epsilon$ is negative (e.g., 2D).
On the other hand, from the expression for the slope parameter $L(\rho)$ in this situation, one can find the linear-$\epsilon$ term is positive, indicating the symmetry energy is hardened towards higher densities.

Moreover, we can investigate the features of the $\epsilon$-expansion by making use of the exact expressions for the (kinetic part of the) relevant quantities in the general dimension $d$.
In doing so, a natural concern is how to determine the nucleon Fermi momentum in $d$D.
Two schemes are possible:
\begin{enumerate}
\item[(a)] The Fermi momentum takes the same value irrespective of the dimension, e.g., $k_{\rm{F}}$ is always about 200-300\,MeV, either in 3D or $d$D.
In this sense, we treat the Fermi momentum $k_{\rm{F}}$ as a fixed quantity, and then one does not need to do perturbative calculations on $k_{\rm{F}}$.
Consequently the factor $\sigma$ is unnecessary for the relevant calculations.
\item[(b)] 
On the other hand, we can treat the density $[\rho]\sim[\rm{fm}^{-3}]$ in 3D as a fixed quantity (the one based on which we make the perturbative calculations as done in the previous paragraphs), and the densities in other dimensions are related to it through $\rho_d=\rho^{d/3}$, thus $k_{\rm{F}}\sim\rho^{1/3}$ holds irrespective of the dimension $d$. In this sense, the $\sigma$ should be kept for corresponding quantities.
\end{enumerate}

Compared with the total EOS, the tendency of the kinetic part when $d$ is perturbed is relatively clear.
Next, we shall extrapolate the kinetic EOS of SNM and the kinetic symmetry energy to other dimensions and compare them with the corresponding exact results.
For $\epsilon=-1$ (i.e., 2D), one has
\begin{align}\label{dkdk}
E_0^{\rm{kin}}(\rho)\approx&\frac{3\overline{k}_{\rm{F}}^2}{10M}\left(1-2\sigma-\frac{2}{15}\right)=\frac{3\overline{k}_{\rm{F}}^2}{10M}\left(\frac{13}{15}-2\sigma\right).
\end{align}
If $\overline{k}_{\rm{F}}$ is assumed to be the same as the one in dimensions 2, i.e., we do not do perturbations on the Fermi momentum, and in this situation the $\sigma$ coefficient should be removed (i.e., set to be zero) in the calculation. Consequently, the ratio between the above approximation and the exact kinetic EOS of SNM in 2D namely $k_{\rm{F}}^2/4M$ becomes $26/25\approx1.04$ (based on the assumption that $\overline{k}_{\rm{F}}=k_{\rm{F}}$), indicating that the perturbation starting from $d=3$ to $d=2$ is effective. 
Similarly, when we extrapolate to $d=4$ (with $\epsilon=1$) or $d=1$ (with $\epsilon=-2$), the ratios between the approximation and the exact one are given as $51/50\approx1.02$ or $33/25\approx1.32$ for 4D and 1D, respectively. There is no doubt that as $|\epsilon|$ deviates far from zero, the perturbative approximation becomes worse.
For the kinetic symmetry energy, we now have
\begin{equation}
E_{\rm{sym}}^{\rm{kin}}(\rho)=\frac{\overline{k}_{\rm{F}}^2}{6M}
\left[1+\left(2\sigma-\frac{1}{3}\right)\epsilon\right]
\to\frac{\overline{k}_{\rm{F}}^2}{6M}\left(1-\frac{\epsilon}{3}\right),~~\mbox{as}\,\,\sigma\to0.
\end{equation}
Then for the 2D situation, this gives $2\overline{k}_{\rm{F}}^2/9M$, which is close to the exact one $k_{\rm{F}}^2/4M$ in 2D.
The ratio is $8/9\approx0.89$, under the assumption that $\overline{k}_{\rm{F}}=k_{\rm{F}}$.
Similarly, for the 4D (with $\epsilon=1$) and the 1D (with $\epsilon=-2$) cases, we have the ratios $(\overline{k}_{\rm{F}}^2/9M)/(k_{\rm{F}}^2/8M)=8/9\approx0.89$ and $(5\overline{k}_{\rm{F}}^2/18M)/(k_{\rm{F}}^2/2M)=5/9\approx0.56$, respectively, using the formula (\ref{4qEsymFF_1}), showing again the effectiveness of the $\epsilon$-expansion for the kinetic EOS.

If we adopt the assumption that the Fermi momenta in different dimensions are not equal but related via the corresponding relation between the density and the Fermi momentum, then the coefficient $\sigma$ should be kept.
For example, the density in dimensions 2 could be estimated via the conventional density $\rho$ in dimensions 3 as $\rho_2=\rho^{2/3}$.
Now the Fermi momentum in 2D is given by $k_{\rm{F}}=(\pi\rho_2)^{1/2}$, consequently we have the ratio for the kinetic EOS of SNM,
\begin{equation}\label{ia-1}
\left.\frac{3\overline{k}_{\rm{F}}^2}{10M}\left(\frac{13}{15}-2\sigma\right)\right/\frac{k_{\rm{F}}^2}{4M}
=\frac{6}{5}\pi^{1/3}\left(\frac{3}{2}\right)^{2/3}\left(\frac{13}{15}-2\sigma\right)\approx0.89,
\end{equation}
where the perturbative 3D kinetic EOS of SNM with $\epsilon=-1$ gives about 8.56\,MeV and the exact 2D kinetic EOS of SNM is about 9.60\,MeV.
In addition, for the 4D with $\epsilon=1$) and 1D situations (with $\epsilon=-2$), we have $3\overline{k}_{\rm{F}}^2/10M\cdot(2\sigma+17/15)\approx35.63\,\rm{MeV}$ and $k_{\rm{F}}^2/3M\approx36.20\,\rm{MeV}$ (where $k_{\rm{F}}=(8\pi^2\rho_4)^{1/4}$ with $\rho_4=\rho^{4/3}$), and $3\overline{k}_{\rm{F}}^2/10M\cdot(11/15-4\sigma)\approx-4.97\,\rm{MeV}$ and $k_{\rm{F}}^2/6M\approx1.26\,\rm{MeV}$ (where $k_{\rm{F}}=\pi\rho_1/4$ with $\rho_1=\rho^{1/3}$), leading to the ratios about 0.98 and $-3.96$, respectively. Although the extrapolation in this scheme to 1D is unreasonable, the absolute magnitude is still not so different.
Furthermore, for the kinetic symmetry energy, we then have the ratio as
\begin{equation}\label{ia-2}
\pi^{1/3}\left(\frac{3}{2}\right)^{5/2}\left(\frac{4}{3}-2\sigma\right)
\approx1.09,
\end{equation}
where the perturbative 3D kinetic symmetry with $\epsilon=-1$ gives about 10.49\,MeV and the corresponding exact one gives 9.60\,MeV (this is the same as the kinetic EOS of SNM).
We have found again the $\epsilon$-expansion is effective.
As a reference, we also list the relevant results for the 4D and 1D cases: for the 4D situation, $\overline{k}_{\rm{F}}^2/6M\cdot(2/3+2\sigma)\approx14.06\,\rm{MeV}$ and $k_{\rm{F}}^2/8M\approx13.57\,\rm{MeV}$ 
gives the ratio about 1.04, while for the 1D situation $\overline{k}_{\rm{F}}^2/6M\cdot(5/3-4\sigma)\approx8.70\,\rm{MeV}$ and $k_{\rm{F}}^2/2M\approx3.77\,\rm{MeV}$ and thus the ratio is found to be about 2.31.

Although the $\epsilon$-expansion to linear order of $\epsilon$ gives reasonable predictions (compared with the exact results) on the kinetic EOS of SNM as well as the kinetic symmetry energy, in order to solidify the method further investigations are certainly necessary.  It will be interesting to perform expansions to higher orders of $\epsilon$ (the breakdown with $\epsilon=-2$ actually indicates the failure of the linear approximation already), or to include effective nucleon-nucleon interactions into the calculations.
We make some initial attempts in these directions in this work as we shall discuss later, and in the remaining of the present work, we adopt the scheme (b) above related to the issue of the perturbative treatment on the density.

\setcounter{equation}{0}
\section{Toy-model calculations}\label{SEC_TOY}

\subsection{A phenomenological single-nucleon potential $U_J(\rho,\delta,|\v{k}|)$ in neutron-rich matter and the corresponding EOS in 3D}

In this section, we adopt a toy model for the single-nucleon potential and study the relevant EOSs in various dimensions (via the $\epsilon$-expansion approach).
When constructing the single-nucleon potential as well as the corresponding EOS of ANM, the following empirical facts about the 3D EOS of ANM should be respected \cite{LiBA2008},
\begin{enumerate}
\item[(a)] Reasonable saturation density and the binding energy, i.e., $\rho_0\approx0.16\,\rm{fm}^{-3}$ and $E_0(\rho_0)\approx-16\,\rm{MeV}$.
\item[(b)] Reasonable incompressibility coefficient $K_0$ of SNM, e.g., $K_0\approx 230\pm20\,\rm{MeV}$\,\cite{Blaizot1980,Youngblood1999PRL,Piekarewicz2010JPG,Stone2014PRC,
ChenLW2012JPG,Khan2012PRL,Garg2018PPNP,LiBA2021PRC,XuJ2021PRC}.
If necessary the higher order coefficients like the skewness $J_0$ could also be taken into consideration when constructing the single-nucleon potential\,\cite{CaiBJ2017NST}.
\item[(c)] Reasonable scalar nucleon effective mass in SNM at $\rho_0$, e.g., $
M_{\rm{s}}^{\ast}/M\approx0.8$\,\cite{LiBA2018PPNP,ZhangZ2021}.
\item[(d)] The kinetic energy dependence of the nucleon potential constrained by the nucleon optical potential at $\rho_0$ from fitting nucleon-nucleus scattering data\,\cite{LiXH2013PLB,LiXH2015PLB,Hama1990}.
\item[(e)] The symmetry energy as well as its slope parameter at the saturation density about $
E_{\rm{sym}}(\rho_0)\approx30\pm4\,\rm{MeV},L\approx50\pm20\,\rm{MeV}$\,\cite{LiBA2013PLB,LiBA2017AIP,LiBA2021Universe,
LiBA2019EPJA}. 
\item[(f)] At small densities the nucleon experiences an attractive potential,  while as the density increases the nucleon tends to experience a repulsive potential. 
\end{enumerate}
The phenomenological single-nucleon potential constructed in this way may provide a convenient starting point for studying astrophysical problems when the detailed information related to finite nuclei is irrelevant\,\cite{Steiner2012PRL,Hebeler2013ApJ,Bedaque2015PRL}.
For non-relativistic applications, the EOS of ANM could be treated as an expansion on $k_{\rm{F}}/M$ or equivalently on $\rho/\rho_0$\,\cite{Fritsch2005}, the single-nucleon potential based on this idea adopted here takes the following form,
\begin{align}
{U_J}(\rho,\delta,|\v{k}|)=&\alpha\left(\frac{\rho}{\rho_0}\right)+\left(\beta+\beta_{\rm{IS}}e^{\Gamma_{4/3}\tau_3^J\delta}\right)\left(\frac{\rho}{\rho_0}\right)^{4/3}\notag\\
&+\left(\lambda+\lambda_{\rm{IS}}e^{\Gamma_{5/3}\tau_3^J\delta}\right)\left(\frac{\rho}{\rho_0}\right)^{5/3}
+\phi\left(\frac{\rho}{\rho_0}\right)^2+
\mu\left(\frac{\rho}{\rho_0}\right)\left(
1+\frac{\theta|\v{k}|^3}{\Lambda^3}\right)^{-1}\label{5q5J}.
\end{align}
There are totally 7 model parameters related to the SNM, i.e, $\alpha,\beta,\lambda,\phi,\mu,\theta$ and $\Lambda$ and 2 model parameters $\beta_{\rm{IS}}$ and $\lambda_{\rm{IS}}$ describing the isospin dependence of ${U_J}(\rho,\delta,|\v{k}|)$.
The factor $\Gamma_{\vartheta}$ here is taken to be $\vartheta-1$ in order to fulfill the constraint from the HVH theorem, see the relation (\ref{Gammay}).

The single-nucleon potential in SNM reads from (\ref{5q5J}) as,
\begin{equation}\label{3qU0}
U_0(\rho,|\v{k}|)=\alpha\left(\frac{\rho}{\rho_0}\right)
 +\beta\left(\frac{\rho}{\rho_0}\right)^{4/3}
+\lambda\left(\frac{\rho}{\rho_0}\right)^{5/3}
+\phi\left(\frac{\rho}{\rho_0}\right)^{2} +\mu
\left(\frac{\rho}{\rho_0}\right)\left/\left[{1+\theta\left(\frac{|\v{k}|}{\Lambda}\right)^3}\right]\right..
\end{equation}
Here the model parameters like $\alpha,\beta$, etc., are fixed by the empirical knowledge on the EOS in dimensions 3, and the density $\rho$ appears in the dimensionless combination $\rho/\rho_0$.
In the present work, the 3D density $\rho$ is fixed and the density in other dimensions is connected to it via $\rho_d=\rho^{d/3}$ automatically, see the explanation (b) above (\ref{dkdk}). In fact, using the (relatively well known) 3D knowledge on the EOS to infer the EOS in other dimensions (near 3) is the main purpose of the $\epsilon$-expansion approach.
In this sense, the combinations like $\alpha/\rho_0$ will be kept constant when we perturb the dimension $d$, and the nontrivial contribution comes from the momentum-dependent term (characterized by $\theta$) of the potential. The perturbative calculations including those on the density $\rho$ are left for future studies.
Based on (\ref{3qU0}), one can obtain the EOS of SNM,
\begin{equation}\label{3qE0}
E_0(\rho)=\frac{3k_{\rm{F}}^2}{10M}+\frac{1}{\rho}\int_0^{\rho}U_0\left(f,k_{\rm{F}}^f\right)\d
f,~~k_{\rm{F}}^f=\left(\frac{3\pi^2f}{2}\right)^{1/3}.
\end{equation}
The terms proportional to the density in the single-nucleon potential give the following contribution,
\begin{equation}
\frac{\alpha}{2}\left(\frac{\rho}{\rho_0}\right)+
\frac{3\beta}{7}\left(\frac{\rho}{\rho_0}\right)^{4/3}
+\frac{3\lambda}{8}\left(\frac{\rho}{\rho_0}\right)^{5/3}
+\frac{\phi}{3}\left(\frac{\rho}{\rho_0}\right)^2,
\end{equation}
while the momentum-dependent term leads to
\begin{align}
\frac{\mu}{\rho\rho_0}\int_0^{\rho}\frac{f}{1+\theta(k_{\rm{F}}^f/\Lambda)^3}\d
f=&\frac{4\mu\Lambda^6}{9\pi^4\theta^2\rho\rho_0}\left[\frac{3\pi^2\rho\theta}{2\Lambda^3}
-\ln\left(1+\frac{3\pi^2\rho\theta}{2\Lambda^3}\right)\right] \notag\\
=&\mu\left(\frac{\rho}{\rho_0}\right)\left[\theta\left(\frac{k_{\rm{F}}}{\Lambda}\right)^3-\ln\left(1+\theta\left(\frac{k_{\rm{F}}}{\Lambda}\right)^3\right)\right]\left/\theta^2\left(\frac{k_{\rm{F}}}{\Lambda}\right)^6\right.
.\end{align}
By combining these contributions, one obtains the EOS of SNM as,
\begin{align}\label{3qEoS}
E_0(\rho)=&\frac{3}{10M}\left(\frac{3\pi^2}{2}\right)^{2/3}\rho^{2/3}
+\frac{\alpha}{2}\left(\frac{\rho}{\rho_0}\right)+
\frac{3\beta}{7}\left(\frac{\rho}{\rho_0}\right)^{4/3}\notag\\
&+\frac{3\lambda}{8}\left(\frac{\rho}{\rho_0}\right)^{5/3}
+\frac{\phi}{3}\left(\frac{\rho}{\rho_0}\right)^2
+\mu\left(\frac{\rho}{\rho_0}\right)\left[\theta\left(\frac{k_{\rm{F}}}{\Lambda}\right)^3-\ln\left(1+\theta\left(\frac{k_{\rm{F}}}{\Lambda}\right)^3\right)\right]\left/\theta^2\left(\frac{k_{\rm{F}}}{\Lambda}\right)^6\right..
\end{align}

At small densities, $k_{\rm{F}}/\Lambda\ll1$, and according to $ \ln(1+x)\approx x-x^2/2-x^3/3$ as $x\to0$, 
\begin{equation}
\mu\left(\frac{\rho}{\rho_0}\right)\left[\theta\left(\frac{k_{\rm{F}}}{\Lambda}\right)^3-\ln\left(1+\theta\left(\frac{k_{\rm{F}}}{\Lambda}\right)^3\right)\right]\left/\theta^2\left(\frac{k_{\rm{F}}}{\Lambda}\right)^6\right.\approx\mu\left(\frac{\rho}{\rho_0}\right)\left[x-\ln(1+x)\right]/x^2\approx\left[\frac{1}{2}-\frac{1}{3}\theta\left(\frac{k_{\rm{F}}}{\Lambda}\right)^3\right]\mu\left(\frac{\rho}{\rho_0}\right),\end{equation}
where $x$ is dimensionless,
\begin{equation}
x=\theta\left(\frac{k_{\rm{F}}}{\Lambda}\right)^3,
\end{equation}
one obtains the following approximation for the EOS of SNM (the term from the momentum-dependent part induces an effective density-dependent term),
\begin{align}
E_0(\rho)\approx&\frac{3}{10M}\left(\frac{3\pi^2}{2}\right)^{2/3}\rho^{2/3}
+\frac{\alpha+\mu}{2}\left(\frac{\rho}{\rho_0}\right)+
\frac{3\beta}{7}\left(\frac{\rho}{\rho_0}\right)^{4/3}
+\frac{3\lambda}{8}\left(\frac{\rho}{\rho_0}\right)^{5/3}
+\frac{1}{3}\left[\phi-\theta\mu\left(\frac{k_{\rm{F}}(\rho_0)}{\Lambda}\right)^3\right]\left(\frac{\rho}{\rho_0}\right)^2.\end{align}

From (\ref{3qEoS}) one could obtain all the characteristics such as the slope $L_0(\rho)$, the incompressibility coefficient $K_0(\rho)$, the skewness coefficient $J_0(\rho)$ and the kurtosis coefficient $I_0$ of the SNM EOS,
\begin{align}\label{3qL0}
L_0(\rho)=&\frac{3}{5M}\left(\frac{3\pi^2}{2}\right)^{2/3}\rho^{2/3}+
\frac{3\alpha}{2}\left(\frac{\rho}{\rho_0}\right)
+\frac{12\beta}{7}\left(\frac{\rho}{\rho_0}\right)^{4/3}
+\frac{15\lambda}{8}\left(\frac{\rho}{\rho_0}\right)^{5/3}
+2\phi\left(\frac{\rho}{\rho_0}\right)^2\notag\\
&+3\mu\left(\frac{\rho}{\rho_0}\right)\left[\theta\left(\frac{k_{\rm{F}}}{\Lambda}\right)^3\right]^{-1}\left\{\left[\theta\left(\frac{k_{\rm{F}}}{\Lambda}\right)^3\right]^{-1}
\ln\left[1+\theta\left(\frac{k_{\rm{F}}}{\Lambda}\right)^3\right]
-1\left/\left[1+\theta\left(\frac{k_{\rm{F}}}{\Lambda}\right)^3\right]\right.\right\},\\
K_0(\rho)=&-\frac{3}{5M}\left(\frac{3\pi^2}{2}\right)^{2/3}\rho^{2/3}+\frac{12\beta}{7}\left(\frac{\rho}{\rho_0}\right)^{4/3}
+\frac{15\lambda}{4}\left(\frac{\rho}{\rho_0}\right)^{5/3}
+6\phi\left(\frac{\rho}{\rho_0}\right)^2\notag\\
&+9\mu\left(\frac{\rho}{\rho_0}\right)\left[\theta\left(\frac{k_{\rm{F}}}{\Lambda}\right)^3\right]^{-1}\left\{\left[2+3\theta\left(\frac{k_{\rm{F}}}{\Lambda}\right)^3\right]\left/\left[1+\theta\left(\frac{k_{\rm{F}}}{\Lambda}\right)^3\right]^2\right.
-2\left[\theta\left(\frac{k_{\rm{F}}}{\Lambda}\right)^3\right]^{-1}\ln\left[1+\theta\left(\frac{k_{\rm{F}}}{\Lambda}\right)^3\right]
\right\}\label{3qK0},\\
J_0(\rho)=&\frac{12}{5M}\left(\frac{3\pi^2}{2}\right)^{2/3}\rho^{2/3}
-\frac{24\beta}{7}\left(\frac{\rho}{\rho_0}\right)^{4/3}
-\frac{15\lambda}{4}\left(\frac{\rho}{\rho_0}\right)^{5/3}
+27\mu\left(\frac{\rho}{\rho_0}\right)\left[\theta\left(\frac{k_{\rm{F}}}{\Lambda}\right)^3\right]^{-1}\notag\\
&\times\left\{
2\left[\theta\left(\frac{k_{\rm{F}}}{\Lambda}\right)^3\right]^{-1}\ln\left[1+\theta\left(\frac{k_{\rm{F}}}{\Lambda}\right)^3\right]
-\left[11\left[\theta\left(\frac{k_{\rm{F}}}{\Lambda}\right)^3\right]^2+15\theta\left(\frac{k_{\rm{F}}}{\Lambda}\right)^3+6\right]\left/\left[1+\theta\left(\frac{k_{\rm{F}}}{\Lambda}\right)^3\right]^3\right.\right\}\label{3qJ0},
\end{align}
and,
\begin{align}
I_0(\rho)=&-\frac{84}{5M}\left(\frac{3\pi^2}{2}\right)^{2/3}\rho^{2/3}
+\frac{120\beta}{7}\left(\frac{\rho}{\rho_0}\right)^{4/3}+
15\lambda\left(\frac{\rho}{\rho_0}\right)^{5/3}+81\mu\left(\frac{\rho}{\rho_0}\right)\left[\theta\left(\frac{k_{\rm{F}}}{\Lambda}\right)^3\right]^{-1}\notag\\
&\times\left\{\left[25\left[\theta\left(\frac{k_{\rm{F}}}{\Lambda}\right)^3\right]^3+52\left[\theta\left(\frac{k_{\rm{F}}}{\Lambda}\right)^3\right]^2+42\theta\left(\frac{k_{\rm{F}}}{\Lambda}\right)^3+12\right]\right.\left/2\left[1+\theta\left(\frac{k_{\rm{F}}}{\Lambda}\right)^3\right]^4\right.\notag\\
&\hspace*{6.cm}\left.-2\left[\theta\left(\frac{k_{\rm{F}}}{\Lambda}\right)^3\right]^{-1}\ln\left[1
+\theta\left(\frac{k_{\rm{F}}}{\Lambda}\right)^3\right]\right\}
\label{3qI0}.
\end{align}
The chemical potential is obtained by taking the single-nucleon energy at the Fermi surface, i.e.,
\begin{align}
\mu_0(\rho)=&\frac{1}{2M}\left(\frac{3\pi^2}{2}\right)^{2/3}\rho^{2/3}+\alpha\left(\frac{\rho}{\rho_0}\right)
+\beta\left(\frac{\rho}{\rho_0}\right)^{4/3}
+\lambda\left(\frac{\rho}{\rho_0}\right)^{5/3}
+\phi\left(\frac{\rho}{\rho_0}\right)^{2} +\mu
\left(\frac{\rho}{\rho_0}\right)\left/\left[1+\theta\left(\frac{k_{\rm{F}}}{\Lambda}\right)^3\right]\right..
\end{align}
Using the formula $P_0(\rho)=\rho[\mu_0(\rho)-E_0(\rho)]$ or directly from the definition of the pressure, one obtains (which could also be obtained directly through $P_0(\rho)=L_0(\rho)\rho/3$ using the expression for $L_0(\rho)$ of (\ref{3qL0}))
\begin{align}\label{3qPress}
P_0(\rho)=&\frac{1}{5M}\left(\frac{3\pi^2}{2}\right)^{2/3}\rho^{5/3} +\rho\left[
\frac{\alpha}{2}\left(\frac{\rho}{\rho_0}\right)+\frac{4\beta}{7}\left(\frac{\rho}{\rho_0}\right)^{4/3}
+\frac{5\lambda}{8}\left(\frac{\rho}{\rho_0}\right)^{5/3}+\frac{2\phi}{3}\left(\frac{\rho}{\rho_0}\right)^2\right]\notag\\
&+\mu\rho\left(\frac{\rho}{\rho_0}\right)\left[1\left/\left[1+\theta\left(\frac{k_{\rm{F}}}{\Lambda}\right)^3\right]\right.-1\left/\theta\left(\frac{k_{\rm{F}}}{\Lambda}\right)^3\right.+\left[\theta\left(\frac{k_{\rm{F}}}{\Lambda}\right)^3\right]^{-2}
\ln\left[1+\theta\left(\frac{k_{\rm{F}}}{\Lambda}\right)^3\right]\right].
\end{align}

The density dependence of the scalar nucleon Landau effective mass could be similarly obtained
by taking the momentum at the Fermi surface,
\begin{equation}\label{3qlanmass1}
{
\frac{M_{\rm{s}}^{\ast}}{M}=\left[1-\frac{3\mu
M\theta}{\Lambda^3\rho_0}\left(\frac{3\pi^2}{2}\right)^{1/3}\rho^{4/3}\left(1+\frac{3\pi^2\rho\theta
}{2\Lambda^3}\right)^{-2}\right]^{-1},}
\end{equation}
which is a function of density alone,
\begin{equation}\label{psi-d}
\frac{M_{\rm{s}}^{\ast}}{M}=1\left/\left(1-\frac{\Xi\rho^{4/3}}{(1+\nu\rho)^2}\right)\right.,~~
\nu=\frac{3\pi^2\theta}{2\Lambda^3},~~\Xi=\frac{3\mu
M\theta}{\Lambda^3\rho_0}\left(\frac{3\pi^2}{2}\right)^{1/3}.
\end{equation}
Assuming that the density is small, one obtains 
$
{M_{\rm{s}}^{\ast}}/{M}\approx1+
\nu\rho^{4/3}-2\nu\Xi\rho^{7/3}+3\nu \Xi^2\rho^{10/3}$.

\renewcommand*\tablename{\footnotesize Tab.}
\begin{table}[h!]\label{3qtheta}
\centering
\begin{tabular}{c*{1}|{c}}\hline
parameter&value\\ \hline\hline $\theta$&1.3\\ \hline $\Lambda$&750.0\,MeV\\
\hline
\end{tabular}
\caption{Parameters $\theta$ and $\Lambda$.}\label{tab_c1}
\end{table}

\renewcommand*\tablename{\footnotesize Tab.}
\begin{table}[h!]
\centering
\begin{tabular}{c |c |c |c |c| c}\hline
&$\alpha$&$\beta$&$\lambda$&$\phi$&$\mu$\\\hline
\hline
I &$-109.5032$&210.7910&$-47.7253$&9.2475&$-122.1256$ \\\hline
II &$-126.1699$&245.7911&$-61.0587$&4.2475&$-122.1256$ \\
\hline
\end{tabular}\caption{Two sets of parameters in which the $\rho_0\approx0.16\,\rm{fm}^{-3}$, $E_0(\rho_0)\approx-16.0\,\rm{MeV}
$, and $M_{\rm{s}}^{\ast}(\rho_0)/M\approx0.8$ are fixed. Unit: MeV.}\label{3qpara_aa}
\end{table}

\renewcommand*\tablename{\footnotesize Tab.}
\begin{table}[h!]
\centering
\begin{tabular}{c|c|c|c}\hline
&$K_0$&$J_0$&$I_0$\\
\hline\hline
I &230.0&$-380.0$&1666.4 \\\hline
II&210.0&$-450.0$&2066.4 \\
\hline
\end{tabular}\caption{The values of $K_0,J_0$ and $I_0$. Unit: MeV.}\label{3qpara_bb}
\end{table}

\renewcommand*\figurename{\footnotesize Fig.}
\begin{figure}[h!]
\centering
 \hspace*{-.5cm} 
\includegraphics[height=6.8cm]{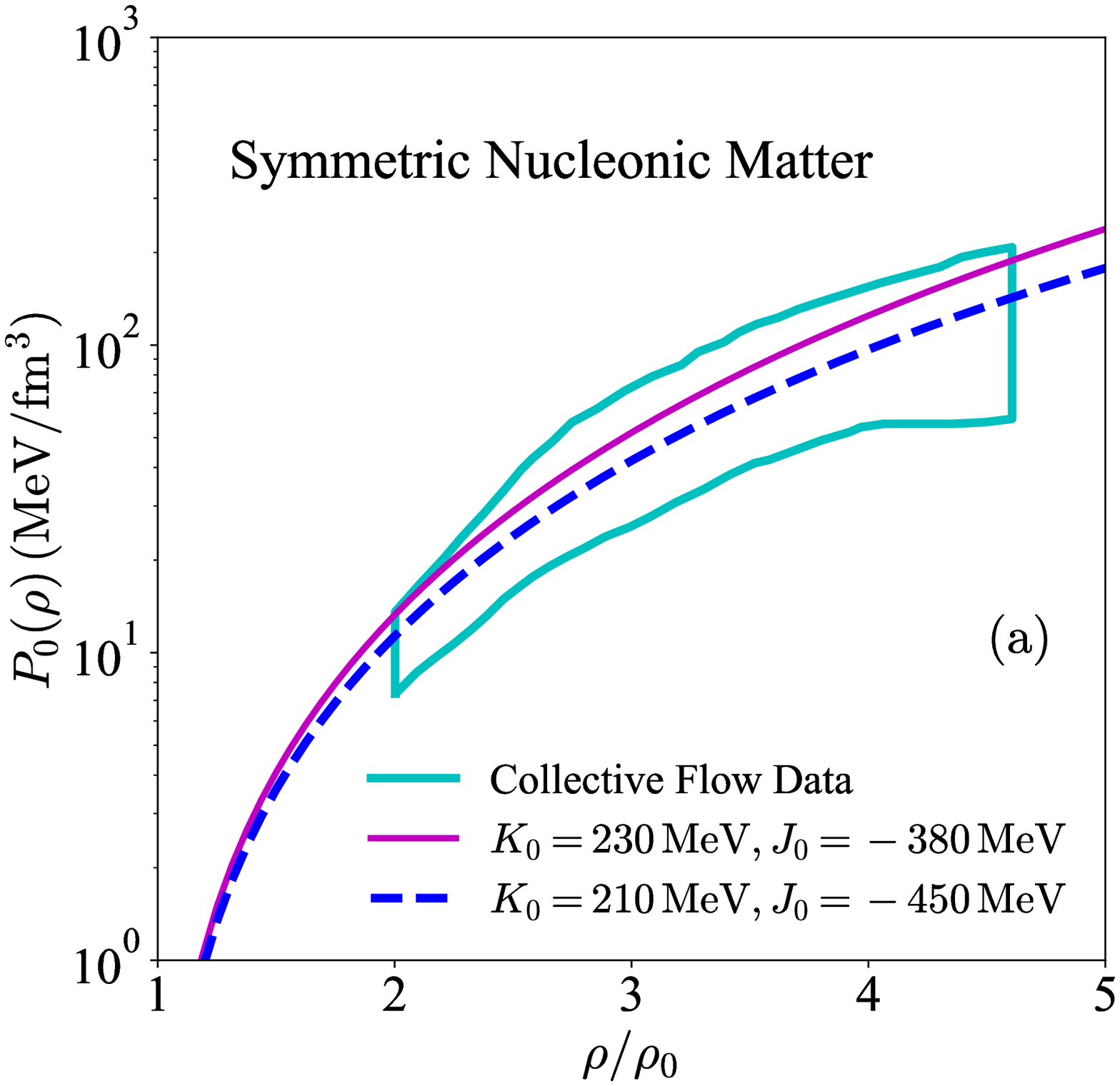}\qquad
\includegraphics[height=6.8cm]{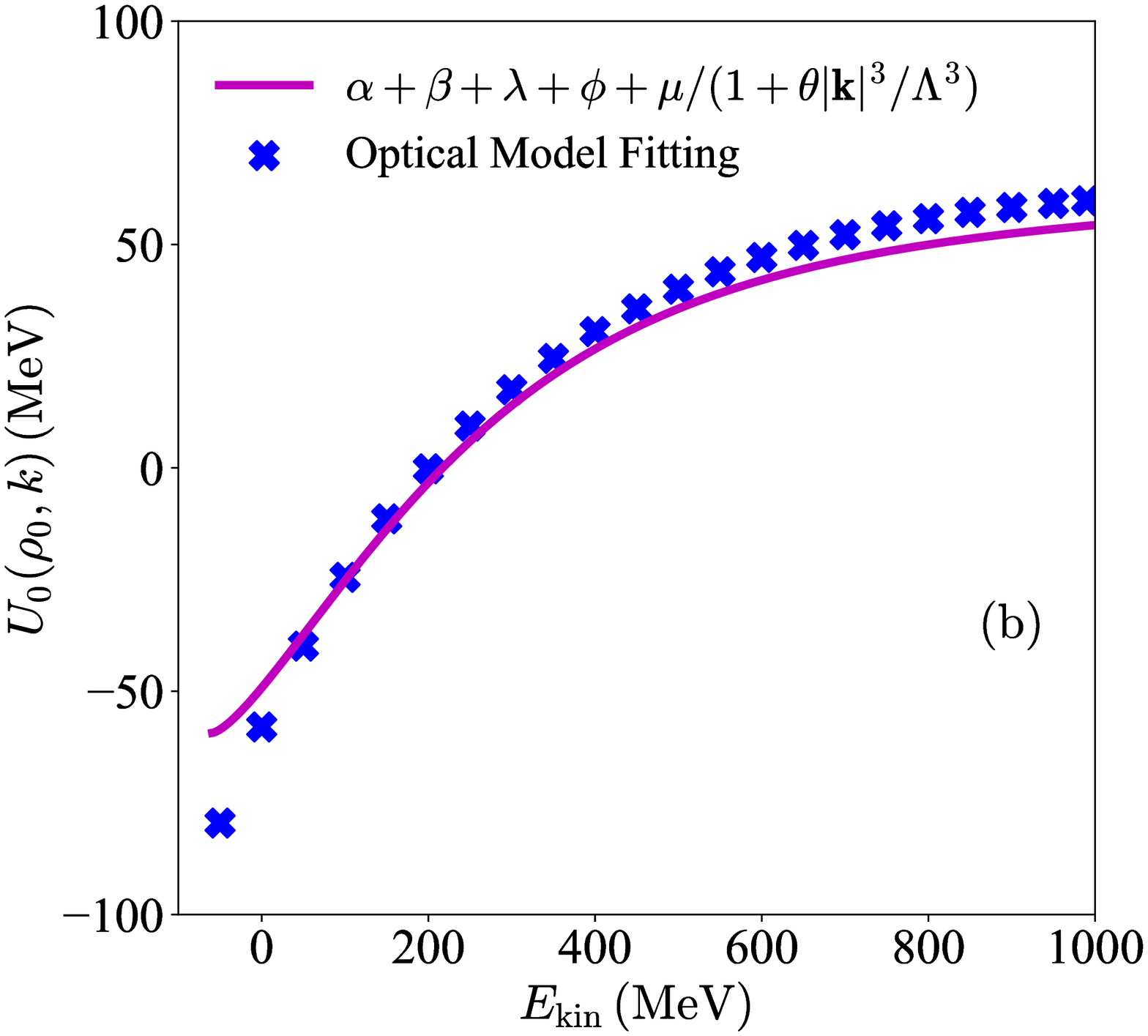}
  \caption{The EOS of SNM either using the hard set or the soft set could pass through the constraint from collective flow data (left) and the single-nucleon potential $U_0(\rho_0,\v{k})$ as a function of the nucleon kinetic energy (right).}\label{fig_toyFlow}
\end{figure}

\begin{figure}[h!]
\centering
\hspace*{-.8cm}\includegraphics[width=7.cm]{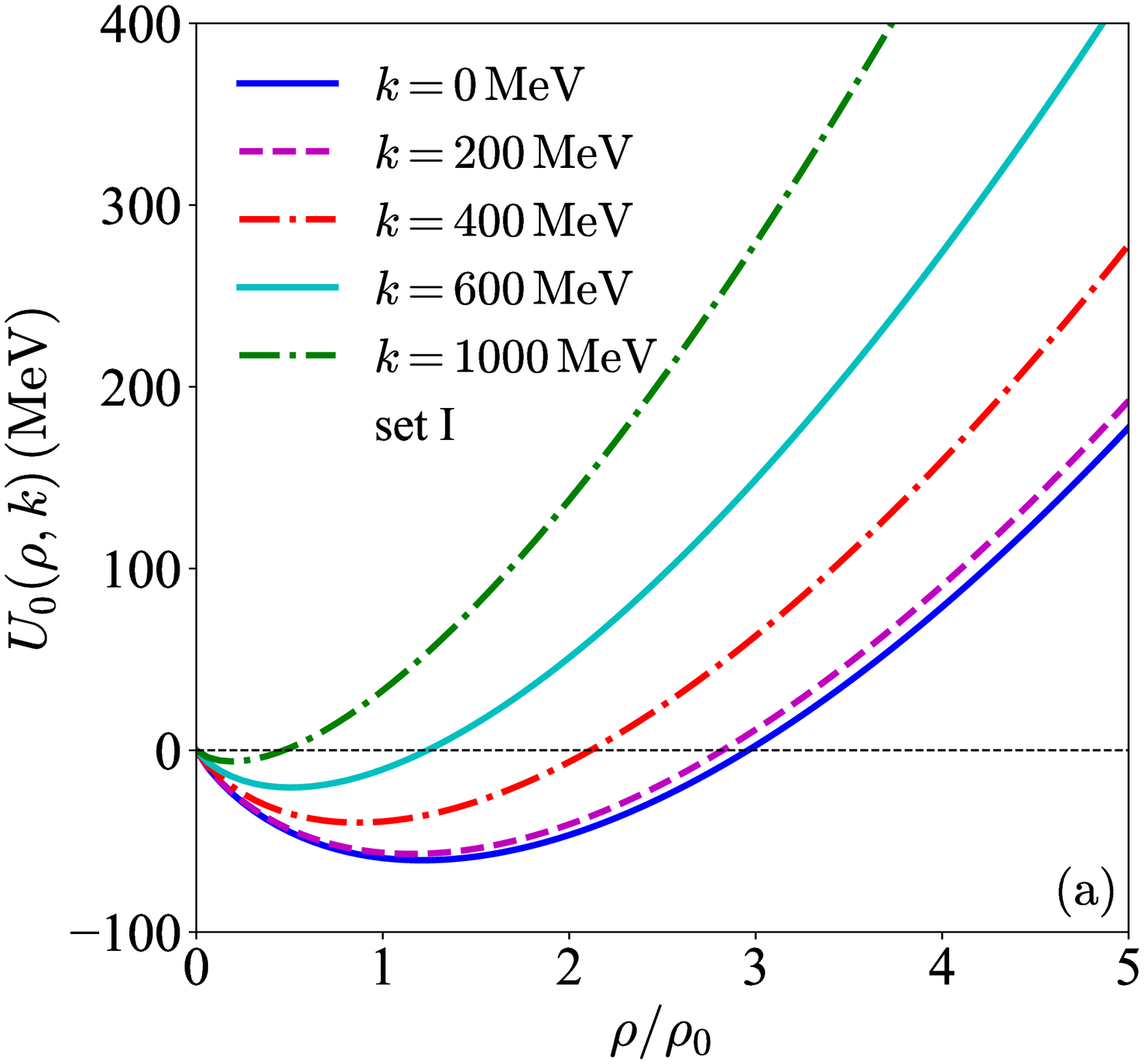}\qquad
\includegraphics[width=7.cm]{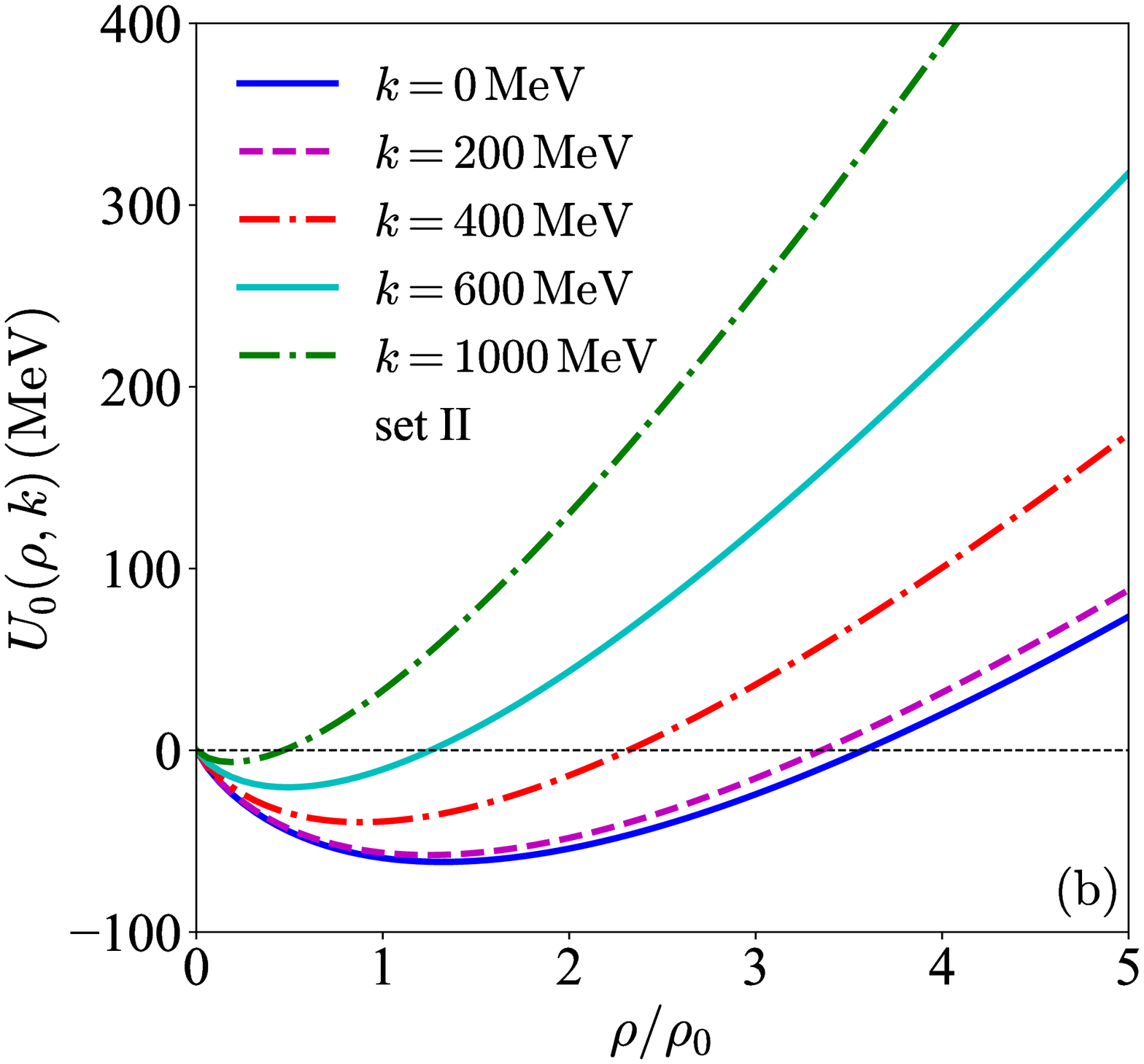}
  \caption{Single-nucleon potential $U_0(\rho,\v{k})$ as a function of density at different momentum.}\label{fig_toyU0-rho}
\end{figure}

The above derivation gives the whole series of the analytical formula for the model considered. 
The 7 parameters for the SNM EOS are determined by the empirical constraints $
\rho_0\approx0.16\,\rm{fm}^{-3},
~E_0(\rho_0)\approx-16.0\,\rm{MeV},
~K_0= K^{\rm{SNM}}\approx230\,\rm{MeV},
~J_0\approx-380\,\rm{MeV}$ and the Landau effective mass as
$M_{\rm{s}}^{\ast}(\rho_0)/M\approx0.8$.
Moreover the parameters $\theta=1.3$ and $\Lambda=750\,\rm{MeV}$ are pre-fixed, see Tab.\,\ref{tab_c1}, consequently $x\approx0.056\rho/\rho_0$.
Besides the default set of $
K_0\approx230.0\,\rm{MeV}$ and $J_0\approx-380\,\rm{MeV}$, we also include another interaction parameter set with $K_0\approx210.0\,\rm{MeV}$ and $J_0\approx-450\,\rm{MeV}$, see Tab.\,\ref{3qpara_aa} and Tab.\,\ref{3qpara_bb} for the relevant quantities and parameters.
It is obvious from Tab.\,\ref{3qpara_bb} that the ``soft'' EOS on the other hand has a larger kurtosis coefficient $I_0$, i.e., the ``softness'' and the ``hardness'' of these parameter sets are only relative.

\begin{figure}[h!]
\centering
\includegraphics[width=7.cm]{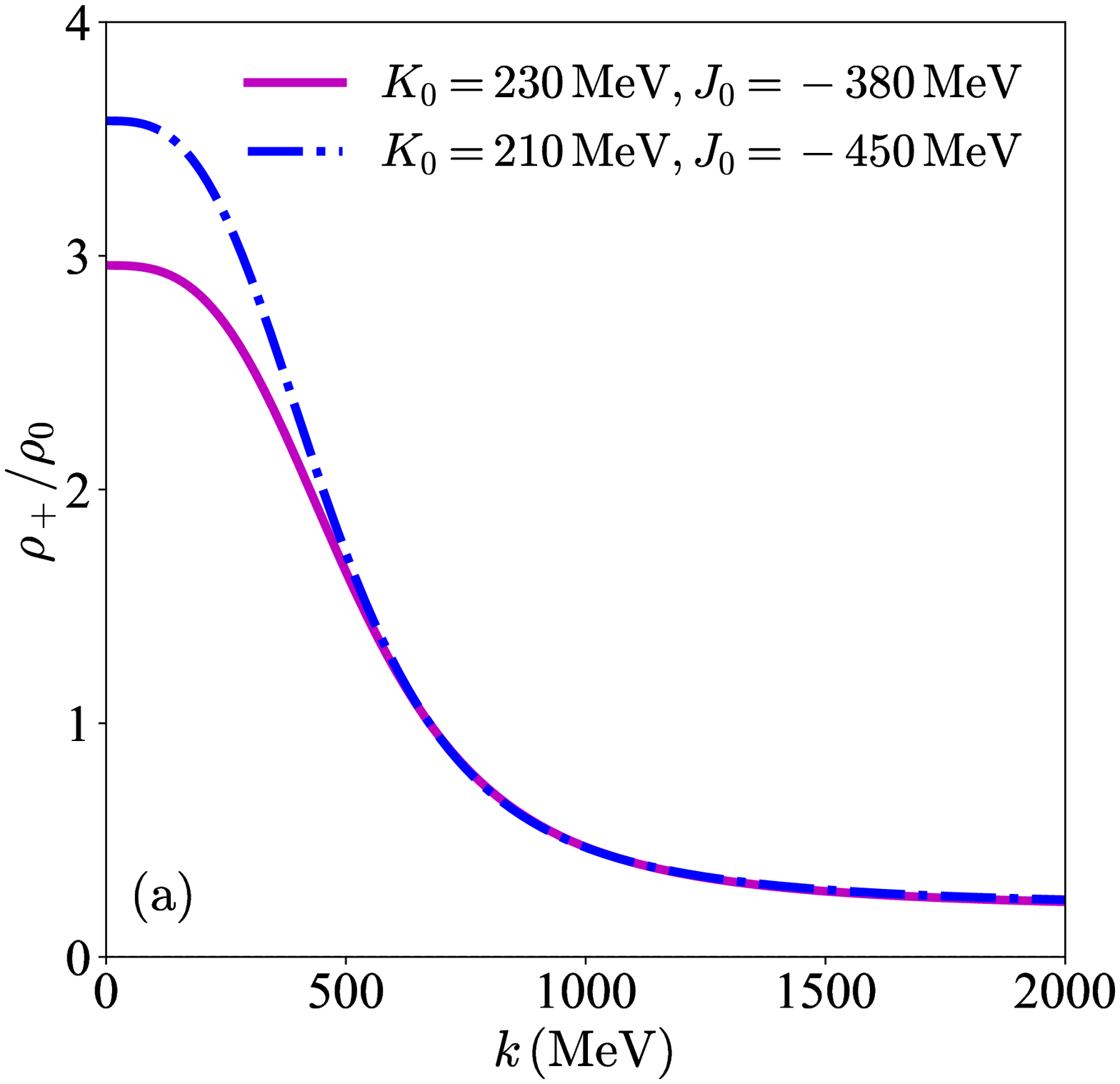}\qquad
\includegraphics[width=7.cm]{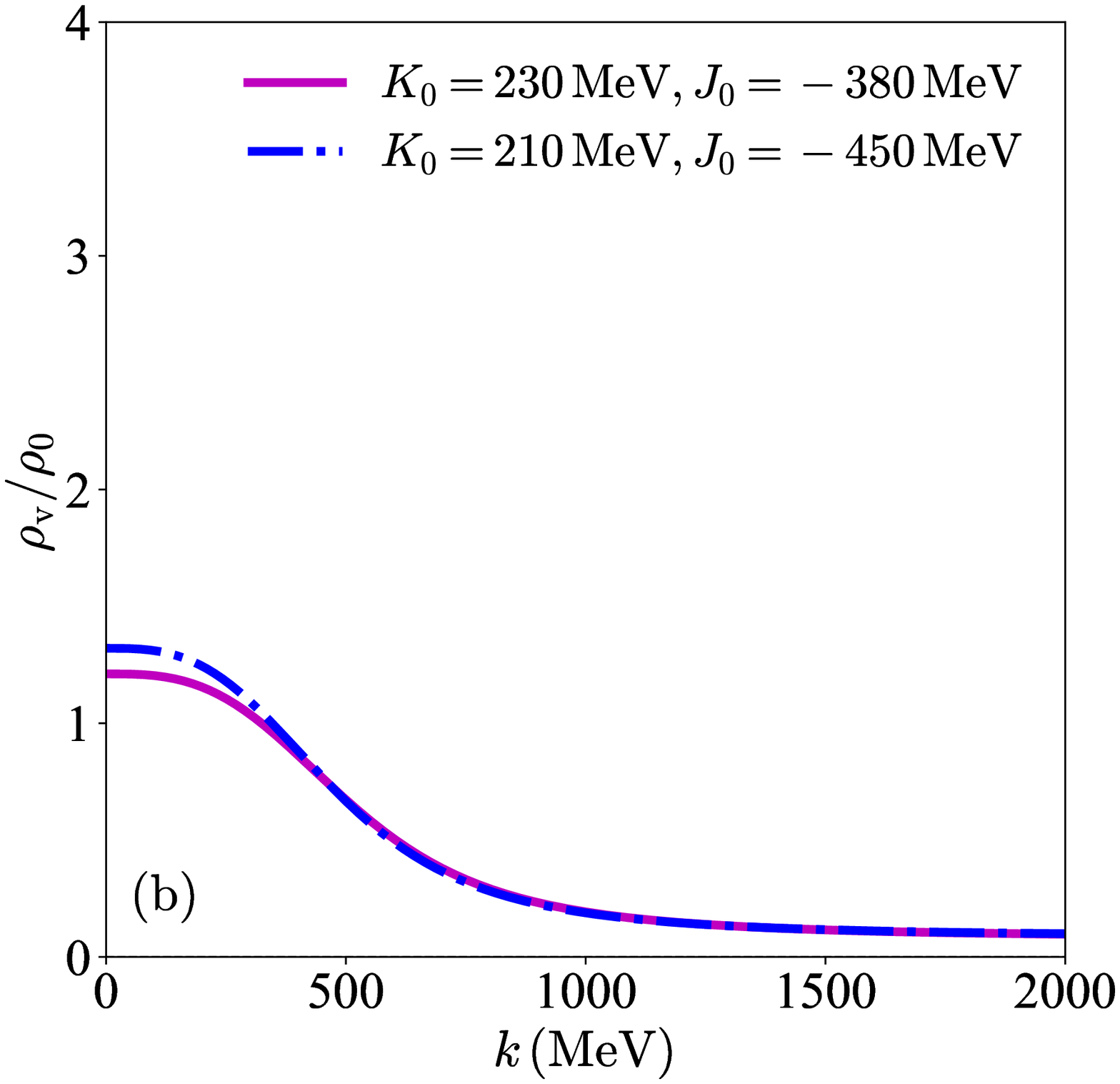}
  \caption{Momentum dependence of the crossing density (left) and the density $\rho_{\rm{v}}$ (right).}\label{fig_toyU0-rhoplus}
\end{figure}

One can find that the pressure (\ref{3qPress}) in SNM could safely pass through its empirical constraining band from analyzing the collective flow data in relativistic heavy-ion collisions\,\cite{Pawel2002Science}, see the left panel of Fig.\,\ref{fig_toyFlow}. In addition, the expression for the single-nucleon potential at the saturation density in SNM is very simple, i.e.,
\begin{equation}
U_0(\rho_0,\v{k})=\alpha+\beta+\lambda+\phi+\mu\left/\left(1+\frac{\theta|\v{k}|^3}{\Lambda^3}\right)\right..
\end{equation}
In the right window of Fig.\,\ref{fig_toyFlow} the kinetic energy dependence of the single-nucleon potential $U_0(\rho_0,\v{k})$ is shown,  with the corresponding kinetic energy defined as $E_{\rm{kin}}=
[\v{k}^2+M^2]^{1/2}+U_0(\rho_0,\v{k})-M$\,\cite{WangR2018,Feldmeier1991}.
Also shown is the experimental constraint on the single-nucleon potential obtained from the optical model analysis of nucleon-nucleus scattering data\,\cite{Hama1990}.
Since the $\alpha+\beta+\lambda+\phi$ in the two parameter sets are the same, i.e., about 62.8\,MeV, it is unnecessary to distinguish the two parameter sets.
Although the momentum dependence of the $U_0(\rho_0,\v{k})$ is very simple it can successfully model the constraint from the optical model analyses.
Similarly shown in Fig.\,\ref{fig_toyU0-rho} is the density dependence of the single-nucleon potential at different momentum.
As the density increases the $U_0(\rho,k)$ eventually increases from negative values to positive at the crossing density $\rho_+$.
The crossing density $\rho_+$ itself decreases as the momentum increases, i.e., in a system with high-momentum nucleons the effective interaction between nucleons starts to be repulsive earlier than a system with low-momentum particles as density increases.
The crossing density $\rho_+$ is obtained as follows,
\begin{equation}
\alpha\left(\frac{\rho_+}{\rho_0}\right)+\beta
\left(\frac{\rho_+}{\rho_0}\right)^{4/3}+\lambda\left(\frac{\rho_+}{\rho_0}\right)^{5/3}+\phi\left(\frac{\rho_+}{\rho_0}\right)^2+\mu\left(\frac{\rho_+}{\rho_0}\right)\left/\left(1+\frac{\theta|\v{k}|^3}{\Lambda^3}\right)\right.=0.
\end{equation}
See the left panel of Fig.\,\ref{fig_toyU0-rhoplus} for the momentum dependence of the crossing density with the two parameter sets.
One can see that as the momentum increases the difference between the $\rho_+$'s obtained from the hard and soft EOSs becomes small.
Another characteristic density is the one at which the $U_0$ takes the minimum, i.e., $\rho_{\rm{v}}$, see the relevant discussion in section \ref{SEC_2D} and the right panel of Fig.\,\ref{fig_toyU0-rhoplus}.
In section \ref{SEC_2D} we have shown for the 2D SNM EOS that generally $\rho_+/\rho_0\geq9/4$ and $\rho_{\rm{v}}/\rho_0\geq9/16$ (see Fig.\,\ref{fig_U0rho_p}), the features shown in Fig.\,\ref{fig_toyU0-rhoplus} are thus consistent with those analyses.

\renewcommand*\figurename{\footnotesize Fig.}
\begin{figure}[h!]
\centering
\includegraphics[width=7.cm]{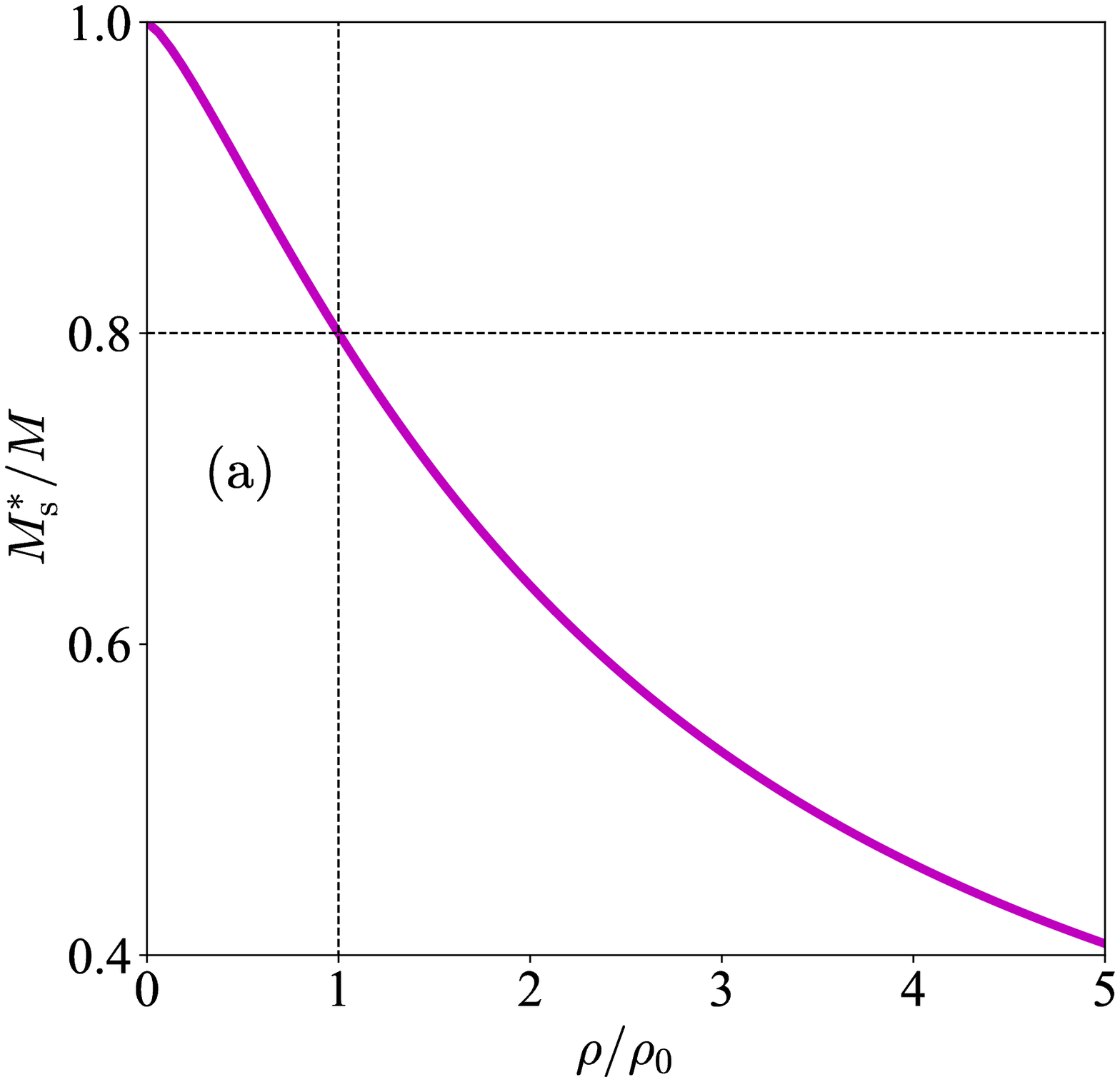}\qquad
\includegraphics[width=7.2cm]{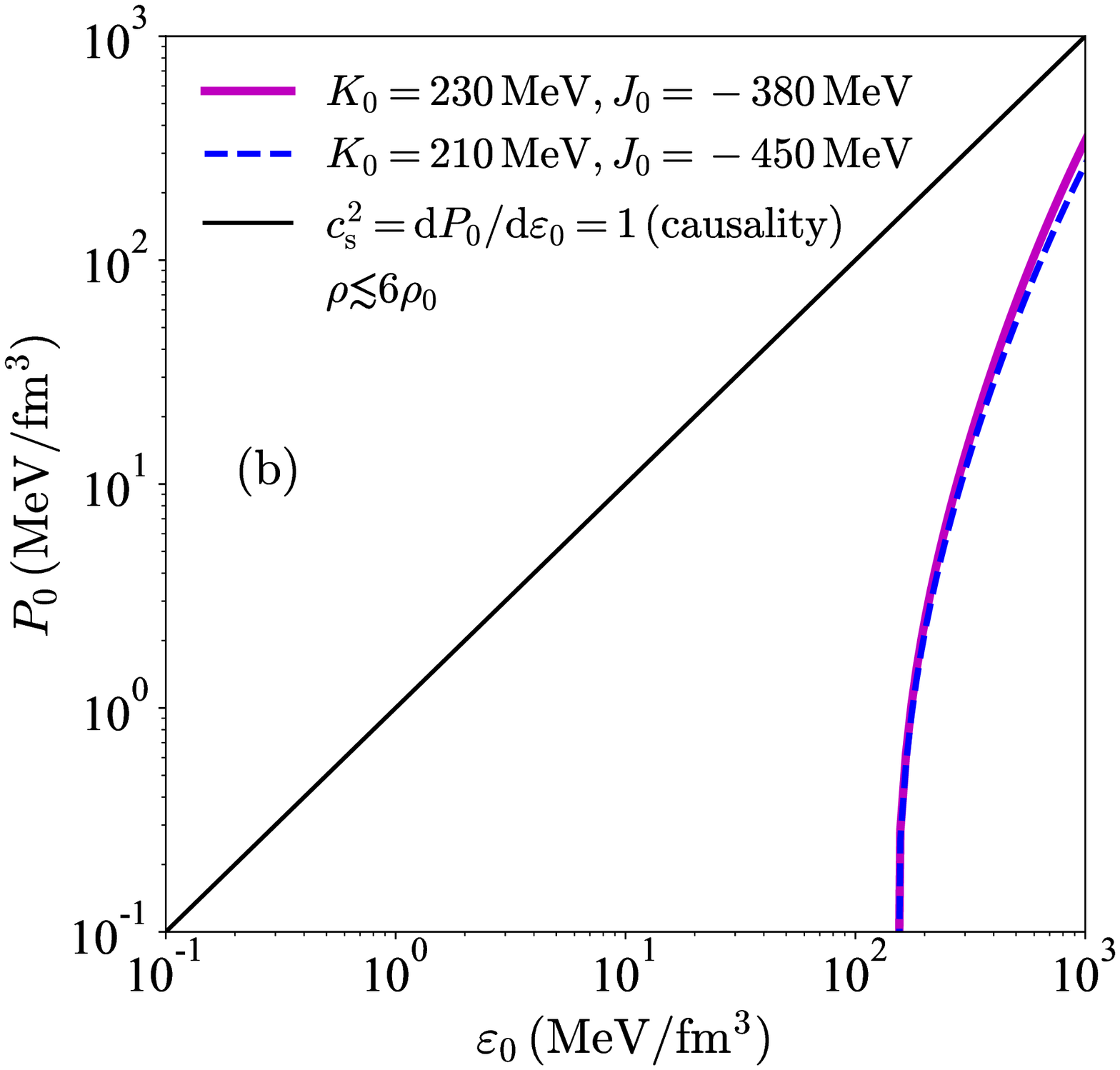}
  \caption{Density dependence of the scalar effective mass (light) and the relation between $P_0(\rho)$ and $\varepsilon_0(\rho)$ (right).}\label{fig_toyP0eps0}
\end{figure}

Since the Landau effective mass only depends on the momentum-dependent part of the single-nucleon potential, the $M_{\rm{s}}^{\ast}/M$ obtained with the hard and soft EOSs thus are the same, given by
\begin{equation}
\frac{M_{\rm{s}}^{\ast}(\rho,\v{k})}{M}=1\left/\left[
1-3\mu\theta\left(\frac{\rho}{\rho_0}\right)\frac{M}{\Lambda^3}|\v{k}|
\left/\left(1+\theta\left(\frac{|\v{k}|}{\Lambda}\right)^3\right)\right.
\right]\right..
\end{equation}
Its density dependence has the following asymptotic behavior, 
\begin{equation}
\frac{M_{\rm{s}}^{\ast}}{M}\sim\frac{1}{1+\Phi\rho^{1/3}},~~\Phi>0,\end{equation}
and consequently $
\lim_{\rho\to\infty}M_{\rm{s}}^{\ast}/M=1$. The density dependence of the effective mass is shown in the left panel of Fig.\,\ref{fig_toyP0eps0}.
It should be pointed out that there is no empirical constraints on the effective mass at very large densities.
However, if one wants to make the effective mass decrease as $\rho$ increases, one could introduce/improve the overall density factor $\Pi(\rho/\rho_0)
$ of the momentum-dependent term. This factor is simply $\rho/\rho_0$ as adopted in (\ref{3qU0}).
Specifically, if one uses $
\Pi(\rho/\rho_0)=(\rho/\rho_0)^y
$, the asymptotic behavior of the Landau effective mass at large densities is given by $
{M_{\rm{s}}^{\ast}}/{M}\sim(1+\Phi'\rho^{y-({r+2})/{3}})^{-1}$, where $\Phi'>0$ and $r$ is the index appearing in the momentum-dependent term (e.g., it is given by $r=3$ via $|\v{k}|^3/\Lambda^3$ in (\ref{3qU0})).
If one forces $
y\geq(r+2)/3$, the Landau effective mass of a nucleon will decrease as the density increases.
Finally, the relation between the $P_0(\rho)$ and the energy density $\varepsilon_0(\rho)$ is shown in the right panel of Fig.\,\ref{fig_toyP0eps0}, where the black line corresponds to the limit set by the causality condition on the sound velocity $c_{\rm{s}}$ in nuclear matter, i.e.,
\begin{equation}
c_{\rm{s}}^2=\d P_0/\d\varepsilon_0|_{s}\leq1.
\end{equation}
It is seen that the causality is respected to very high energy densities relevant for the problems discussed in the present work.

\begin{figure}[h!]
\centering
\includegraphics[width=7.cm]{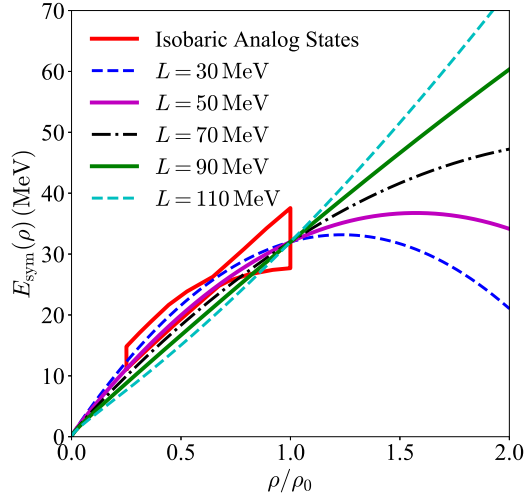}
\caption{Symmetry energy with different $L$ and fixed $E_{\rm{sym}}(\rho_0)=32\,\rm{MeV}$.}\label{fig_toyEsym}
\end{figure}

Furthermore, the symmetry energy can be obtained straightforwardly using the HVH theorem, i.e.,
\begin{align}
E_{\rm{sym}}(\rho)
=\frac{1}{6M}\left(\frac{3\pi^2}{2}\right)^{2/3}\rho^{2/3}+
\frac{1}{6}\beta_{\rm{IS}}\left(\frac{\rho}{\rho_0}\right)^{4/3}+\frac{1}{3}\lambda_{\rm{IS}}
\left(\frac{\rho}{\rho_0}\right)^{5/3}
-\frac{1}{2}\mu\left(\frac{\rho}{\rho_0}\right)\theta\left(\frac{k_{\rm{F}}}{\Lambda}\right)^3\left/\left[1+\theta\left(\frac{k_{\rm{F}}}{\Lambda}\right)^3\right]^2\right..
\end{align}
Similarly, the slope parameter of the symmetry energy is obtained as,
\begin{equation}\label{def_L_tt}
{
L(\rho)=\frac{1}{3M}\left(\frac{3\pi^2}{2}\right)^{2/3}\rho^{2/3}+
\frac{2}{3}\beta_{\rm{IS}}\left(\frac{\rho}{\rho_0}\right)^{4/3}+\frac{5}{3}\lambda_{\rm{IS}}
\left(\frac{\rho}{\rho_0}\right)^{5/3}-3\mu\left(\frac{\rho}{\rho_0}\right)
\theta\left(\frac{k_{\rm{F}}}{\Lambda}\right)^3\left/\left[1+\theta\left(\frac{k_{\rm{F}}}{\Lambda}\right)^3\right]^3\right..}
\end{equation}
In this situation one could use the empirical values for  $E_{\rm{sym}}(\rho_0)\equiv
S$ and $L(\rho_0)\equiv{L}$ to determine the parameters $\beta_{\rm{IS}}$ and $\lambda_{\rm{IS}}$.
Consequently, the isospin-related parameters are found to be \begin{align}
\beta_{\rm{IS}}=&6[5E_{\rm{sym}}(\rho_0)-{L}-3E^{\rm{kin}}_{\rm{sym}}(\rho_0)-5X+Y],\\
\lambda_{\rm{IS}}=&3[2E^{\rm{kin}}_{\rm{sym}}(\rho_0)+4X-Y-4E_{\rm{sym}}(\rho_0)+{L}],
\end{align} where
\begin{align}
E_{\rm{sym}}^{\rm{kin}}(\rho_0)=&{k_{\rm{F}}^2(\rho_0)}/{6M},~~\\
X=&-
\frac{1}{2}\mu\theta\left(\frac{k_{\rm{F}}(\rho_0)}{\Lambda}\right)^3\left/\left[1+\theta\left(\frac{k_{\rm{F}}(\rho_0)}{\Lambda}\right)^3\right]^2\right.,~~\\
Y=&-
3\mu\theta\left(\frac{k_{\rm{F}}(\rho_0)}{\Lambda}\right)^3\left/\left[1+\theta\left(\frac{k_{\rm{F}}(\rho_0)}{\Lambda}\right)^3\right]^3\right.
.\end{align} 
In Fig.\,\ref{fig_toyEsym} the symmetry energy functions with different $L$ values and a fixed $E_{\rm{sym}}(\rho_0)=32\,\rm{MeV}$ are shown.
For a comparison, the constraint on the symmetry energy from the isobaric analog states (IAS)\,\cite{Pawel2014} is also shown.
It is obvious that at large densities the symmetry energy may increase or decrease depending on the value of $L$. For example, with $L=50\,\rm{MeV}$, the symmetry energy starts to decrease at about 1.6\,$\rho_0$.
It is necessary to point out that the density dependence of the symmetry energy given here is so simple that when generalized to large densities one should be more cautious.
All the other higher order coefficients could be obtained from the expressions given above, e.g., the curvature of the symmetry energy,
\begin{align}
K_{\rm{sym}}=&6E^{\rm{kin}}_{\rm{sym}}(\rho_0)+6{L}-20E_{\rm{sym}}(\rho_0)\notag\\
&-\mu\left[10\left[\theta\left(\frac{k_{\rm{F}}(\rho_0)}{\Lambda}\right)^3\right]^3-16\left[\theta\left(\frac{k_{\rm{F}}(\rho_0)}{\Lambda}\right)^3\right]^2+\theta\left(\frac{k_{\rm{F}}(\rho_0)}{\Lambda}\right)^3\right]\left/\left[1+\theta\left(\frac{k_{\rm{F}}(\rho_0)}{\Lambda}\right)^3\right]^4\right..
\end{align}
If one takes $\rho_0\approx0.16\pm0.02\,\rm{fm}^{-3},E_{\rm{sym}}(\rho_0)\approx30\pm4\,\rm{MeV},L\approx60\pm30\,\rm{MeV}$, then $
K_{\rm{sym}}\approx-166\pm197\,\rm{MeV}$, which is also reasonable with respect to the results currently available in the literature\,\cite{ZhouY2019PRD,CaiBJ2021PRC-IN,LiBA2020PRC-Ksym,XieWJ2019,
XieWJ2020,Choi2021,Mondal2017,Holt2018PLB,Tews2018,Essick2021}.

\begin{figure}[h!]
\centering
\includegraphics[height=7.cm]{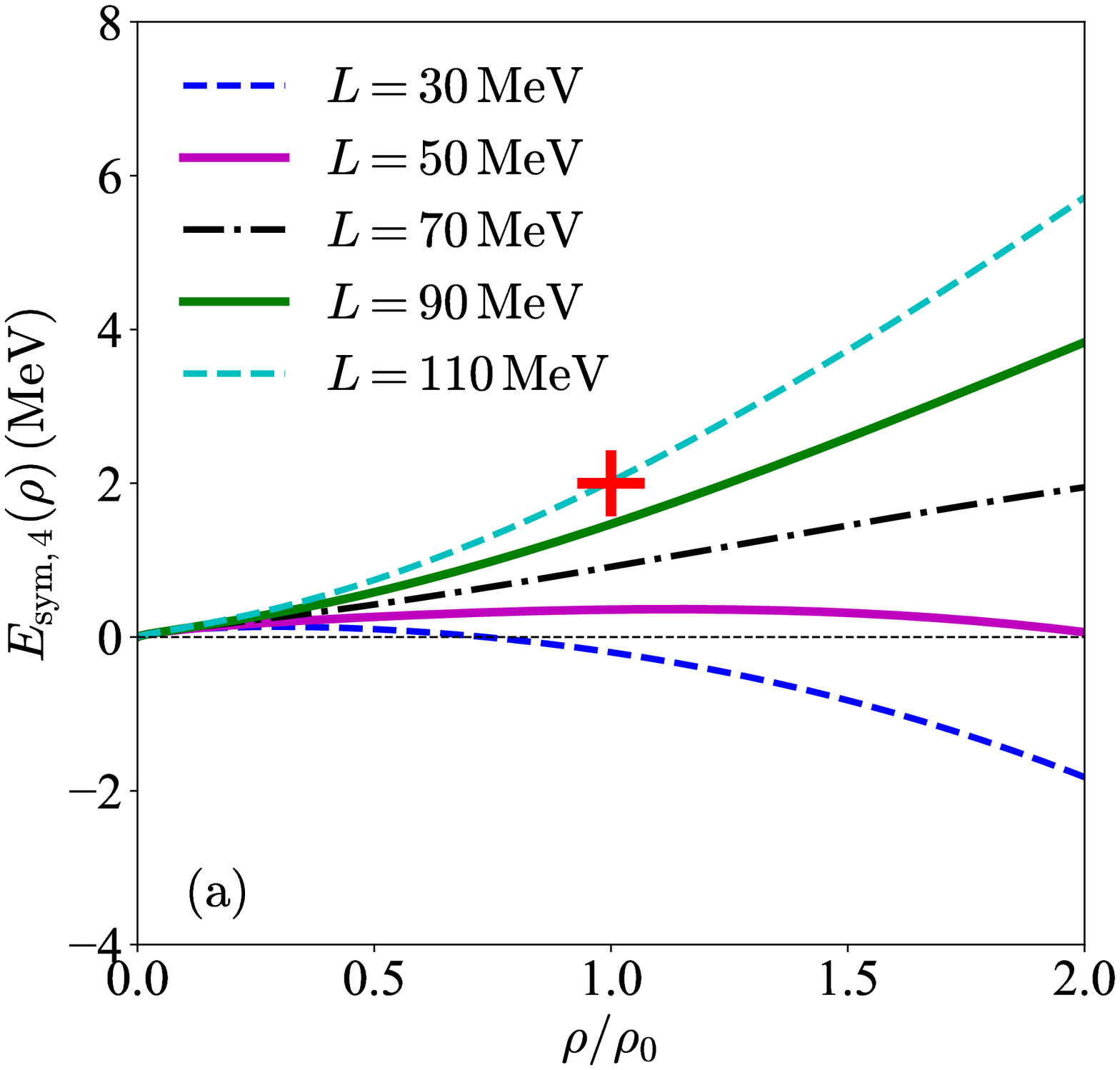}\qquad
\includegraphics[height=7.cm]{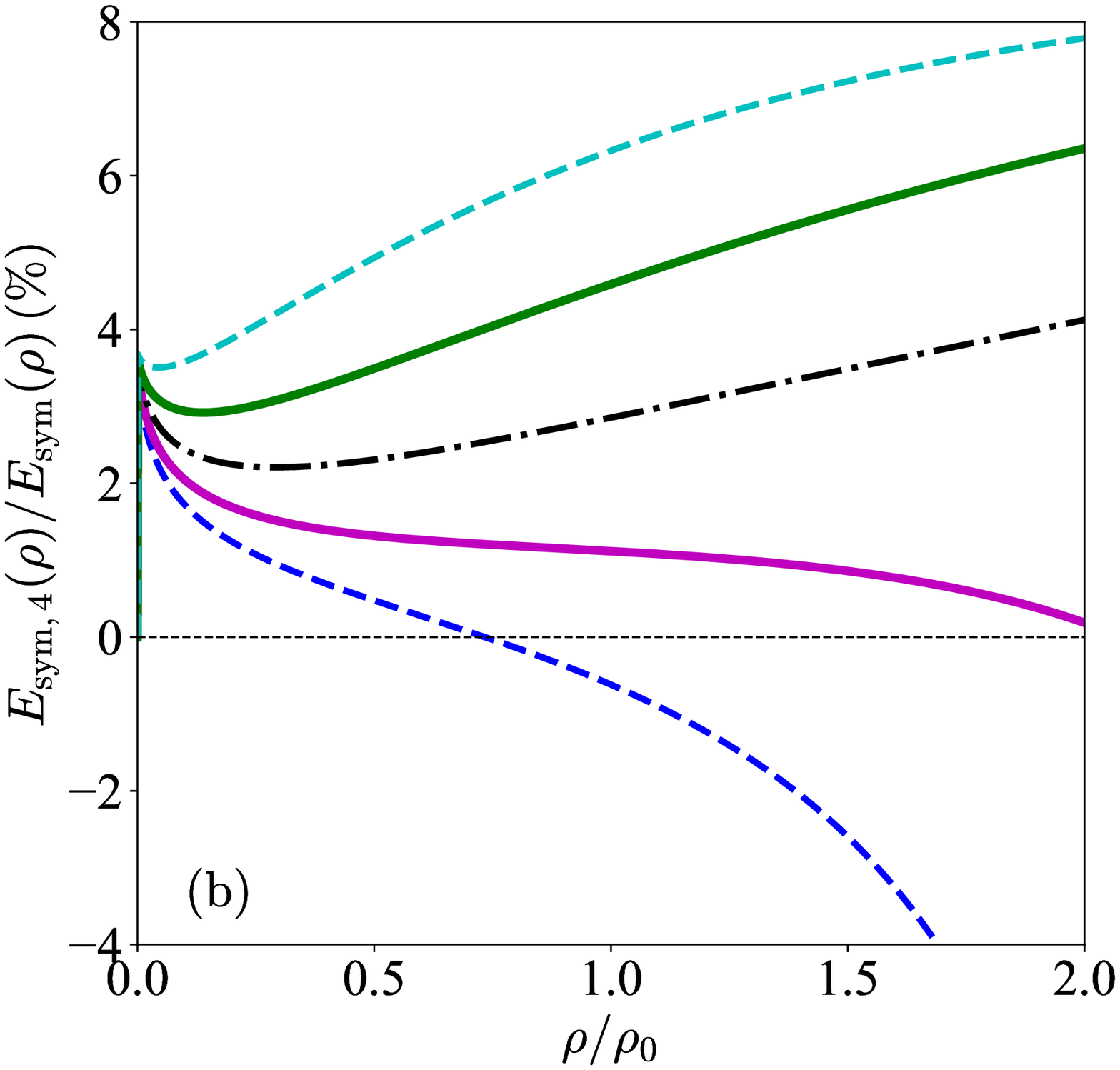}
\caption{Fourth-order symmetry energy with different $L$ and fixed $E_{\rm{sym}}(\rho_0)=32\,\rm{MeV}$ (left) and the ratio $E_{\rm{sym,4}}(\rho)/E_{\rm{sym}}(\rho)$ (right).}\label{fig_toyEsym4}
\end{figure}

In addition, the fourth-order symmetry energy could also be obtained via the HVH theorem as,
\begin{align}
{
E_{\rm{sym},4}(\rho)=\frac{k_{\rm{F}}^2}{162M}+\frac{1}{648}\beta_{\rm{IS}}\left(\frac{\rho}{\rho_0}\right)^{4/3}+\frac{1}{81}\lambda_{\rm{IS}}
\left(\frac{\rho}{\rho_0}\right)^{5/3}-\frac{1}{4}\mu\left(\frac{\rho}{\rho_0}\right)\left[\theta\left(\frac{k_{\rm{F}}}{\Lambda}\right)^3\right]^3\left/\left[1+\theta\left(\frac{k_{\rm{F}}}{\Lambda}\right)^3\right]^4\right.}.
\end{align}
In this toy model, the fourth-order symmetry energy is totally determined by the symmetry energy and the EOS of SNM via the parameters $\beta_{\rm{IS}}, ~\lambda_{\rm{IS}}$ and the parameter $\mu$.
See Fig.\,\ref{fig_toyEsym4} for the density dependence of the $E_{\rm{sym},4}(\rho)$ (left) and that of the ratio $E_{\rm{sym,4}}(\rho)/E_{\rm{sym}}(\rho)$ (right), adopting different values of $L$ while fixing the $E_{\rm{sym}}(\rho_0)$ at 32\,MeV.
We have found that the fourth-order symmetry energy $E_{\rm{sym,4}}(\rho_0)$ at saturation density is generally $\lesssim2\,\rm{MeV}$ (indicated by the red ``+''). This value is also consistent with available empirical findings\,\cite{Bombaci1991,CaiBJ2012PRC-4th,PuJ2017PRC,Boquera2017PRC}.

\begin{figure}[h!]
\centering
\includegraphics[width=7.cm]{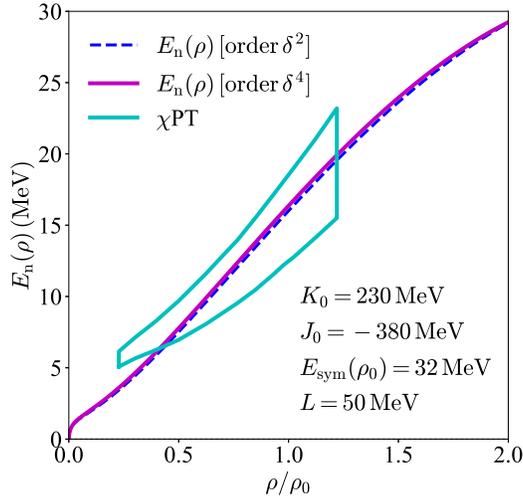}
\caption{Toy model EOS for PNM obtained in the parabolic approximation and the one including the fourth-order symmetry energy in comparison with the chiral perturbation theory prediction.}\label{fig_toyEnapp}
\end{figure}

Finally, the EOS of pure neutron matter (PNM) obtained in the parabolic approximation and the one including the fourth-order symmetry energy can then be obtained, i.e., $E_{\rm{n}}(\rho)\approx E_0(\rho)+E_{\rm{sym}}(\rho)$ and $E_{\rm{n}}(\rho)\approx E_0(\rho)+E_{\rm{sym}}(\rho)+E_{\rm{sym,4}}(\rho)$, respectively.
The expression for the latter is given by
\begin{align}
E_{\rm{n}}(\rho)\approx&\frac{383}{810}\frac{k_{\rm{F}}^2}{M}
+\frac{\alpha}{2}\left(\frac{\rho}{\rho_0}\right)
+\left(\frac{3\beta}{7}+\frac{109\beta_{\rm{IS}}}{648}\right)\left(\frac{\rho}{\rho_0}\right)^{4/3}+\left(\frac{3\lambda}{8}+\frac{28\lambda_{\rm{IS}}}{81}\right)\left(\frac{\rho}{\rho_0}\right)^{5/3}\notag\\
&+\frac{\phi}{3}\left(\frac{\rho}{\rho_0}\right)^2
+\mu\left(\frac{\rho}{\rho_0}\right)\left[\frac{x-\ln(1+x)}{x^2}
-\frac{1}{2}\frac{x}{(1+x)^2}-\frac{1}{4}\frac{x^3}{(1+x)^4}\right],
\end{align}
with $x=\theta(k_{\rm{F}}/\Lambda)^3$.
For small values of $x$, the above $E_{\rm{n}}(\rho)$ becomes,
\begin{align}
E_{\rm{n}}(\rho)\approx&\frac{383}{810}\frac{k_{\rm{F}}^2}{M}
+\frac{\alpha}{2}\left(\frac{\rho}{\rho_0}\right)
+\left(\frac{3\beta}{7}+\frac{109\beta_{\rm{IS}}}{648}\right)\left(\frac{\rho}{\rho_0}\right)^{4/3}+\left(\frac{3\lambda}{8}+\frac{28\lambda_{\rm{IS}}}{81}\right)\left(\frac{\rho}{\rho_0}\right)^{5/3}\notag\\
&+\frac{\phi}{3}\left(\frac{\rho}{\rho_0}\right)^2
+\frac{1}{2}\mu\left(\frac{\rho}{\rho_0}\right)\left[1-\frac{5}{3}\theta\left(\frac{k_{\rm{F}}}{\Lambda}\right)^3+\frac{5}{2}\left[\theta\left(\frac{k_{\rm{F}}}{\Lambda}\right)^3\right]^2-\frac{39}{10}\left[\theta\left(\frac{k_{\rm{F}}}{\Lambda}\right)^3\right]^3+\frac{19}{3}\left[\theta\left(\frac{k_{\rm{F}}}{\Lambda}\right)^3\right]^4\right].
\end{align}
Numerically, we found that the fourth-order symmetry energy induces a contribution $\lesssim0.5\,\rm{MeV}$ to the PNM EOS at densities $\rho\approx2\rho_0$, see Fig.\,\ref{fig_toyEnapp}. For a comparison, the constraint on the PNM EOS from the chiral perturbation theory ($\chi\rm{PT}$)\,\cite{Tews2013} is also shown. It is seen that the constructed toy model satisfies the constraint from the chiral perturbation theory.

So far we have constructed a phenomenological single-nucleon potential $U_J(\rho,\delta,\v{k})$ in the form of (\ref{5q5J}). The resulting EOS of ANM can safely fulfills the empirical constraints (a)-(f) mentioned at the beginning of this subsection.
In particular, the EOS of SNM is consistent with the collective flow constraint on the pressure\,\cite{Pawel2002Science} and the optical potential constraint on the momentum dependence of the isoscalar potential in SNM\,\cite{Hama1990}. In addition, the EOS of PNM from our phenomenological potential is consistent with the microscopic calculations using chiral effective interactions\,\cite{Tews2013}. Moreover, the symmetry energy around $\rho_0$ is also consistent with existing constraints from analyzing various terrestrial experiments (e.g, the analysis of isobaric analog states\,\cite{Pawel2014}) and neutron star observations. 
Besides, the potential (\ref{5q5J}) thus designed is also consistent with the HVH theorem, i.e., the isovector symmetry potential and the second-order symmetry potential should be related as in Eq. (\ref{U2-U1}).
Of course, here we have no desire to construct a celebrated effective model for the EOS of ANM. Indeed, advanced nuclear many-body theories either phenomenological or microscopic can predict numerically more accurately the EOS of neutron-rich matter at certain densities and/or isospin asymmetries. However, analytical expressions for the $U_J(\rho,\delta,\v{k})$ and all the characteristics of the corresponding EOS with only 9 parameters that can describe satisfactorily essentially all known constraints on the EOS are invaluable for many purposes. As we shall show next, serving as a reliable basis in the $\epsilon$-expansion approach  these expressions facilitate the study of nuclear EOSs in spaces with dimensions different from 3.

\subsection{Perturbing the EOS of spatial dimension $d_{\rm{f}}=3$}

We now study the EOS of ANM in spaces with dimensions near the reference dimension $d_{\rm{f}}=3$ based on the $\epsilon$-expansion formalism with the 3D toy model EOS developed in the last subsection.
The EOS of SNM in dimension $d+\epsilon$ with $\epsilon$ being the perturbative dimension is given by the formula (\ref{ddef_E0}), where the term originated from the momentum-dependence of the single-nucleon potential $U_0$ (see (\ref{3qU0})) could be written out as,
\begin{equation}
\frac{\sigma}{\rho}\int_0^{\rho}\left(
\left.\frac{\partial
U_0}{\partial|\v{k}|}\right|_{|\v{k}|=\overline{k}^f_{\rm{F}}}\cdot\overline{k}^f_{\rm{F}}\right)\d
f=-\frac{\sigma}{\rho}\frac{3\mu \nu}{\rho_0}\int_0^{\rho}\d f\frac{f^2}{(1+\nu f)^2},\end{equation}
where $\nu=3\pi^2\theta/2\Lambda^3$. The integration in the above equation can be carried out analytically, leading to
\begin{equation}
\int_0^{\rho}\d f\frac{f^2}{(1+\nu f)^2}
=\frac{1}{\nu^3}\left[\frac{x(x+2)}{x+1}-\ln(x+1)\right],
\end{equation}
where $x=\theta({k_{\rm{F}}}/{\Lambda})^3$ as introduced earlier is a function of density through $k_{\rm{F}}$.
Consequently,
\begin{align}
\frac{\sigma}{\rho}\int_0^{\rho}\left(
\left.\frac{\partial
U_0}{\partial|\v{k}|}\right|_{|\v{k}|=\overline{k}^f_{\rm{F}}}\cdot\overline{k}^f_{\rm{F}}\right)\d
f=&-3\sigma\mu\left(\frac{\rho}{\rho_0}\right)\frac{1}{x^2}\left[\frac{x^2+2x}{x+1}-2\ln(x+1)\right]\\
\approx&-3\sigma\mu\left(\frac{\rho}{\rho_0}\right)\left(\frac{1}{3}x-\frac{1}{2}x^2+\frac{3}{5}x^3\right),~~\mbox{for small $x$},\label{dkj-a}\end{align}
where the second line (approximation) holds for small values of $x$ (corresponding to low densities). In fact the limit $x\approx1$ is equivalent to a density about $\rho\approx2.85\,\rm{fm}^{-3}\approx18\rho_0$ using $\theta=1.3$ and $\Lambda=750\,\rm{MeV}$ (see Tab.\,\ref{tab_c1}). Moreover, the saturation density $\rho_0$ corresponds to $x$ about $x\approx 0.056$.
Thus, the above small-$x$ approximation is very reasonable/effective for our discussions of situations with densities up to several times of $\rho_0$. Another interesting feature indicated by (\ref{dkj-a}) is that the correction is positive since $\mu$ is negative (see Tab.\,\ref{3qpara_aa}).
This is consistent with the general analysis given in the paragraph after formula (\ref{ddef_KSNM}), i.e., the momentum-dependence of the nuclear potential tends to reduce (enhance) the EOS of SNM in lower (higher) dimensions compared with the 3D case. In fact, it is a direct consequence of the well-constrained momentum dependence of the $U_0$ (e.g., see the right panel of Fig.\,\ref{fig_toyFlow}).

Combining (\ref{dkj-a}) with other relevant terms gives the EOS of SNM in perturbative dimension $\epsilon$ as,
\begin{align}\label{EOS_zeta_p}
E_0(\rho)=&\frac{3}{10M}\left(\frac{3\pi^2}{2}\right)^{2/3}\rho^{2/3}
+\frac{\alpha}{2}\left(\frac{\rho}{\rho_0}\right)+
\frac{3\beta}{7}\left(\frac{\rho}{\rho_0}\right)^{4/3}
+\frac{3\lambda}{8}\left(\frac{\rho}{\rho_0}\right)^{5/3}
+\frac{\phi}{3}\left(\frac{\rho}{\rho_0}\right)^2\notag\\
&+\mu\left(\frac{\rho}{\rho_0}\right)\left[\theta\left(\frac{k_{\rm{F}}}{\Lambda}\right)^3-\ln\left(1+\theta\left(\frac{k_{\rm{F}}}{\Lambda}\right)^3\right)\right]\left/\theta^2\left(\frac{k_{\rm{F}}}{\Lambda}\right)^6\right.\notag\\
&+\epsilon
\left[\frac{3\overline{k}_{\rm{F}}^2}{5M}\left(\sigma+\frac{1}{15}\right)-3\sigma\mu\left(\frac{\rho}{\rho_0}\right)
\frac{1}{x^2}\left[\frac{x^2+2x}{x+1}-2\ln(x+1)\right]\right].
\end{align}
Similarly, the pressure could be worked out as,
\begin{align}\label{P0_zeta_p}
P_0(\rho)
=&\frac{1}{5M}\left(\frac{3\pi^2}{2}\right)^{2/3}\rho^{5/3} +\rho\left[
\frac{\alpha}{2}\left(\frac{\rho}{\rho_0}\right)+\frac{4\beta}{7}\left(\frac{\rho}{\rho_0}\right)^{4/3}
+\frac{5\lambda}{8}\left(\frac{\rho}{\rho_0}\right)^{5/3}+\frac{2\phi}{3}\left(\frac{\rho}{\rho_0}\right)^2\right]\notag\\
&+\mu\rho\left(\frac{\rho}{\rho_0}\right)\left[1\left/\left[1+\theta\left(\frac{k_{\rm{F}}}{\Lambda}\right)^3\right]\right.-1\left/\theta\left(\frac{k_{\rm{F}}}{\Lambda}\right)^3\right.+\left[\theta\left(\frac{k_{\rm{F}}}{\Lambda}\right)^3\right]^{-2}
\ln\left[1+\theta\left(\frac{k_{\rm{F}}}{\Lambda}\right)^3\right]\right]\notag\\
&+\epsilon\rho\left[\frac{\overline{k}_{\rm{F}}^2}{10M}\left(2\sigma-\frac{1}{5}\right)
+3\sigma\mu\left(\frac{\rho}{\rho_0}\right)\left[
\frac{1}{x^2}\left[\frac{x^2+2x}{x+1}-2\ln(x+1)\right]
-\frac{x}{(x+1)^2}\right]
\right],
\end{align}
where the relevant momentum-dependent term in the pressure is,
\begin{align}
&-\frac{\sigma}{\rho}\int_0^{\rho}\left( \left.\frac{\partial
U_0}{\partial|\v{k}|}\right|_{|\v{k}|=\overline{k}^f_{\rm{F}}}\cdot\overline{k}^f_{\rm{F}}\right)\d
f+\left.\sigma\frac{\partial U_0}{\partial
|\v{k}|}\right|_{|\v{k}|=\overline{k}_{\rm{F}}}\cdot\overline{k}_{\rm{F}}\notag\\
=&3\sigma\mu\left(\frac{\rho}{\rho_0}\right)\left[
\frac{1}{x^2}\left[\frac{x^2+2x}{x+1}-2\ln(x+1)\right]
-\frac{x}{(x+1)^2}\right]\\
\approx&3\sigma\mu\left(\frac{\rho}{\rho_0}\right)\left(-\frac{2}{3}x+\frac{3}{2}x^2-\frac{12}{5}x^3\right),~~\mbox{for small $x$}.
\end{align}

\begin{figure}[h!]
\centering
\includegraphics[width=7.cm]{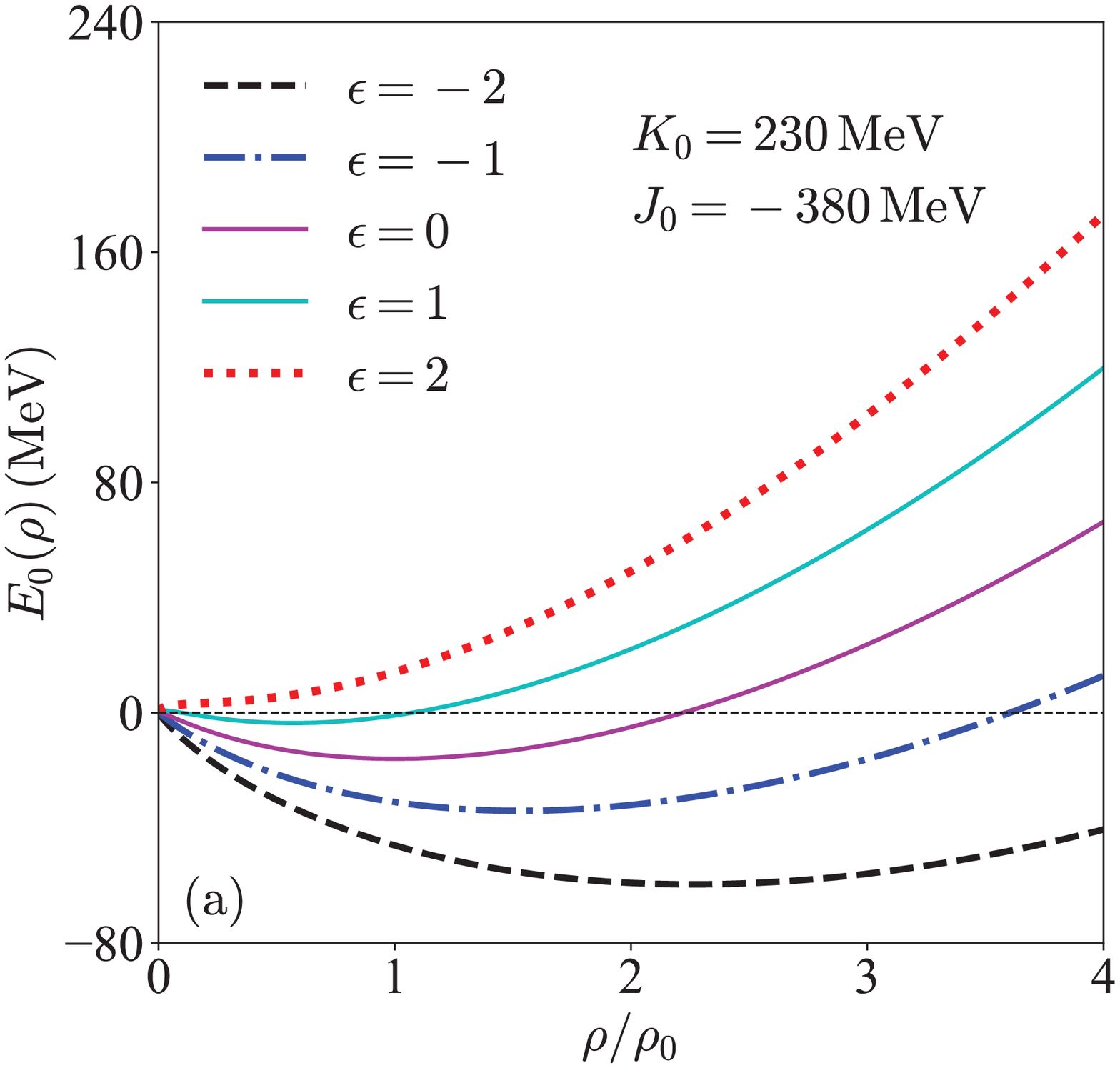}\qquad
\includegraphics[width=7.cm]{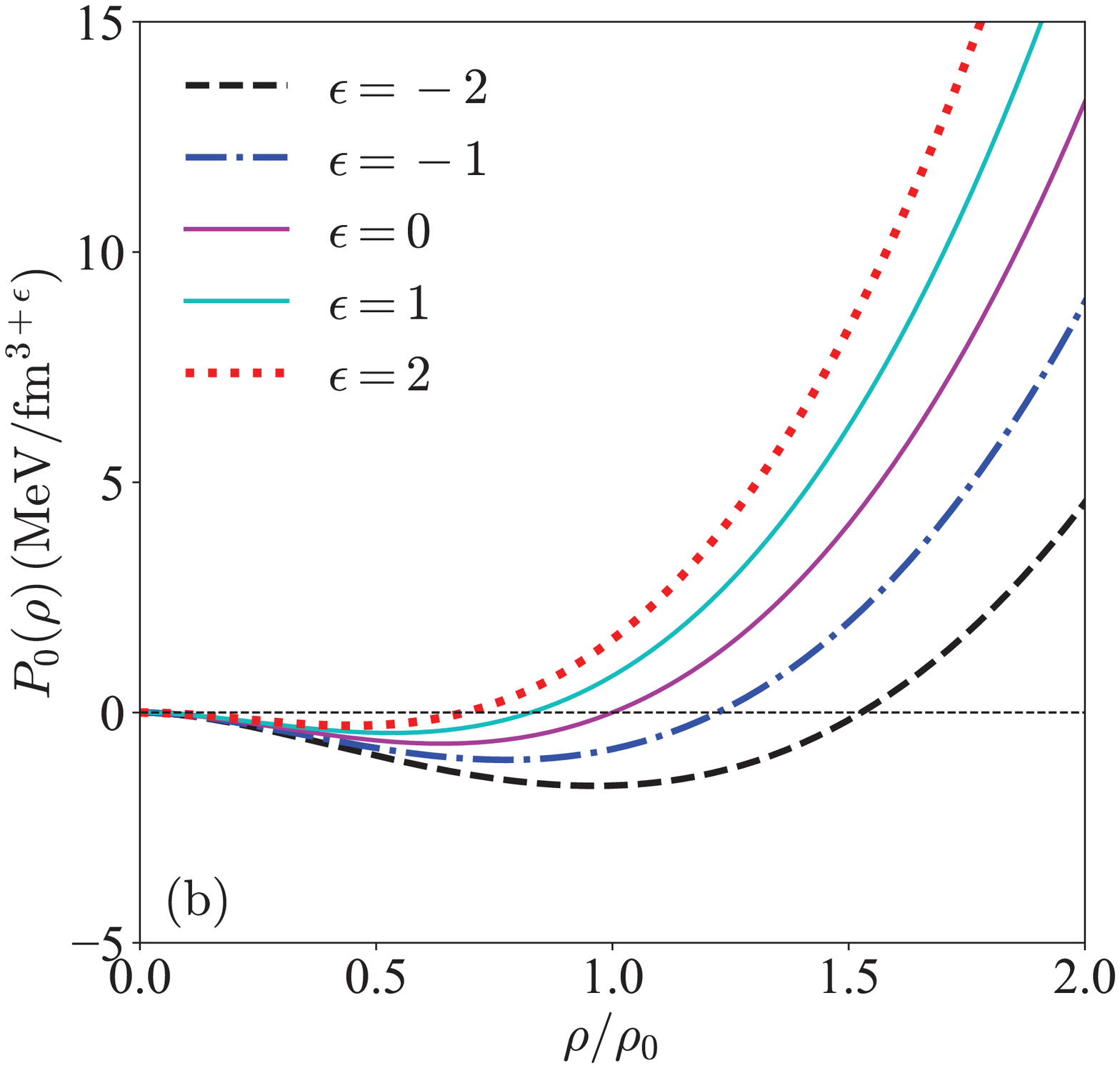}
\caption{EOS of SNM and the corresponding pressure adopting different values of $\epsilon$.}\label{fig_toypertE0}
\end{figure}

For small $x$ and $\epsilon=-1$ (in 2D), one has
\begin{align}
E_0^{(\rm{2D})}(\rho)\approx&\frac{3\overline{k}_{\rm{F}}^2}{10M}\left(\frac{13}{15}-2\sigma\right)
+\frac{1}{2}\mu\left(\frac{\rho}{\rho_0}\right)\left(1+2x\sigma\right)\notag\\
&+\frac{\alpha}{2}\left(\frac{\rho}{\rho_0}\right)+
\frac{3\beta}{7}\left(\frac{\rho}{\rho_0}\right)^{4/3}
+\frac{3\lambda}{8}\left(\frac{\rho}{\rho_0}\right)^{5/3}
+\frac{\phi}{3}\left(\frac{\rho}{\rho_0}\right)^2,
\end{align}
from which it is clear that the correction to the EOS of SNM from the momentum-dependent interaction term (characterized by $2x\sigma$) is generally smaller than that from the kinetic part (characterized by $2\sigma+2/15$), due to the smallness of $x\approx0.056$ near the saturation density.
If one introduces an effective coupling ``constant'' 
\begin{equation}
\mu_{\rm{eff}}(\rho)=\mu(1+2x\sigma)
\end{equation}
and an effective mass related to the $\sigma$ as,
\begin{equation}
M_{\sigma}^{\rm{eff}}=M\left(\frac{13}{15}-2\sigma\right)^{-1},
\end{equation}
then the EOS of SNM in 2D has the same form as that in 3D, i.e.,
\begin{equation}
E_0^{(\rm{2D})}(\rho)=\frac{3\overline{k}_{\rm{F}}^2}{10M_{\sigma}^{\rm{eff}}}
+\frac{1}{2}\mu_{\rm{eff}}(\rho)\left(\frac{\rho}{\rho_0}\right)
+\frac{\alpha}{2}\left(\frac{\rho}{\rho_0}\right)+
\frac{3\beta}{7}\left(\frac{\rho}{\rho_0}\right)^{4/3}
+\frac{3\lambda}{8}\left(\frac{\rho}{\rho_0}\right)^{5/3}
+\frac{\phi}{3}\left(\frac{\rho}{\rho_0}\right)^2,
\end{equation}
see Eq. (\ref{3qE0}) and the few equations following it.
The effective mass $M_{\sigma}^{\rm{eff}}\approx2.6M$ in 2D is much larger than the bare mass in 3D, indicating the nucleon is now becoming too heavy to move easily (i.e., it is more bounded). However, the coupling ``constant'' $\mu_{\rm{eff}}$ is density dependent, i.e., the density dependence of the SNM EOS in 2D (or generally in other dimensions) is modified due to the $\mu_{\rm{eff}}$-term.

In Fig.\,\ref{fig_toypertE0} we show the nucleon specific energy $E_0(\rho)$ (left) and pressure $P_0(\rho)$ (right) in SNM adopting 5 perturbative dimensions, i.e., $\epsilon=-2,-1,0,1$ and $2$ (corresponding to $d$ from 1 to 5). It is seen that as the dimension decreases, the EOS of SNM becomes much more bounded and as $d$ increases, the $E_0(\rho)$ becomes less negative.
In particular, if the $d$ is larger than some critical dimension, the $E_0(\rho)$ becomes totally positive, indicating the system is unbounded and there would be no saturation density in these dimensions. Noticing that our results given here are from perturbating the EOS in 3D, if one adopts the exact scheme, the qualitative conclusion will not change, i.e., the nucleon specific energy in SNM eventually increases as the $d$ increases and there would be no saturation density above certain critical dimension.
This phenomenon actually can be traced back to a deep principle behind the high-dimensional geometry: According to the relation between the Fermi momentum $k_{\rm{F}}$ and the density $\rho$, i.e., (\ref{def_kF_d}), one can equivalently write the density in terms of $k_{\rm{F}}$ as $\rho=k_{\rm{F}}^d/[{2^{d-2}\pi^{d/2}\Gamma(d/2+1)}]$.  The volume of a high-dimensional sphere with radius $k_{\rm{F}}$ is $(2\pi)^d\rho/4$. For a fixed Fermi momentum $k_{\rm{F}}$, the volume of the sphere approaches zero as $d\to\infty$\,\cite{Blum2020}. Inversely, if one fixes the density $\rho$, then the Fermi momentum $k_{\rm{F}}$ should approach infinity as $d\to\infty$, indicating that the kinetic EOS of SNM namely $dE_{\rm{F}}/(d+2)\sim E_{\rm{F}}=k_{\rm{F}}^2/2M$ (as $d\to\infty$, see (\ref{def_kF_d})) will always denominate over the potential part, irrespective of the form of the latter. 

\begin{figure}[h!]
\centering
\includegraphics[width=7.cm]{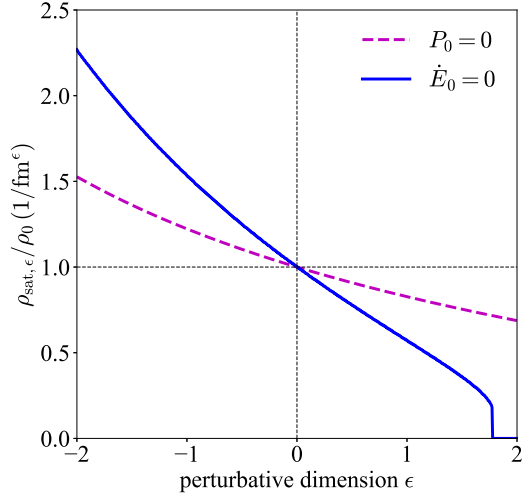}
\caption{Saturation density $\rho_{\rm{sat},\epsilon}$ as a function of $\epsilon$.}\label{fig_toy-sat}
\end{figure}

Another interesting feature of Fig.\,\ref{fig_toypertE0} is that perturbatively the saturation density inferred from $E_0(\rho)$ (as the minimum of the curves in the left panel) does not necessarily correspond to the vanishing point of $P_0(\rho)$ (the right panel). The reason is simple: As one does perturbative calculations, the $d$-dependences of the $E_0(\rho)$ and $P_0(\rho)$ are different. More specifically, we have $E_0^{\rm{kin}}(\rho)=dk_{\rm{F}}^2/2(d+2)M$ while $3P^{\rm{kin}}_0(\rho)/\rho=L^{\rm{kin}}_0(\rho)=k_{\rm{F}}^2/(d+2)M$ (see formula (\ref{deff_L0})).
When the dimension $d$ is perturbed around $d_{\rm{f}}$,  the self-consistent relation $P_0^{\rm{kin}}(\rho)=\rho^2\d E_0^{\rm{kin}}(\rho)/\d\rho$ may be lost. 
Similar phenomenon also occurs for the isospin-dependent part of the EOS of ANM.
If one takes the exact calculations in $d$D instead of the perturbation on $\epsilon$, then the self-consistency will naturally be maintained (e.g., it could be straightforwardly shown using the single-nucleon potential $U_J(\rho,\delta)=\sum_i[\alpha_i+\alpha_i^{\rm{IS}}\exp((\theta_i-1)\tau_3^J\delta)](\rho/\rho_0)^{\vartheta_i}$ with the exact solution of the constraining equations for certain empirical characteristics).
On the other hand, the (perturbative) potential of the EOS of SNM fulfills the self-consistent relation, i.e.,
\begin{equation}
\rho^2\frac{\d}{\d\rho}\left[\frac{\sigma}{\rho}\int_0^{\rho}\left(
\left.\frac{\partial
U_0}{\partial|\v{k}|}\right|_{|\v{k}|=\overline{k}^f_{\rm{F}}}\cdot\overline{k}^f_{\rm{F}}\right)\d
f\right]=\rho\sigma\left.\frac{\partial U_0}{\partial
|\v{k}|}\right|_{|\v{k}|=\overline{k}_{\rm{F}}}\cdot\overline{k}_{\rm{F}}-\sigma\int_0^{\rho}\left( \left.\frac{\partial
U_0}{\partial|\v{k}|}\right|_{|\v{k}|=\overline{k}^f_{\rm{F}}}\cdot\overline{k}^f_{\rm{F}}\right)\d
f,
\end{equation}
see the relevant terms in (\ref{ddef_E0}) and (\ref{ddef_P0}).
In Fig.\,\ref{fig_toy-sat} we show the saturation density $\rho_{\rm{sat},\epsilon}$ as a function of the perturbative dimension $\epsilon$ through the two approaches, namely one from $P_0=0$ and the other from $\dot{E}_0\equiv \d E_0/\d\rho=0$.
As mentioned previously, the saturation densities from these two approaches are generally different due to the perturbative calculation, but the qualitative feature is the same: as the dimension $d$ is downward perturbed, the saturation density is effectively increased. In particular, the critical dimension above which there is no saturation point for $E_0(\rho)$ is found to be about 4.77 for the toy model.

We now switch to calculating the symmetry energy in $d$D perturbatively, see (\ref{ddef_Esym}). Since the symmetry potential in the toy model has no momentum dependence, we immediately obtain the following expression
\begin{align}
E_{\rm{sym}}(\rho)=&\frac{\overline{k}_{\rm{F}}^2}{6M}+\left.\frac{\overline{k}_{\rm{F}}}{6}
\frac{\partial U_0}{\partial
|\v{k}|}\right|_{|\v{k}|=\overline{k}_{\rm{F}}}+\frac{1}{2}U_{\rm{sym}}(\rho,\overline{k}_{\rm{F}})\notag\\
&+\epsilon\left[\frac{\overline{k}_{\rm{F}}^2}{6M}\left(2\sigma-\frac{1}{3}\right)
+\frac{\overline{k}_{\rm{F}}}{6}\frac{\partial
U_0}{\partial
|\v{k}|}\left(\left.{\displaystyle\sigma\overline{k}_{\rm{F}}\frac{\partial^2
U_0}{\partial
|\v{k}|^2}}\right/{\displaystyle\frac{\partial
U_0}{\partial
|\v{k}|}}+\sigma-\frac{1}{3}\right)\right]_{|\v{k}|=\overline{k}_{\rm{F}}}.
\end{align}
After evaluating the derivatives, one obtains
\begin{align}
E_{\rm{sym}}(\rho)
=&\frac{\overline{k}_{\rm{F}}^2}{6M}+
\frac{1}{6}\beta_{\rm{IS}}\left(\frac{\rho}{\rho_0}\right)^{4/3}+\frac{1}{3}\lambda_{\rm{IS}}
\left(\frac{\rho}{\rho_0}\right)^{5/3}-
\frac{1}{2}\mu\left(\frac{\rho}{\rho_0}\right)\theta\left(\frac{k_{\rm{F}}}{\Lambda}\right)^3\left/\left[1+\theta\left(\frac{k_{\rm{F}}}{\Lambda}\right)^3\right]^2\right.\notag\\&+\epsilon\left[\frac{\overline{k}_{\rm{F}}^2}{6M}\left(2\sigma-\frac{1}{3}\right)
-\frac{1}{2}\mu\left(\frac{\rho}{\rho_0}\right)\frac{x}{(1+x)^2}\left(\frac{2\sigma(1-2x)}{1+x}
+\sigma-\frac{1}{3}\right)\right].
\end{align}
For small $x$, the $\epsilon$-term becomes,
\begin{align}
\epsilon\left[\frac{\overline{k}_{\rm{F}}^2}{6M}\left(2\sigma-\frac{1}{3}\right)
-\frac{1}{2}\mu\left(\frac{\rho}{\rho_0}\right)
\left[
\left(3\sigma-\frac{1}{3}\right)x-\left(12\sigma-\frac{2}{3}\right)x^2+(27\sigma-1)x^3
\right]
\right].\label{djk-2}
\end{align}

Since $3\sigma-1/3>0$, $2\sigma-1/3$>0 and $\mu<0$, the factor in the square bracket of (\ref{djk-2}) is positive definite,  indicating that the symmetry energy in lower dimensions (e.g., 2D) with negative $\epsilon$ is reduced compared with its 3D counterpart. This conclusion is consistent with the analysis given in section \ref{SEC_EXP} (see discussion given after Eq.\,(\ref{edf})). Specifically, we have $\overline{k}_{\rm{F}}^2/6M\cdot(2\sigma-1/3)\approx1.8\,\rm{MeV}$ and $-2^{-1}\mu(\rho/\rho_0)(3\sigma-1/3)x\approx1.3\,\rm{MeV}$ for $\rho=\rho_0$, and thus the linear $\epsilon$ term is about $3.1\epsilon\,\rm{MeV}$. It is totally different from the one considering the momentum dependence of the symmetry potential  giving about $-2.1\epsilon\,\rm{MeV}$, as discussed after the Eq.\,(\ref{edf}).

\renewcommand*\tablename{\footnotesize Tab.}
\begin{table}[h!]
\centering
\begin{tabular}{c|c|c|c|c}\hline
&linear-$\epsilon$ (potential)&linear-$\epsilon$ (total)&2D $E_{\rm{sym}}(\rho_0)$&reference\\
\hline\hline
optical model fitting &$-3.9\epsilon\,\rm{MeV}$&$-2.1\epsilon\,\rm{MeV}$&enhanced&Ref.\,\cite{LiXH2013PLB} \\\hline
Skyrme&$+0.4\epsilon\,\rm{MeV}$&$+2.2\epsilon\,\rm{MeV}$&reduced&Ref.\,\cite{WangR2018}\\\hline
MDI&$-2.9\epsilon\,\rm{MeV}$&$-1.1\epsilon\,\rm{MeV}$&enhanced&Ref.\,\cite{Das2003PRC}\\\hline
ImMDI&$2.6\epsilon\,\rm{MeV}+0.14y\epsilon$&$4.4\epsilon\,\rm{MeV}+0.14y\epsilon$&undetermined&Ref.\,\cite{XuJ2015PRC}\\\hline
toy model&$+1.3\epsilon\,\rm{MeV}$&$+3.1\epsilon\,\rm{MeV}$&reduced&$/$ \\
\hline
\end{tabular}\caption{Linear-$\epsilon$ correction to the symmetry energy from different models, the kinetic contribution is $1.8\epsilon\,\rm{MeV}$.}\label{Tab_2DEsym}
\end{table}

\begin{figure}[h!]
\centering
\includegraphics[width=7.cm]{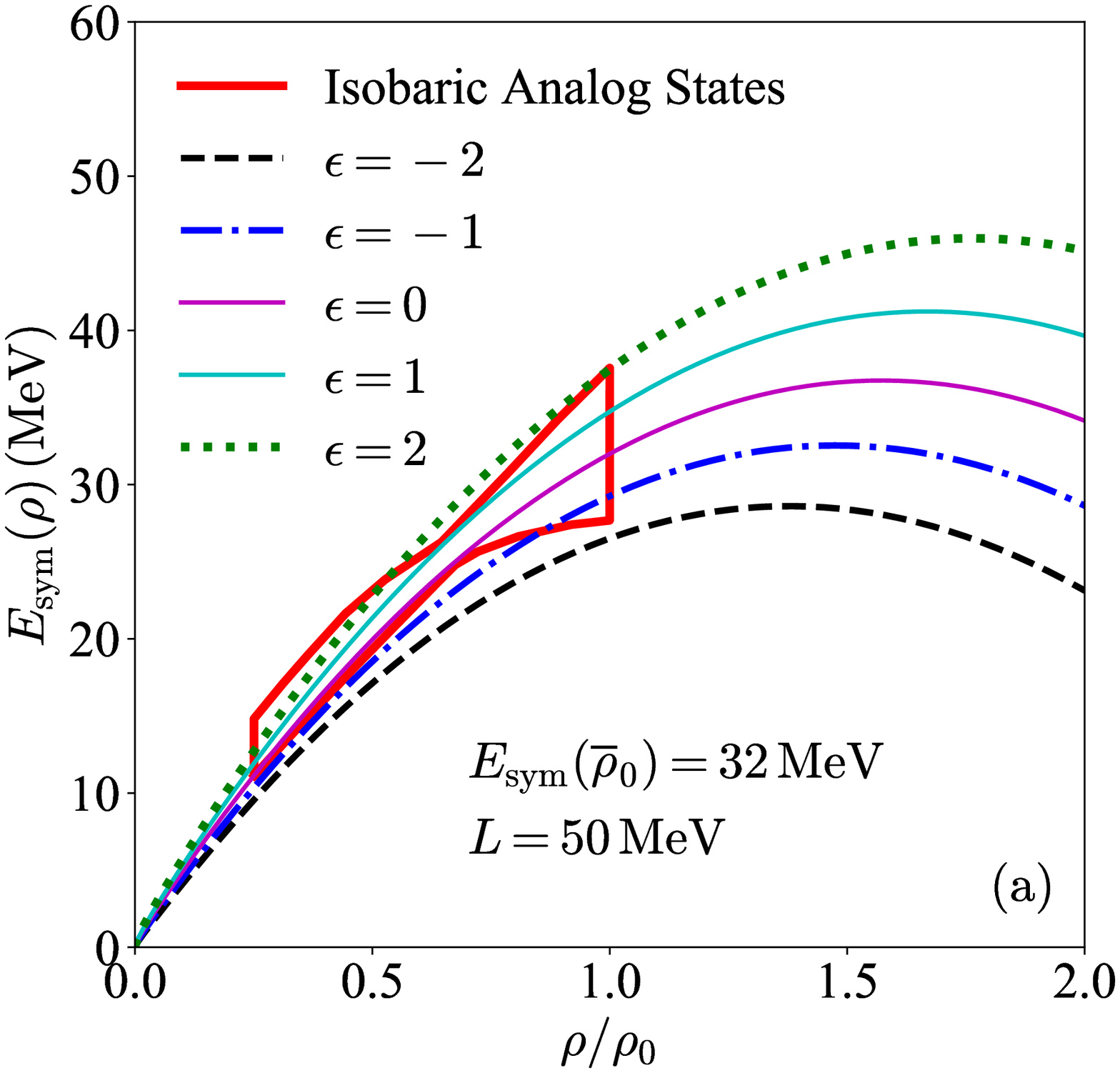}\qquad
\includegraphics[width=7.cm]{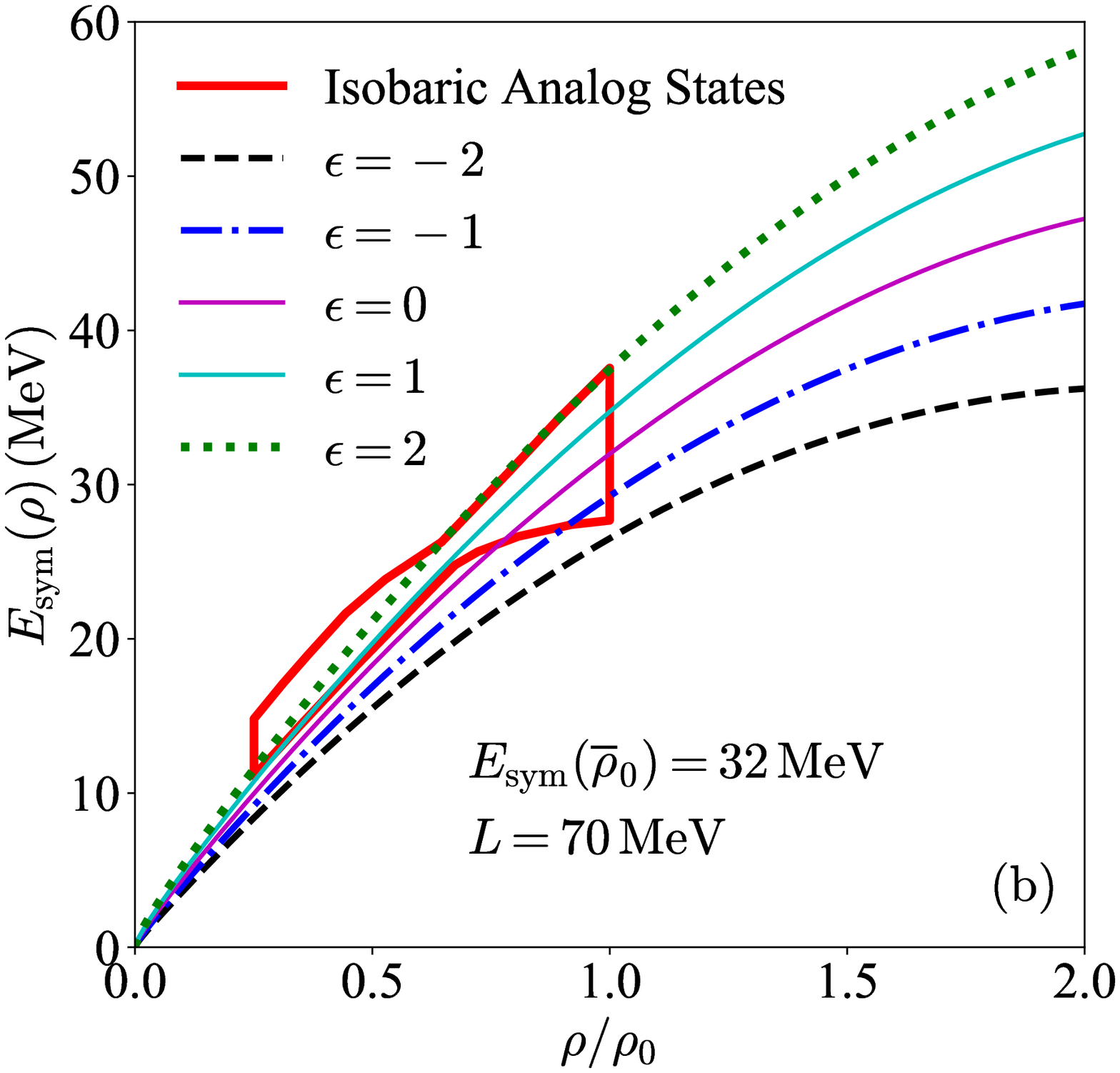}
\caption{Symmetry energy under different $L$ for five perturbative dimensions $\epsilon$, $E_{\rm{sym}}(\overline{\rho}_0)=32\,\rm{MeV}$ is fixed.}\label{fig_toypertEsym}
\end{figure}

Besides using our toy model potentials discussed above, we mention another two examples in the following.
In the first example, we adopt the Skyrme-type single-nucleon potential to order $|\v{k}|^6k_{\rm{F}}^3$ and $k_{\rm{F}}^9$ from Ref.\,\cite{WangR2018}. In such a model, the analytical expressions for the $U_0$ and $U_{\rm{sym}}$ are given and the relevant parameters were fitted to fulfill similar empirical constraints as we discussed earlier. The potential-related part of the linear-$\epsilon$ correction in (\ref{ddef_Esym}) can thus be evaluated straightforwardly to be about $0.4\,\rm{MeV}$ for the SP6s (SP6m or SP6h) set\,\cite{WangR2018}. Consequently, the linear-$\epsilon$ correction to the total symmetry energy is about $2.2\epsilon\,\rm{MeV}$ which is close to our toy model calculation. In the second example, we adopt an isospin- and momentum-dependent interaction (MDI) potential that has been used extensively to explore heavy-ion reaction dynamics within transport models\,\cite{Das2003PRC,ChenLW2005PRL}. With the MDI potential, the potential-related part of the linear-$\epsilon$ correction in (\ref{ddef_Esym}) is found to be about $-2.9\epsilon\,\rm{MeV}$, leading to a total correction of about $-1.1\epsilon\,\rm{MeV}$, i.e., the 2D symmetry energy would be enhanced.
Furthermore, for a comparison the prediction from using an improve MDI (ImMDI) potential \cite{XuJ2015PRC} can also be obtained similarly. In the ImMDI potential model, the momentum-dependence of $U_0$ is improved to better fit the experimentally constrained high energy behavior of the nucleon optical potential.
The $y$ parameter (in unit of MeV) characterizes the momentum dependence of the $U_{\rm{sym}}$, i.e., $\partial U_{\rm{sym}}/\partial |\v{k}|>0$ if $y>0$. Tab.\,\ref{Tab_2DEsym} gives a summary of the linear-$\epsilon$ correction to the symmetry energy from different models investigated in this work.
We notice that the sign of the total correction using the ImMDI potential is not deterministic. For comparisons, in section \ref{SEC_Esym2D-O2} we shall extend the perturbative calculations to order $\epsilon^2$ for the symmetry energy and investigate the corresponding effects in 2D, see Tab.\,\ref{Tab_2DEsym-o2}.

\renewcommand*\tablename{\footnotesize Tab.}
\begin{table}[h!]
\centering
\begin{tabular}{c|c|c|c|c}\hline
&linear-$\epsilon$ (potential)&linear-$\epsilon$ (total)&2D $L\equiv L(\rho_0)$&reference\\
\hline\hline
Skyrme (SP6s)&$-8.1\epsilon\,\rm{MeV}$&$-12.7\epsilon\,\rm{MeV}$&enhanced&Ref.\,\cite{WangR2018}\\\hline
Skyrme (SP6m)&$-9.0\epsilon\,\rm{MeV}$&$-13.6\epsilon\,\rm{MeV}$&enhanced&Ref.\,\cite{WangR2018}\\\hline
Skyrme (SP6h)&$-9.9\epsilon\,\rm{MeV}$&$-14.5\epsilon\,\rm{MeV}$&enhanced&Ref.\,\cite{WangR2018}\\\hline
MDI&$-20.1\epsilon\,\rm{MeV}$&$-24.7\epsilon\,\rm{MeV}$&enhanced&Ref.\,\cite{Das2003PRC}\\\hline
ImMDI&$18.4\epsilon\,\rm{MeV}+0.80y\epsilon$&$13.8\epsilon\,\rm{MeV}+0.80y\epsilon$&undetermined&Ref.\,\cite{XuJ2015PRC}\\\hline
toy model&$+0.5\epsilon\,\rm{MeV}$&$-4.1\epsilon\,\rm{MeV}$&enhanced& $/$\\
\hline
\end{tabular}\caption{Linear-$\epsilon$ correction to the slope of the symmetry energy from different models, the kinetic contribution is $-4.6\epsilon\,\rm{MeV}$.}\label{Tab_2DL}
\end{table}

The relevant contribution to the slope parameter parameter $L$ of the symmetry energy in the toy model could also be evaluated (see formula (\ref{ddef_L})) by inspecting the following expression
\begin{align}
&\left(2\sigma-\frac{2}{3}\right)\left(\frac{\overline{k}_{\rm{F}}^2}{3M}\right)\epsilon 
+\left[\frac{\sigma}{6}\frac{\partial^3
U_0}{\partial|\v{k}|^3}\cdot\overline{k}_{\rm{F}}^3+
\frac{1}{6}\left(3\sigma-\frac{2}{3}\right)\frac{\partial^2
U_0}{\partial|\v{k}|^2}\cdot\overline{k}_{\rm{F}}^2+
\frac{1}{6}\left(\sigma-\frac{2}{3}\right)
\frac{\partial
U_0}{\partial|\v{k}|}\cdot \overline{k}_{\rm{F}}\right]\epsilon\notag\\
=&\left(2\sigma-\frac{2}{3}\right)\left(\frac{\overline{k}_{\rm{F}}^2}{3M}\right)\epsilon 
-\mu\left(\frac{\rho}{\rho_0}\right)\left[
\frac{1}{2}\left(\sigma-\frac{2}{3}\right)\frac{x}{(x+1)^2}+
\left(3\sigma-\frac{2}{3}\right)\frac{x(1-2x)}{(x+1)^3}
+\frac{\sigma x(10x^2-16x+1)}{(x+1)^4}
\right]\epsilon\notag\\
\approx&\left[\left(2\sigma-\frac{2}{3}\right)\left(\frac{\overline{k}_{\rm{F}}^2}{3M}\right)
-\frac{1}{2}\mu\left(\frac{\rho}{\rho_0}\right)\left[\left(9\sigma-2\right)x-8(9\sigma-1)x^2\right]\right]\epsilon,~~\rm{for~small~}x.
\end{align}
As the first term (characterized by $2\sigma-2/3$) in the square bracket is negative, while the second term (characterized by the $\mu$ constant) is positive since $\mu$ is negative,  the overall sign of the linear $\epsilon$-term depends on the model parameter $\mu$. 
Numerically we have $\overline{k}_{\rm{F}}^2/3M\cdot(2\sigma-2/3)\approx-4.6\,\rm{MeV}$, and $-2^{-1}\mu(\rho/\rho_0)(9\sigma-2)x\approx0.5\,\rm{MeV}$ at $\rho=\rho_0$ in the toy model calculation, leading to a linear-$\epsilon$ correction of about $-4.1\epsilon\,\rm{MeV}$.
Since the $\mu$ parameter in the current model characterizes the momentum-dependence of the single-nucleon potential, we find that the latter could effectively affect the behavior of $L$ when a perturbed expansion $\epsilon$ is applied.
This should be compared with the perturbed $L$ parameter without momentum dependence, see formula (\ref{ddef_L_nonm}),  or the first term in the above expression in which a negative $\epsilon$ (e.g., 2D) will always positively increase the slope parameter, i.e., hardening the symmetry energy as the density increases.
We see once again the non-trivial role played by the momentum dependence of the symmetry potential.

In Fig.\,\ref{fig_toypertEsym}, we show two cases (corresponding to two different values of the $L$ parameter) in which the symmetry energy is calculated perturbatively around $d=3$ from $d=1$ to $d=5$. It is seen that as the dimension $d$ decreases, the symmetry energy decreases correspondingly. Of course, the momentum dependence of the symmetry potential $U_{\rm{sym}}$ missing in our toy model potential, as mentioned and briefly discussed in section \ref{SEC_EXP} (see the paragraph after formula (\ref{ddef_KSNM}) and the analysis after (\ref{edf})), may introduce important effects on the density dependence of $E_{\rm{sym}}(\rho)$ when considering a finite $\epsilon$.
These effects are left for further studies. As listed in Tab.\,\ref{Tab_2DL}, here we compare the toy model prediction on the $L$ parameter with the results from using the Skyrme and MDI potentials. For the Skyrme-type pseudo potential developed in Ref.\,\cite{WangR2018}, we find that the linear-$\epsilon$ correction of (\ref{ddef_L}) is about $-12.7\epsilon\,\rm{MeV}$, $-13.6\epsilon\,\rm{MeV}$, or $-14.5\epsilon\,\rm{MeV}$ for the SP6s, SP6m, or the SP6h parameter set, respectively. For the MDI model\,\cite{Das2003PRC,ChenLW2005PRL}, the linear-$\epsilon$ correction is about $-24.7\epsilon\,\rm{MeV}$.
Unlike the circumstance for the magnitude of the symmetry energy listed in Tab.\,\ref{Tab_2DEsym}, except for the ImMDI potential (which strongly depends on the $y$ parameter) the sign of the linear-$\epsilon$ correction for the $L$ parameter is definite, i.e., the symmetry energy in 2D tends to be hardened (i.e., the correction is positive) at higher densities (since $\epsilon<0$).

\begin{figure}[h!]
\centering
\includegraphics[width=7.cm]{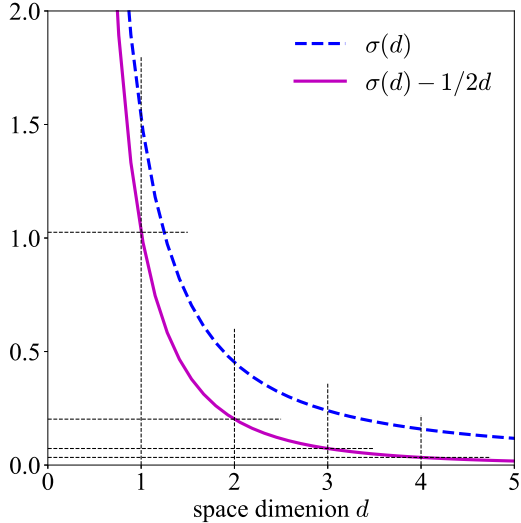}
\caption{The $d$-dependence of the function $\sigma(d)$ and $\sigma(d)-1/2d$.}\label{fig_sig_df}
\end{figure}

Finally, we want to point out that all the above perturbative calculations on the EOS of ANM are based on the conventional 3D. However, there is no fundamental reasons why this should be the choice.
In fact, there exist examples in which the basic dimension to be perturbed is not 3\,\cite{Nishida2006,Wilson1974}.
In situations where one needs to consider perturbation based on other dimensions, the correction $\sigma$ should be generalized correspondingly,
\begin{equation}\label{def_sigf}
\sigma(d_{\rm{f}})=\frac{1}{2d_{\rm{f}}^2}\left(d_{\rm{f}}\Psi\left(\frac{d_{\rm{f}}}{2}\right)
+2(d_{\rm{f}}+2)\ln2+d_{\rm{f}}\ln\pi+2\right)-\frac{1}{d_{\rm{f}}^2}\ln\left(2^{d_{\rm{f}}}\pi^{d_{\rm{f}}/2}\Gamma\left(\frac{d_{\rm{f}}}{2}+1\right)\right),
\end{equation}
where $\Psi(x)$ is defined as the derivative of the $\ln\Gamma(x)$, i.e., $\Psi(x)\equiv \d\ln \Gamma(x)/\d x$, and $\sigma(3)\equiv\sigma\approx0.2396$, see (\ref{defff_sig}).
As an example, we work out the perturbed symmetry energy to linear $\epsilon$-order by neglecting the 2nd derivative of $U_0$ with respect to $|\v{k}|$,
\begin{align}
E_{\rm{sym}}(\rho)
\approx& \frac{k_{\rm{F}}^2}{2d_{\rm{f}}M}
+\left.\frac{k_{\rm{F}}}{2d_{\rm{f}}}\frac{\partial U_0}{\partial |\v{k}|}\right|_{|\v{k}|=k_{\rm{F}}}+\frac{1}{2}U_{\rm{sym}}(\rho,k_{\rm{F}})\notag\\
&+\epsilon\left[
\left(\sigma_{\rm{f}}-\frac{1}{2d_{\rm{f}}}\right)\frac{k_{\rm{F}}^2}{d_{\rm{f}}M_{\rm{s}}^{\ast}}+\frac{k_{\rm{F}}}{2}\sigma_{\rm{f}}
\left(\frac{\partial U_{\rm{sym}}}{\partial|\v{k}|}
-\frac{1}{d_{\rm{f}}}\frac{\partial U_0}{\partial|\v{k}|}\right)_{|\v{k}|=k_{\rm{F}}}
\right],
\end{align}
where $\sigma_{\rm{f}}\equiv \sigma(d_{\rm{f}})$ and the second line is a direct generalization of (\ref{edf}).
If we assume that the sign of the momentum-dependence of the single potential is the same as that in 3D, i.e., $\partial U_{\rm{sym}}/\partial |\v{k}|<0$ and $\partial U_0/\partial|\v{k}|>0$, then the first term in the $\epsilon$-order contribution, i.e., $(\sigma_{\rm{f}}-1/2d_{\rm{f}})k_{\rm{F}}^2/d_{\rm{f}}M_{\rm{s}}^{\ast}$ eventually becomes smaller (but still positive) as $d_{\rm{f}}$ increases, see Fig.\,\ref{fig_sig_df} for the $d$-dependence of the function $\sigma(d)$ and that of $\sigma(d)-1/2d$.
While the second term in the $\epsilon$-order contribution eventually saturates to $2^{-1}k_{\rm{F}}\sigma_{\rm{f}}\partial U_{\rm{sym}}/\partial |\v{k}|$ as $d_{\rm{f}}$ increases, indicting that in the large $d$-limit the density dependence of the (first-order) perturbed symmetry energy is dominated by the momentum-dependence of the symmetry potential, 
i.e., the symmetry energy tends to be reduced compared with the one in $d_{\rm{f}}$D if it was upwardly perturbed (with $\epsilon>0$) since $\partial U_{\rm{sym}}/\partial|\v{k}|<0$ as assumed,
see the relevant discussion in section \ref{SEC_EXP}.
In particular, for $d_{\rm{f}}=1$, we can rewrite the above expression as,
\begin{align}
E_{\rm{sym}}(\rho)
\approx& \frac{k_{\rm{F}}^2}{2M}
+\left.\frac{k_{\rm{F}}}{2}\frac{\partial U_0}{\partial |\v{k}|}\right|_{|\v{k}|=k_{\rm{F}}}+\frac{1}{2}U_{\rm{sym}}(\rho,k_{\rm{F}})\notag\\
&+\epsilon\left[
\frac{k_{\rm{F}}^2}{2M_{\rm{s}}^{\ast}}\left(\sigma_1-1\right)
+\sigma_1\frac{k_{\rm{F}}}{2}\left(\frac{k_{\rm{F}}}{M}+\frac{\partial U_{\rm{sym}}}{\partial|\v{k}|}\right)_{|\v{k}|=k_{\rm{F}}}
\right].
\end{align}
If we assume the combination $[k_{\rm{F}}/M+\partial U_{\rm{sym}}/\partial|\v{k}|]_{|\v{k}|=k_{\rm{F}}}$ behaves similarly as in 3D, then this term is positive\,\cite{LiXH2013PLB}. Combining it with the first term namely $(\sigma_1-1)k_{\rm{F}}^2/2M_{\rm{s}}^{\ast}$ which is positive since $\sigma_1>1$ (as shown in Fig.\,\ref{fig_sig_df}), one then finds that the overall $\epsilon$-order correction to the symmetry energy is positive if $\epsilon>0$, indicating that the $E_{\rm{sym}}(\rho)$ should be enhanced if its is upwardly perturbed.
From these discussions, we can find once again that the dimensionality could change essentially the conclusion to certain problems though more accurate (quantitative) conclusions need further investigations.
Finally, the relatively smaller $\sigma_{\rm{f}}$ for $d_{\rm{f}}\approx2$ or $3$ than $\sigma_1\approx1.5253$ indicates the $\epsilon$-expansion around the reference dimension $d_{\rm{f}}=1$ may introduce difficulties in certain problems.

\setcounter{equation}{0}
\section{Kinetic EOS of neutron-rich matter with high-momentum nucleons induced by short-range correlations}\label{SEC_HMT}

\subsection{Single-nucleon momentum distribution function in the presence of short-range correlations}

In this section, we incorporate the short-range correlations (SRCs) induced high momentum tail (HMT) into the calculations of the kinetic EOS of ANM in a space of general dimension $d$. 
Here we first describe briefly the single-nucleon momentum distribution function $n_{\v{k}}^J$ in ANM in 3D.
Based on predictions of microscopic nuclear many-body theories and relevant experimental
findings\,\cite{Hen2014,Duer2018,Schmookler2019,Schmidt2020}, the single-nucleon momentum distribution in ANM can be parametrized as\,\cite{CaiBJ2015PRC,CaiBJ2016PRC-1,CaiBJ2016PRC-2,CaiBJ2016PLB},
\begin{equation}\label{MDGen}
n^J_{\v{k}}(\rho,\delta)=\left\{\begin{array}{ll}
\Delta_J,~~&0<|\v{k}|<k_{\rm{F}}^J,\\
\displaystyle{C}_J\left({k_{\rm{F}}^{J}}/{|\v{k}|}\right)^4,~~&k_{\rm{F}}^J<|\v{k}|<\phi_Jk_{\rm{F}}^J.
\end{array}\right.
\end{equation}
Here, $\Delta_J$ is the depletion of the Fermi sphere with respect to the step function in the FFG model.
The three parameters $\Delta_J$, ${C}_J$ and $\phi_J$ are constrained by the
fraction of nucleons in the HMT 
\begin{equation}\label{def_xJHMT}
\Delta_J+x_J^{\rm{HMT}}=1,~~
x_J^{\rm{HMT}}=\left.\int_{k_{\rm{F}}^J}^{\phi_Jk_{\rm{F}}^J}
n_{\v{k}}^J\d\v{k}\right/{\displaystyle\int_0^{\phi_Jk_{\rm{F}}^J}
n_{\v{k}}^J\d\v{k}}=3C_{{J}}\left(1-\frac{1}{\phi_{{J}}}\right),
\end{equation} and the normalization condition 
$
[{2}/{(2\pi)^3}]\int_0^{\infty}n^J_{\v{k}}(\rho,\delta)\d\v{k}=\rho_J=(k_{\rm{F}}^{J})^3/3\pi^2$.
Only two of the three parameters $\Delta_J$, $C_J$ and
$\phi_J$ are independent. Using the last two as independent parameters and assuming they have the same isospin dependence, i.e, 
$Y_J=Y_0(1+Y_1\tau_3^J\delta)$ with $Y=C$ or $\phi$\,\cite{CaiBJ2015PRC}, the associated parameters were correspondingly constrained to be about $C_0\approx0.161\pm0.015,C_1\approx-0.25\pm0.07,\phi_0\approx2.38\pm0.56$ and $\phi_1\approx-0.56\pm0.10$ using information from the SRC experiments\,\cite{Hen2014,Hen2017RMP}, see, e.g., Ref.\,\cite{LiBA2018PPNP} for a recent review on related issues\,\cite{XuC2013,Hen2015PRC,LiBA2015PRC-1,Carbone2012,
Carbone2014,Vidana2011PRC,Lovato2011PRC,Souza2020,Souza2020PRC,
Lourenco2022,YongGC2017PRC,YongGC2018PLB,
Wang2017PRC,Guo2021PRC,Hagel2021,Hong2022,Lu2022,Burrello2022}.

When generalizing the $n_{\v{k}}$ to arbitrary dimensions, we need to make certain assumptions on the parameters involved. The contact coefficient $C_J$ thus far explored in nuclear system using different theoretical approaches, or in ultra-cold gases adopting different types of trapped potentials,  at zero or finite temperatures all show good universality\,\cite{Bulgac2005,Kuhnle2010PRL,Stewart2010PRL,Drut2011PRL,Kuhnle2011PRL,Wild2012PRL,Sagi2012PRL,
Frohlich2012PRL,Hoinka2013PRL,Vignolo2013PRL,
Rossi2018PRL,Yao2018PRL,Mukherjee2019PRL, Carcy2019PRL, Jensen2020PRL},  e.g., see Chap.\,18 of Ref.\,\cite{Pit2016} for a review. In this work, we thus assume that the $C_J$ (including its $C_0$ and $C_1$ terms) is also (near) universal with respect to dimension $d$. On the other hand, to our best knowledge, there is no fundamental constraints on the high momentum cutoffs in lower or higher dimensions (different from 3). We thus assume that they are also universal and investigate the kinetic EOS of ANM based on these assumptions. Although the SRC-induced HMT appears only in the kinetic EOS (through the single-nucleon momentum distribution function) in the study here, it should be thought as an effective nucleon-nucleon interaction. Here we focus on its effects on the kinetic EOS of ANM in $d$D via the $\epsilon$-expansion.

For nucleons in ANM in $d$D, the relation (\ref{def_xJHMT}) should be generalized to,
\begin{equation}\label{def_DCPanm}
\Delta_J+\frac{dC_J}{d-4}\left(\phi_J^{d-4}-1\right)=1,
\end{equation}
where the second term on the left hand side is the corresponding $x_J^{\rm{HMT}}$ in $d$D.
Consequently, we obtain the zero-momentum depletion $\Delta_J$ in the first four dimensions as,
\begin{align}
\Delta_J^{(\rm{1D})}=&1-\frac{C_J}{3}\left(1-\frac{1}{\phi_J^{3}}\right),\label{kef_DJ1}\\
\Delta_J^{(\rm{2D})}=&1-C_J\left(1-\frac{1}{\phi_J^2}\right),\label{kef_DJ2}\\
\Delta_J^{(\rm{3D})}=&1-3C_J\left(1-\frac{1}{\phi_J}\right),\label{kef_DJ3}\\
\Delta_J^{(\rm{4D})}=&1-4C_J\ln\phi_J.\label{kef_DJ4}
\end{align}
A main feature of these expressions is that the $\phi_J$ in $d=1,2,3$ could be mathematically taken to be infinity, while that in 4D has a natural limit, i.e., $\phi_J\leq\exp(1/4C_J)$.
Similarly, the kinetic EOS of ANM is given by
\begin{equation}\label{kef_EOSANM}
E^{\rm{kin}}(\rho,\delta)=\sum_{J=\rm{n,p}}\frac{d}{d+2}\frac{k_{\rm{F}}^2}{2M}\left(1+\tau_3^J\delta\right)^{1+2/d}
\left[1+C_J\left(\frac{8}{(d-2)(d-4)}+\frac{d+2}{d-2}\phi^{d-2}_J
-\frac{d}{d-4}\phi^{d-4}_J\right)\right].
\end{equation}
The dependence of $E^{\rm{kin}}(\rho,\delta)$ on the dimension $d$ is rather non-trivial due to the complicated combination of $C_J,\phi_J$ and $d$.

\subsection{Kinetic EOS of SNM and the $\epsilon$-expansion}

According to the general formula (\ref{kef_EOSANM}), we can immediately obtain the kinetic part of SNM EOS in $d$D as,
\begin{align}
E_0^{\rm{kin}}(\rho)=&\frac{d}{d+2}\frac{k_{\rm{F}}^2}{2M}
\left[1+C_0\left(\frac{8}{(d-2)(d-4)}+\frac{d+2}{d-2}\phi^{d-2}_0
-\frac{d}{d-4}\phi^{d-4}_0\right)\right]
\equiv\frac{k_{\rm{F}}^2}{2M}\Lambda(d,C_0,\phi_0),\label{def_LAM}
\end{align}
where the last relation defines the function $\Lambda(d,C_0,\phi_0)$, which encapsulates the dependence of the EOS on the dimension $d$. We notice that the $\Lambda(d,C_0,\phi_0)$ apparently diverge with $d=d_{\rm{f}}=2$ and/or $d=d_{\rm{f}}=4$. Nevertheless, away from the poles with $d=d_{\rm{f}}+\epsilon$, the diverging terms all cancel completely and naturally in the $\epsilon$ expansion. For example, for the 2D situation, one can obtain the $\epsilon$-expansion
\begin{equation}
\frac{8}{(d-2)(d-4)}+\frac{d+2}{d-2}\phi^{d-2}_0
-\frac{d}{d-4}\phi^{d-4}_0\approx-1+4\ln\phi_0+\frac{1}{\phi_0^2}+\epsilon\left(1+\ln\phi_0+2\ln^2\phi_0+\frac{\ln\phi_0+1}{\phi_0^2}\right),
\end{equation}
to order $\epsilon=d-2$, using the formula $a^x\approx 1+x\ln a+2^{-1}x^2\ln^2a$ for small $x$.

Expanding the EOS of SNM around the dimension $d_{\rm{f}}$ (where $d_{\rm{f}}=1,2,3,4$) to order $\epsilon$ gives
\begin{align}
E_{0,d_{\rm{f}}=1}^{\rm{kin}}(\rho)\approx&\frac{k_{\rm{F}}^2}{6M}
\left[1+C_0\left(\frac{8}{3}-\frac{3}{\phi_0}+\frac{1}{3\phi_0^3}\right)\right]+\epsilon\cdot\frac{k_{\rm{F}}^2}{2M}\left[
\frac{2}{9}+C_0\left(\frac{16}{9}-\frac{\ln\phi_0}{\phi_0}-\frac{2}{\phi_0}+\frac{\ln\phi_0}{9\phi_0^3}+\frac{2}{9\phi_0^3}\right)\right]\notag\\
&+\epsilon^2\cdot\frac{k_{\rm{F}}^2}{2M}\left[-\frac{2}{27}+C_0\left(\frac{152}{81}-\frac{\ln^2\phi_0}{2\phi_0}-\frac{2\ln\phi_0}{\phi_0}-\frac{2}{\phi_0}+\frac{\ln^2\phi_0}{18\phi_0^3}+\frac{2\ln\phi_0}{9\phi_0^3}+\frac{10}{81\phi_0^3}\right)\right],\\
E_{0,d_{\rm{f}}=2}^{\rm{kin}}(\rho)\approx&
\frac{k_{\rm{F}}^2}{4M}\left[1+C_0\left(4\ln \phi_0+\frac{1}{\phi_0^2}-1\right)\right]
+\epsilon\cdot\frac{k_{\rm{F}}^2}{2M}\left[\frac{1}{8}
+C_0\left(
\ln^2\phi_0+\ln\phi_0-\frac{5}{8}+\frac{\ln\phi_0}{2\phi_0^2}+\frac{5}{8\phi_0^2}
\right)\right]\notag\\
&+\epsilon^2\cdot\frac{k_{\rm{F}}^2}{2M}\left[-\frac{1}{32}+C_0\left(
-\frac{11}{32}+\frac{\ln^3\phi_0}{3}+\frac{\ln^2\phi_0}{2}
+\frac{\ln^2\phi_0}{4\phi_0^2}+\frac{5\ln\phi_0}{8\phi_0^2}+\frac{11}{32\phi_0^2}\right)\right],\label{d200}
\end{align}
and
\begin{align}
E_{0,d_{\rm{f}}=3}^{\rm{kin}}(\rho)\approx &\frac{3}{5}\frac{k_{\rm{F}}^2}{2M}
\left[1+C_0\left(5\phi_0+\frac{3}{\phi_0}-8\right)\right]
+\epsilon\cdot\frac{k_{\rm{F}}^2}{2M}\left[\frac{2}{25}+C_0\left(
3\phi_0\ln\phi_0-2\phi_0+\frac{9\ln\phi_0}{5\phi_0}
+\frac{66}{25\phi_0}-\frac{16}{25}\right)\right]\notag\\
&+\epsilon^2\cdot\frac{k_{\rm{F}}^2}{2M}\left[-\frac{2}{125}
+C_0\left(-\frac{584}{125}+\frac{3\phi_0\ln^2\phi_0}{2}-2\phi_0\ln\phi_0+2\phi_0+\frac{9\ln^2\phi_0}{10\phi_0}+\frac{66\ln\phi_0}{25\phi_0}+\frac{334}{125\phi_0}\right)\right],\\
E_{0,d_{\rm{f}}=4}^{\rm{kin}}(\rho)
\approx& \frac{k_{\rm{F}}^2}{3M}\left[1+C_0\left(3\phi_0^2-4\ln\phi_0-3\right)\right]+\epsilon\cdot\frac{k_{\rm{F}}^2}{2M}\left[\frac{1}{18}+C_0\left(\frac{1}{2}+2\phi_0^2\ln\phi_0-\frac{\phi_0^2}{2}-\frac{4\ln^2\phi_0}{3}-\frac{8\ln\phi_0}{9}\right)\right]\notag\\
&+\epsilon^2\cdot\frac{k_{\rm{F}}^2}{2M}\left[-\frac{1}{108}+C_0\left(-\frac{1}{4}+\phi_0^2\ln^2\phi_0-\frac{\phi_0^2\ln\phi_0}{2}+\frac{\phi_0^2}{4}-\frac{4\ln^3\phi_0}{9}-\frac{4\ln^2\phi_0}{9}
-\frac{\ln\phi_0}{54}\right)\right].
\end{align}
It is obvious that the factors in the square brackets of (\ref{E-0-1-kin}) to (\ref{E-0-4-kin}) reduce to 1 naturally if $\phi_0=1$ is taken (i.e., no high momentum nucleon above $k_{\rm{F}}$).
In addition, the $\epsilon^0$-order terms in the above expansions are the EOS of SNM in dimensions $d$, i.e.,
\begin{align}
E_{0,{(\rm{1D})}}^{\rm{kin}}(\rho)=&\frac{k_{\rm{F}}^2}{6M}
\left[1+C_0\left(\frac{8}{3}-\frac{3}{\phi_0}+\frac{1}{3\phi_0^3}\right)\right],\label{E-0-1-kin}\\
E_{0,{(\rm{2D})}}^{\rm{kin}}(\rho)=&\frac{k_{\rm{F}}^2}{4M}\left[1+C_0\left(4\ln \phi_0+\frac{1}{\phi_0^2}-1\right)\right],\label{E-0-2-kin}\\
E_{0,{(\rm{3D})}}^{\rm{kin}}(\rho)=&\frac{3}{5}\frac{k_{\rm{F}}^2}{2M}
\left[1+C_0\left(5\phi_0+\frac{3}{\phi_0}-8\right)\right],\label{E-0-3-kin}\\
E_{0,{(\rm{4D})}}^{\rm{kin}}(\rho)=&\frac{k_{\rm{F}}^2}{3M}\left[1+C_0\left(3\phi_0^2-4\ln\phi_0-3\right)\right],\label{E-0-4-kin}
\end{align}
where the nucleon Fermi momentum $k_{\rm{F}}$ in dimensions $d$ is related to the nucleon density by (\ref{def_kF_d}).

\begin{figure}[h!]
\renewcommand*\figurename{\footnotesize Fig.}
\centering
 \includegraphics[height=7.cm]{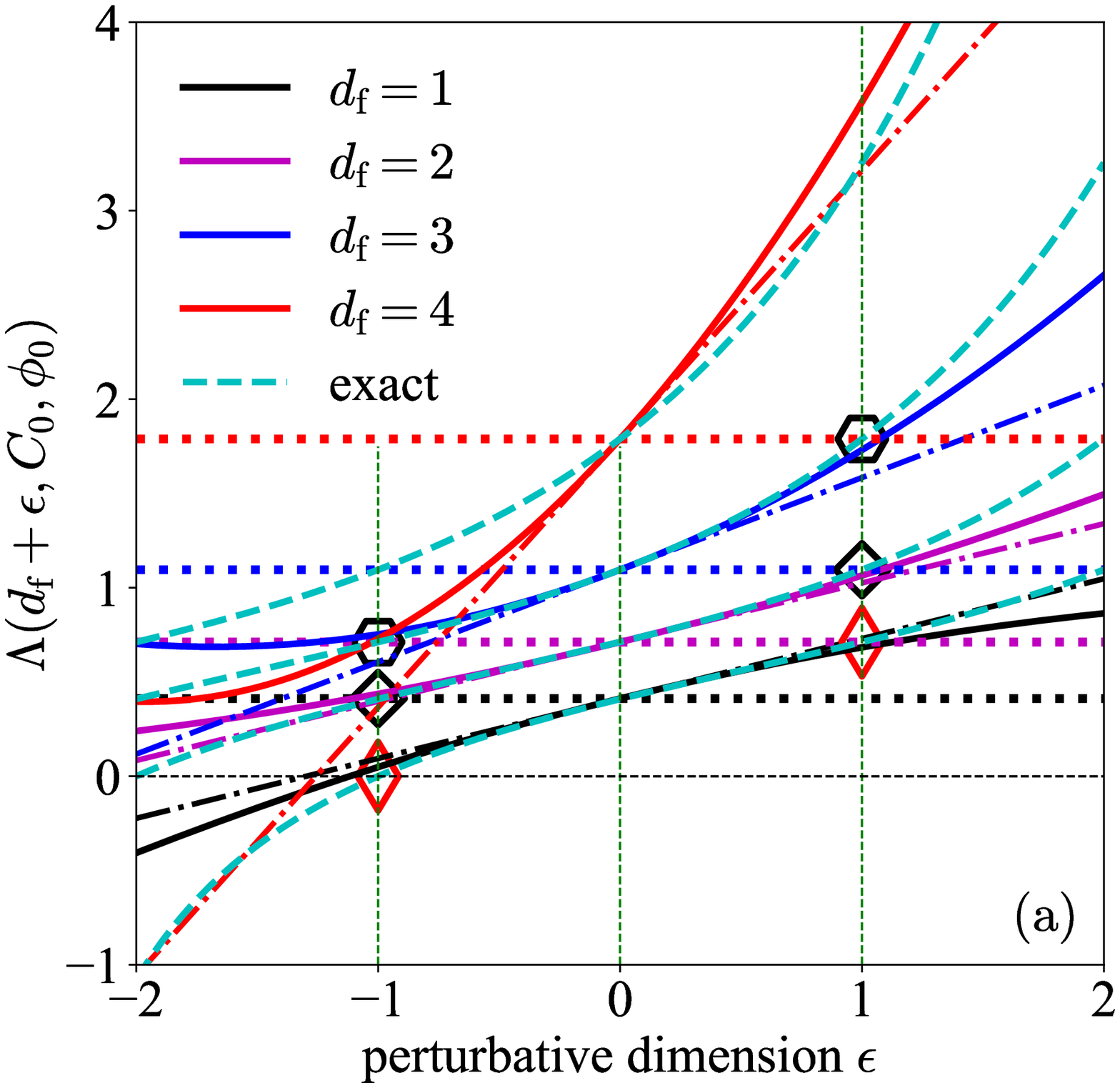}\qquad
 \includegraphics[height=7.cm]{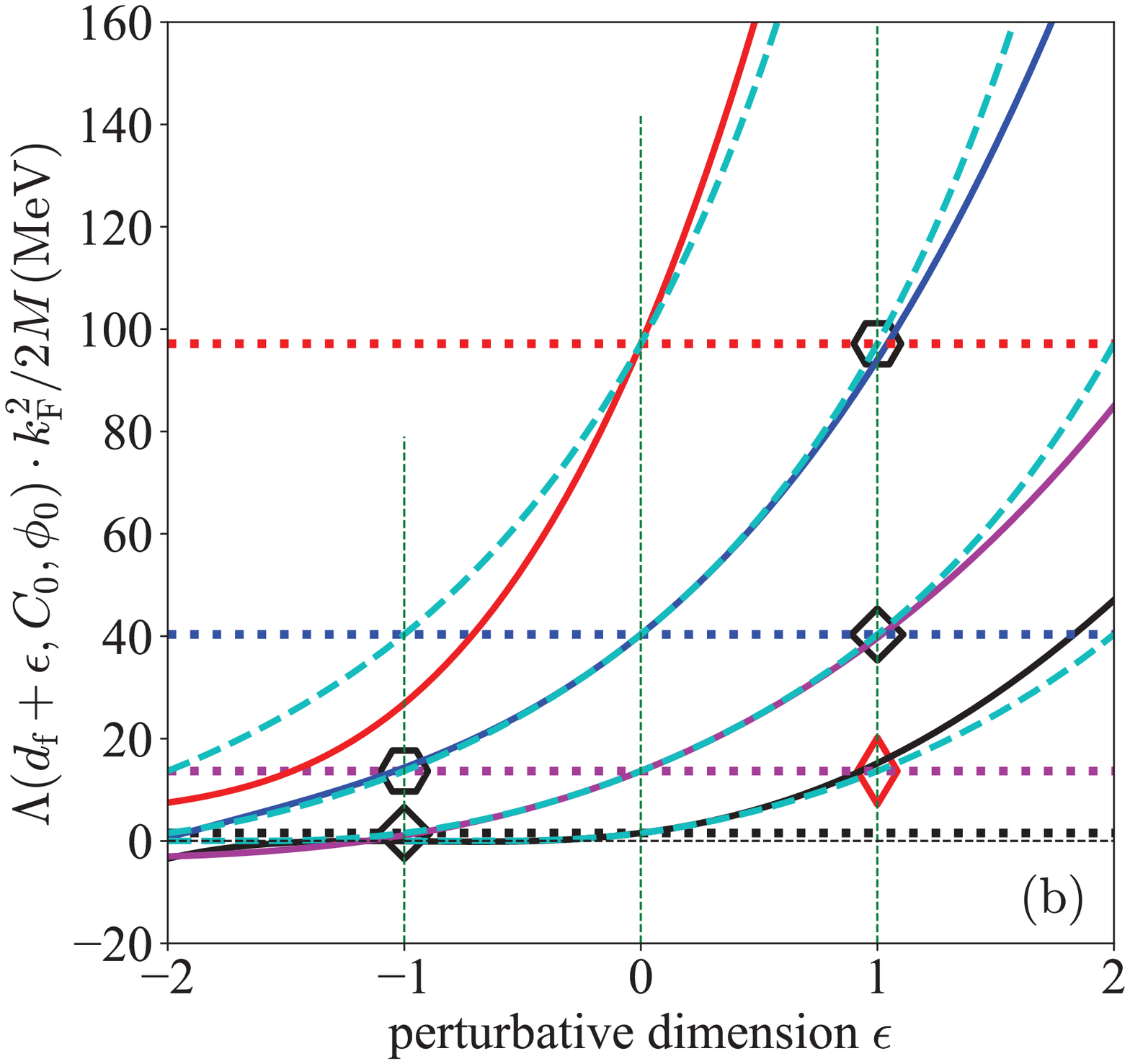}
\caption{\footnotesize Perturbative calculations on $\Lambda(d_{\rm{f}}+\epsilon,C_0,\phi_0)$ (left) and the kinetic EOS $E_0^{\rm{kin}}(\rho_0)$ (right) for different $d_{\rm{f}}$. Only the second-order results are shown in the right panel.}
\label{fig_pert-ep}
\end{figure}

In the left panel of Fig.\,\ref{fig_pert-ep} we show the values of $\Lambda(d,C_0,\phi_0)$ from the perturbative calculations.
The dotted, dash-dotted and solid lines are the results from the zeroth-order, first-order and second-order approximation, respectively. The dashed cyan lines are the exact results for each $d_{\rm{f}}$.
It is seen clearly that the difference between the first-order and the second-order results are smaller compared to that between the zeroth-order and the first-order approximations (e.g., within the range of $-1\lesssim\epsilon\lesssim1$), and this difference increases as $d$ increases. This feature shows the convergence of the $\epsilon$-expansion even though the perturbative dimension $\epsilon$ could take sizable values like 1 or $-1$. In addition, the results from equating $d_{\rm{f}}$ to 1, 2 or 3 are all reasonable. For example, taking $d_{\rm{f}}=1$ and $\epsilon=1$ or $\epsilon=2$ to order $\epsilon$ or order $\epsilon^2$ reasonably gives the corresponding $\Lambda$ function (with the HMT included) in 2D or 3D, see the dash-dotted black line shown in the left panel of Fig.\,\ref{fig_pert-ep}. 
In particular, we have $\Lambda(1+1,C_0,\phi_0)\approx0.728$ to linear order in $\epsilon$ and $\Lambda(1+1,C_0,\phi_0)\approx0.683$ to quadratic order in $\epsilon$, while the exact value for the $\Lambda$ function in 2D is 0.713. Here $C_0\approx0.161$ and $\phi_0\approx2.38$ are adopted, 
see the solid and dot-dashed black lines and the red diamond.
Similarly, starting from $d_{\rm{f}}=2$ and taking $\epsilon=-1$ or $\epsilon=1$ (to order $\epsilon^2$) gives the $\Lambda$ function reasonably well in 1D and 3D, see the dash-dotted magenta line shown in the left panel of  Fig.\,\ref{fig_pert-ep}.
The reference dimension $d_{\rm{f}}=3$ could also be used to make reliable perturbative calculations, as shown by the blue lines in the left panel of Fig.\,\ref{fig_pert-ep}. In particular, taking $d_{\rm{f}}=3$ and $\epsilon=-1$ gives $\Lambda(3-1,C_0,\phi_0)\approx0.607$ and $0.754$, to linear order of $\epsilon$ and quadratic order of $\epsilon^2$, respectively,  and both are close to the exact value 0.713.
These features of the $\epsilon$-expansion indicate that although $\epsilon=1$ or $\epsilon=2$ is not small in conventional sense, the structure of the mathematical expressions makes the expansions effective and reasonable.
On the other hand, one can not effectively obtain the $\Lambda$ function in 3D from $d_{\rm{f}}=4$ with $\epsilon=-1$.
The perturbations behave well for the $\Lambda$ function for $1\lesssim d_{\rm{f}}\lesssim 3$ with the perturbation $\epsilon$
in the range of about $-1\lesssim\epsilon\lesssim1$.
For even better approximation/extrapolation, e.g.,  from $d_{\rm{f}}=1$ to $d=3$ (i.e., with $\epsilon=2$), the Pade approximation\,\cite{Press2007} may be more useful.

In the above, we only considered the perturbative expansion of the function $\Lambda(d,C_0,\phi_0)$. In order to estimate the kinetic EOS of SNM, one also needs to consider the Fermi energy $E_{\rm{F}}=k_{\rm{F}}^2/2M$ in $d$D.
Let's do here the relevant calculations for the 2D situation, where $k_{\rm{F}}^{(\rm{2D})}=\sqrt{\pi\rho_2}$ with $\rho_2=\rho^{2/3}$. Taking the conventional density in 3D as $\rho\approx0.16\,\rm{fm}^{-3}$, then we obtain $k_{\rm{F}}^{(\rm{2D})}\approx189.88\,\rm{MeV}$ and  $E_{\rm{F}}^{(\rm{2D})}\approx19.20\,\rm{MeV}$, 
and consequently the exact kinetic EOS $E_{0,(\rm{2D})}^{\rm{kin}}(\rho_0)$ of SNM is found to be about $E_{\rm{F}}^{(2\rm{D})}\Lambda(2,C_0,\phi_0)\approx13.69\,\rm{MeV}$ (using the exact $\Lambda(2,C_0,\phi_0)\approx0.713$).
We can estimate this value by starting from the kinetic EOS in 1D or 3D to linear order of $\epsilon$.
Specifically, we have for the Fermi momentum $k_{\rm{F}}^{(\rm{1D})}=\pi\rho_1/4\approx84.13\,\rm{MeV}$ where $\rho_1=\rho^{1/3}$, and Fermi momentum in 2D could be obtained as $k_{\rm{F}}^{(\rm{2D})}\approx k_{\rm{F}}^{(\rm{1D})}(1+\sigma_1\epsilon)\approx212.46\,\rm{MeV}$ where $\epsilon=1$ is applied and $\sigma_1\equiv \sigma(1)=-\gamma_{\rm{E}}/2+2^{-1}\ln\pi-\ln(\pi/4)\approx1.5253$ (see definition (\ref{def_sigf})), and thus the approximation for $E_{0,(\rm{2D})}^{\rm{kin}}(\rho_0)$ is about $17.50\,\rm{MeV}$ (adopting $\Lambda(1+\epsilon,C_0,\phi_0)\approx0.728$ for $\epsilon=1$).
Similarly, if we start from the setting in 3D, then the Fermi momentum is approximated as $k_{\rm{F}}^{(\rm{2D})}\approx k_{\rm{F}}^{(\rm{3D})}(1-\sigma)\approx200.02\,\rm{MeV}$ and consequently $E_{0,(\rm{2D})}^{\rm{kin}}(\rho_0)\approx12.93\,\rm{MeV}$, where $\sigma\approx0.2396$ is given in (\ref{defff_sig}) and $\Lambda(3-1,C_0,\phi_0)\approx0.607$ is used.
We have seen both the perturbative expansions based on the 1D and 3D correspondences are effective.
More results are shown in the right panel of Fig.\,\ref{fig_pert-ep}, from which we find again that $1\lesssim d_{\rm{f}}\lesssim3$ with $-1\lesssim\epsilon\lesssim1$ is reasonable.
In fact, one obtains that $E_{0,(\rm{2D})}^{\rm{kin}}(\rho_0)\approx14.36\,\rm{MeV}$ and $E_{0,(\rm{2D})}^{\rm{kin}}(\rho_0)\approx15.21\,\rm{MeV}$ from the corresponding EOS in 3D and 1D, respectively.
Obviously, the approximation for $E_{0,(\rm{2D})}^{\rm{kin}}(\rho_0)$ is better from the 3D correspondence than that from the 1D correspondence. It is partially due to the smaller value of $\sigma(d_{\rm{f}})$ with $d_{\rm{f}}=3$, i.e., $\sigma(3)\approx0.2396$ while $\sigma(1)\approx1.5253$.
Thus in order to investigate the 2D EOS, the starting point from $d_{\rm{f}}=3$ is reasonable (as we already have done in the previous sections).

\begin{figure}[h!]
\renewcommand*\figurename{\footnotesize Fig.}
\centering
 \includegraphics[height=7.cm]{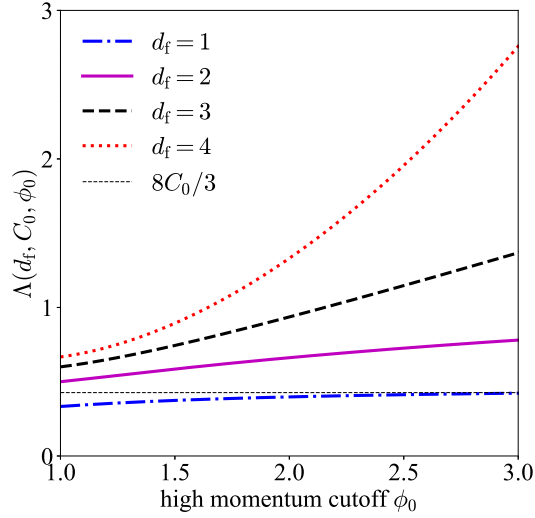}
\caption{\footnotesize Dependence of the function $\Lambda$ on $\phi_0$.}
\label{fig_LL-d}
\end{figure}

As an application of the $\epsilon$-expansion, next we estimate the kinetic EOS of SNM in 2D as $\phi_0\to\infty$ using the corresponding kinetic EOS in 1D. 
First, let us examine the dependence of the kinetic EOS on the high momentum cutoff $\phi_0$.
In particular, in Fig.\,\ref{fig_LL-d}, the dependence of the function $\Lambda(d,C_0,\phi_0)$ on the high momentum cutoff $\phi_0$ is shown. Moreover, we have analytically
\begin{align}\label{def-Lf}
\frac{\partial\Lambda(d,C_0,\phi_0)}{\partial\phi_0}
=&-\frac{C_0d\phi_0^{d-3}}{d+2}\frac{1}{d+2}\left(\frac{d}{d+2}\frac{1}{\phi_0^{2}}-1\right).
\end{align}
Since the high momentum cutoff $\phi_0\geq1$, the derivative (\ref{def-Lf}) should always be larger than zero, indicating that $\Lambda(d,C_0,\phi_0)$ is an increasing function of $\phi_0$, regardless of the dimension $d$ of the space where the nucleons live. The kinetic EOS of SNM diverges if the high momentum cutoff $\phi_0$ (unphysically) approaches infinity, except in 1D. As the dimension $d$ increases the order of divergence also increase, i.e., it changes from $\ln\phi_0$ (logarithmic divergence) in 2D, to $\phi_0$ (linear divergence) in 3D and to $\phi_0^2$  (quadratic divergence) in 4D. More generally, for $d\geq3$, the order of divergence of the kinetic EOS of SNM is $d-2$, as clearly shown in the expression (\ref{def_LAM}).
When $\phi_0\to\infty$ in 1D, the high momentum part of the nucleon distribution function contributes a non-trivial term, i.e., $8C_0/3$ (compared with 1), to the kinetic EOS, as shown in Eq.\,(\ref{E-0-1-kin}) and demonstrated as the dashed line in Fig.\,\ref{fig_LL-d}. 
By taking $\phi_0\to\infty$ in (\ref{d200}), we obtain to linear order of $\epsilon$ that,
\begin{align}\label{dE_eps}
E_{0,d_{\rm{f}}=1,\epsilon}^{\rm{kin}}(\rho)
\approx& \frac{k_{\rm{F}}^{(\rm{1D}),2}(1+2\sigma_1\epsilon)}{6M}
\left[1+\frac{8}{3}C_0+\left(\frac{2}{3}
+\frac{16}{3}C_0
\right)\epsilon\right]+\mathcal{O}(\epsilon^2)\notag\\
&+\sum\rm{term\;approaches\;zero\;as\;}\phi_0\to\infty.
\end{align}
Consequently,
\begin{equation}
E_{0,\rm{(2D)}}^{\rm{kin},\phi_0\to\infty}(\rho)\approx
E_{0,d_{\rm{f}}=1,\epsilon=1}^{\rm{kin}}(\rho)
\approx\frac{k_{\rm{F}}^{(\rm{1D}),2}(1+2\sigma_1)}{6M}
\left(\frac{5}{3}+8C_0
\right).
\end{equation}
Taking $\rho\approx0.16\,\rm{fm}^{-3}$, this gives approximately $E_{0,\rm{(2D)}}^{\rm{kin},\phi_0\to\infty}(\rho)\approx15.04\,\rm{MeV}$. It is slightly large than the exact $E_{\rm{F}}^{(2\rm{D})}\Lambda(2,C_0,\phi_0)\approx13.69\,\rm{MeV}$ if a finite high momentum cutoff $\phi_0\approx2.38$ is used.
The FFG model prediction on the 2D kinetic EOS of SNM is simply $k_{\rm{F}}^{(\rm{2D}),2}/4M=\pi\rho_2/4M\approx9.60\,\rm{MeV}$ where $\rho_2=\rho^{2/3}$.
It is also very interesting to notice that if $\epsilon=2$ is applied in (\ref{dE_eps}), then one obtains $E_{0,\rm{(3D)}}^{\rm{kin},\phi_0\to\infty}(\rho)\approx39.97\,\rm{MeV}$ which is very close to the value of about 40.45\,MeV\,\cite{CaiBJ2015PRC} if $\phi_0\approx2.38$ and $C_0\approx0.161$ are adopted in 3D.
Physically one can not extend the cutoff $\phi_0$ to infinity in conventional 3D situation. This is because above certain momentum the $k^{-4}$ form of the HMT in the single-nucleon momentum distribution is no longer reasonable again due to the higher-order correlations such as the three-nucleon interactions\,\cite{Hen2017RMP,Egiyan2006PRL, Fomin2012PRL,Fomin2017,Ye2018PRC,Sargsian2019PRC}. In this case, other HMT forms in $n_{\v{k}}^0$ should be necessarily considered.

The above results indicate that the $\epsilon$-expansion of the kinetic EOS from certain dimension $d_{\rm{f}}$ (e.g., $d_{\rm{f}}=1$) is reasonable and effective. However, as the expansion order $n$ of $\epsilon$ increases, the extrapolation becomes less believable, since $(1+1)^n=2^n$ could be very large and even become divergent.

\subsection{Momentum fluctuations of nucleons in symmetric nuclear matter in $d$D spaces}

Once the nucleon momentum distribution is specified, one can investigate its moments. In particular, the variance of nucleon momentum $k$ namely $\langle k^2\rangle-\langle k\rangle^2$ characterizing the strength of nucleon momentum fluctuations is very informative. We define the (relative) nucleon momentum fluctuation $\Upsilon_k$ in SNM in $d$D spaces as,
\begin{equation}
\Upsilon_k(d)\equiv\sqrt{\frac{\langle k^2\rangle-\langle k\rangle^2}{\langle k\rangle^2}}=\sqrt{\frac{\langle k^2\rangle}{\langle k\rangle^2}
-1},
\end{equation}
with the moment $\langle k^{\gamma}\rangle=\langle \v{k}^{\gamma}\rangle$ being defined with respect to the momentum distribution $n_{\v{k}}^0$, i.e.,
\begin{equation}
\langle \v{k}^{\gamma}\rangle=\left.\int n_{\v{k}}^0\v{k}^{\gamma}\d^d\v{k}\right/\int n_{\v{k}}^0\d^d\v{k}.
\end{equation}
For our purposes here, $\gamma=1$ and $\gamma=2$ are relevant, 
\begin{align}
\left.\langle k\rangle\right/k_{\rm{F}}=&\frac{d}{d+1}
\left[1+C_0\left(\frac{4}{(d-3)(d-4)}+\frac{d+1}{d-3}\phi_0^{d-3}-\frac{d}{d-4}\phi_0^{d-4}\right)\right],\\
\left.\langle k^2\rangle\right/k_{\rm{F}}^2=&\frac{d}{d+2}
\left[1+C_0\left(\frac{8}{(d-2)(d-4)}+\frac{d+2}{d-2}\phi_0^{d-2}-\frac{d}{d-4}\phi_0^{d-4}\right)\right],
\end{align}
from which the fluctuation factor $\Upsilon_k(d)$ in $d$D could be obtained.

For the nucleons in $d=1,2,3,4$ with HMT, one has,
\begin{align}
\rm{1D}:~~\left.\langle k\rangle\right/k_{\rm{F}}=&\frac{1}{2}\left[1+C_0\left(
\frac{2}{3}+\frac{1}{3\phi_0^3}-\frac{1}{\phi_0^2}
\right)\right],
~~\left.\langle k^2\rangle\right/k_{\rm{F}}^2=\frac{1}{3}\left[1+C_0\left(
\frac{8}{3}+\frac{1}{3\phi_0^3}-\frac{3}{\phi_0}
\right)\right],\\
\rm{2D}:~~\left.\langle k\rangle\right/k_{\rm{F}}=&\frac{2}{4}\left[1+C_0\left(2+\frac{1}{\phi_0^2}-\frac{3}{\phi_0}\right)\right],
~~\left.\langle k^2\rangle\right/k_{\rm{F}}^2=\frac{1}{2}\left[1+C_0\left(4\ln\phi_0+\frac{1}{\phi_0^2}-1\right)\right],\\
\rm{3D}:~~\left.\langle k\rangle\right/k_{\rm{F}}=&\frac{3}{4}\left[1+C_0\left(4\ln\phi_0+\frac{3}{\phi_0}-3\right)\right],
~~\left.\langle k^2\rangle\right/k_{\rm{F}}^2=\frac{3}{5}\left[1+C_0\left(5\phi_0+\frac{3}{\phi_0}-8\right)\right],\\
\rm{4D}:~~\left.\langle k\rangle\right/k_{\rm{F}}=&\frac{4}{5}\left[1+C_0\left(5\phi_0-4\ln\phi_0-5\right)\right],
~~\left.\langle k^2\rangle\right/k_{\rm{F}}^2=\frac{2}{3}\left[1+C_0\left(3\phi_0^2-4\ln\phi_0-3\right)\right].
\end{align}
For example, we have for 3D, 
\begin{equation}
\Upsilon_k^{\rm{HMT}}(3)=\sqrt{\frac{16\phi_0}{15}\frac{5C_0\phi_0^2-8C_0\phi_0+3C_0+\phi_0}{4C_0\phi_0\ln\phi_0-3C_0\phi_0+3C_0+\phi_0}-1}\to\Upsilon_k^{\rm{FFG}}(3)=\sqrt{\frac{1}{15}},~~\rm{as}~~\phi_0\to1.
\end{equation}
For the FFG model, one has $
\Upsilon_k^{\rm{FFG}}(d)=1/\sqrt{d^2+2d}$,  and thus $\Upsilon_k^{\rm{FFG}}(3)\approx0.258$ and $\Upsilon_k^{\rm{FFG}}(2)\approx0.354$.

\begin{figure}[h!]
\renewcommand*\figurename{\footnotesize Fig.}
\centering
 \includegraphics[height=6.8cm]{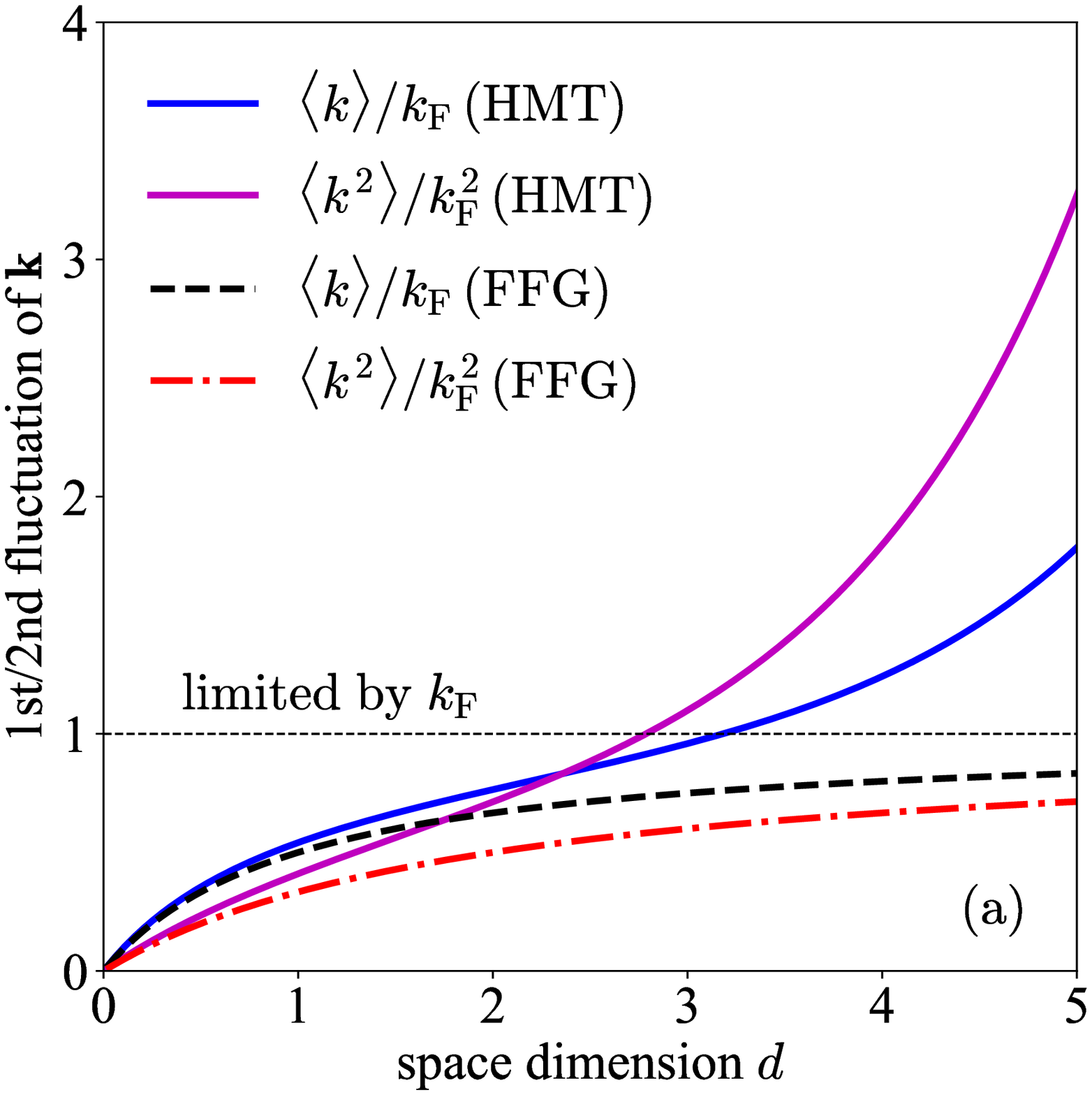}\qquad
 \includegraphics[height=6.8cm]{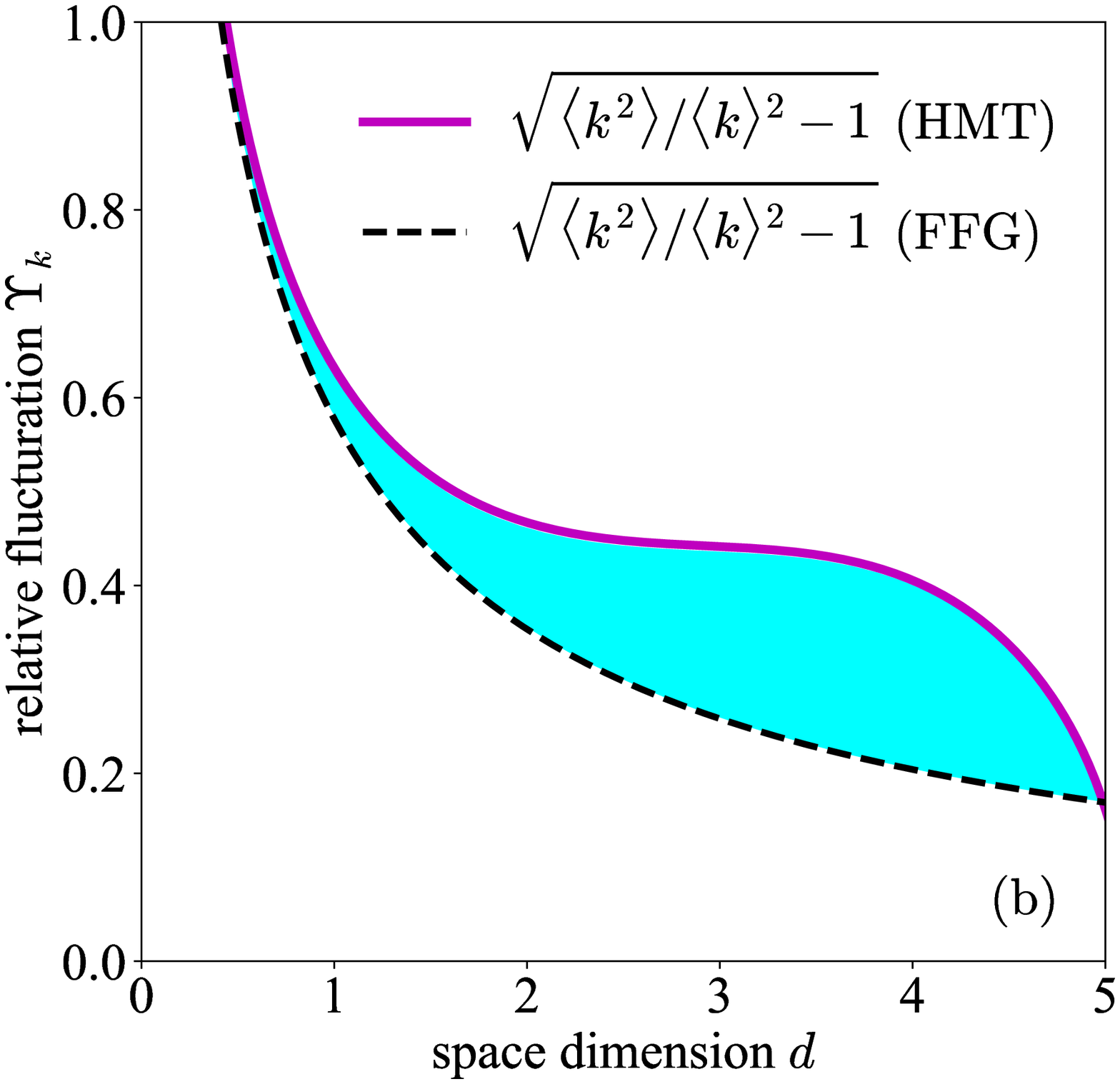}
\caption{\footnotesize Fluctuation of the momentum $\Upsilon_k$ at saturation density $\rho_0$ as a function of $d$.}
\label{fig_fluc-d}
\end{figure}

In Fig.\,\ref{fig_fluc-d} we show the first and the second fluctuations in momentum $\v{k}$ as functions of the spacial dimension $d$ (left) and the relative momentum fluctuation $\Upsilon_k$, both in the FFG and the HMT models.
One can see from the figure: (a) as the dimension $d$ decreases both the first and the second fluctuations decrease, while the relative fluctuation $\Upsilon_k$ becomes larger, i.e., low-dimensional nucleonic matter has larger (relative) momentum fluctuation compared with the high dimensional situations, for either the FFG or the HMT model; (b) when the SRC-induced HMT is considered, the fluctuation at a fixed $d$ increases, i.e., the interactions strengthen the momentum fluctuations as now nucleons are allowed to move into the high-momentum states above the Fermi surface.
Obviously the moment $\langle k^{\gamma}\rangle^{1/\gamma}$ is limited by the Fermi momentum $k_{\rm{F}}$ in the FFG model since no nucleons are allowed to be above the Fermi surface. The finding that fluctuations due to interactions are stronger in low dimensions is well known in solid state physics, as the Mermin-Wagner theorem\,\cite{Mermin1966} states that continuous symmetries cannot be spontaneously broken at finite temperatures in systems with short-range interactions in dimensions $d\leq 2$.
More intuitively, this means that long-range fluctuations can be created with small energy costs.
In addition, we also find that the relative fluctuation $\Upsilon_k$ in low dimensions approaches its FFG counterpart (although both are large quantitatively), i.e., $\Upsilon_k$ approaches the free model prediction.
More quantitatively, we have $\Upsilon_k^{\rm{HMT}}(4)\approx0.406$, $\Gamma_k^{\rm{HMT}}(3)\approx0.442$, $\Gamma_k^{\rm{HMT}}(2)\approx0.467$ and $\Gamma_k^{\rm{HMT}}(1)\approx0.632$. Compared with $\Upsilon_k^{\rm{FFG}}(4)\approx0.204$, $\Upsilon_k^{\rm{FFG}}(3)\approx0.258$, $\Upsilon_k^{\rm{FFG}}(2)\approx0.354$ and $\Gamma_k^{\rm{FFG}}(1)\approx0.577$, the HMT enhancements are about 99\%, 71\%, 32\% and 9.5\%, respectively as $d$ decreases from 4 to 1.

\subsection{Kinetic symmetry energy in $d$D spaces with SRC-induced high-momentum nucleons}

By expanding the kinetic EOS of ANM and extract the coefficient of the $\delta^2$ term, we obtain the kinetic symmetry energy for $d=$1-4 as,
\begin{align}
E_{\rm{sym}}^{\rm{kin},(\rm{1D})}(\rho)
=&\frac{k_{\rm{F}}^2}{2M}\Bigg\{1+\frac{8C_0}{3\phi_0^3}
\Bigg[\left(C_1+1\right)\phi_0^3
+\Bigg(\left(\frac{3C_1}{8}+\frac{9}{8}\right)\phi_1-\frac{3\phi_1^2}{8}-\frac{9C_1}{8}-\frac{9}{8}\Bigg)\phi_0^2
\notag\\
&\hspace{6.cm}-\frac{(\phi_1-1)(-2\phi_1+C_1+1)}{8}\Bigg]\Bigg\},\label{E2-d1}\\
E_{\rm{sym}}^{\rm{kin},(\rm{2D})}(\rho)
=&\frac{k_{\rm{F}}^2}{4M}\Bigg\{
1+\frac{4C_0}{\phi_0^2}\Bigg[
2\left(C_1+\frac{1}{2}\right)\phi_0^2\ln\phi_0
+\Bigg(
(C_1+2)\phi_1-\frac{\phi_1^2}{2}-\frac{C_1}{2}-\frac{1}{4}
\Bigg)\phi_0^2\notag\\
&\hspace{6.cm}-\frac{1}{2}(\phi_1-1)\left(C_1-\frac{3\phi_1}{2}+\frac{1}{2}\right)
\Bigg]
\Bigg\},\label{E2-d2}\\
E_{\rm{sym}}^{\rm{kin},(\rm{3D})}(\rho)=&\frac{k_{\rm{F}}^2}{6M}\Bigg\{1+\frac{9C_0}{\phi_0}\Bigg[\frac{1}{3}+C_1-\left(1+\frac{3C_1}{5}\right)\phi_1+\frac{3\phi_1^2}{5}-\left(\frac{8}{9}+\frac{8C_1}{3}\right)\phi_0\notag\\
&\hspace{6.cm}+
\left(\frac{5}{9}+\frac{5C_1}{3}+\left(\frac{5}{3}+C_1\right)\phi_1\right)\phi_0^2\Bigg]\Bigg\},\label{E2-d3}\\
E_{\rm{sym}}^{\rm{kin},(\rm{4D})}(\rho)=&\frac{k_{\rm{F}}^2}{8M}
\Bigg\{1-16\left(C_1+\frac{1}{4}\right)C_0\ln\phi_0\notag\\
&\hspace*{1.cm}
+16C_0\Bigg[
\Bigg(\frac{\phi_0^2}{2}+\frac{1}{3}\Bigg)\phi_1^2+\left(\phi_0^2-\frac{2}{3}\right)\left(C_1+\frac{3}{2}\right)\phi_1+\frac{3}{4}\left(\phi_0^2-1\right)\left(C_1+\frac{1}{4}\right)
\Bigg]\Bigg\}.\label{E2-d4}
\end{align}
Similarly, we also list the analytical expressions for the fourth-order symmetry energy for $d=1-4$ as,
\begin{align}
E_{\rm{sym,4}}^{\rm{kin},(\rm{1D})}(\rho)
=&\frac{4k_{\rm{F}}^2}{9M}\frac{C_0}{\phi_0^3}
\left[
\left(-\frac{9\phi_0^2}{8}+\frac{15}{8}\right)\phi_1^4
+\frac{9(C_1+3)\phi_1^3}{8}\left(\phi_0^2-\frac{10}{9}\right)
-\frac{27(C_1+1)\phi_1^2}{8}\left(\phi_0^2-\frac{2}{3}\right)
\right.\notag\\
&\hspace*{3.cm}
\left.+\frac{27\phi_1}{8}\left(C_1+\frac{1}{3}\right)\left(\phi_0^2-\frac{1}{3}\right)+\phi_0^3C_1-\frac{9\phi_0^2C_1}{8}+\frac{C_1}{8}
\right],\label{E4-d1}\\
E_{\rm{sym,4}}^{\rm{kin},(\rm{2D})}(\rho)
=&\frac{k_{\rm{F}}^2}{3M}\frac{C_0\phi_1}{\phi_0^2}
\Bigg[
\Bigg(-\frac{3\phi_0^2}{4}+\frac{15}{4}\Bigg)\phi_1^3
+\left(\phi_0^2-3\right)(C_1+2)\phi_1^2\notag\\
&\hspace*{3.cm}
\left.-3\left(C_1+\frac{1}{2}\right)\left(\phi_0^2-\frac{3}{2}\right)\phi_1+3\phi_0^2C_1-\frac{3C_1}{2}
\right],\label{E4-d2}\\
E_{\rm{sym,4}}^{\rm{kin},(\rm{3D})}(\rho)
=&\frac{k_{\rm{F}}^2}{162M}
\Bigg\{
1+\frac{135C_0}{\phi_0}\Bigg[
\frac{27\phi_1^4}{25}-\left(\frac{27C_1}{25}+\frac{9}{5}\right)\phi_1^3
+\left(\frac{9C_1}{5}+\frac{3}{5}\right)\phi_1^2\notag\\
&\hspace*{3.cm}+\left(\phi_0^2-\frac{3}{5}\right)\left(C_1-\frac{1}{9}\right)\phi_1-\frac{\phi_0-1}{9}\left(C_1-\frac{1}{3}\right)\left(\phi_0-\frac{3}{5}\right)
\Bigg]
\Bigg\},\label{E4-d3}\\
E_{\rm{sym,4}}^{\rm{kin},(\rm{4D})}(\rho)
=&\frac{k_{\rm{F}}^2}{128M}\left\{1
+\frac{32C_0\ln\phi_0}{3}\left(C_1-\frac{3}{8}\right)
+192C_0\left[\frac{2\phi_1^4}{9}-\left(\frac{8C_1}{27}+\frac{4}{9}\right)\phi_1^3+\left(\phi_0^2+\frac{2}{3}\right)\left(C_1+\frac{1}{4}\right)\phi_1^2\right.\right.\notag\\
&\hspace{3cm}\left.\left.
+\frac{\phi_1}{2}\left(C_1-\frac{1}{6}\right)\left(\phi_0^2-\frac{2}{3}\right)-\frac{\phi_0^2-1}{24}\left(C_1-\frac{3}{8}\right)\right]
\right\}.\label{E4-d4}
\end{align}
As a reference, here we also give analytical expressions for the kinetic symmetry energy as well as the fourth-order kinetic symmetry energy in a space of general dimensions $d$ (not necessarily integers) as
\begin{align}
&E_{\rm{sym}}^{\rm{kin},d}(\rho)=\frac{1}{2d(d-4)(d-2)(d+2)}\frac{k_{\rm{F}}^2}{2M}
\Bigg\{2(d+2)\left[8C_0(dC_1+1)+(d-4)(d-2)\right]\notag\\
&\hspace*{2.cm}+C_0(d-4)(d-2)\Big[
\phi_1^2d^4+\left(2C_1-5\phi_1+2\right)\phi_1d^3\notag\\
&\hspace{2.cm}+2\left(C_1-2C_1\phi_1+3\phi_1^2\right)d^2+2(2C_1-4\phi_1+1)d+4
\Big]\phi_0^{d-2}\notag\\
&\hspace*{2.cm}-C_0d(d-2)\Big[\phi_1^2d^4+(2C_1-9\phi_1+2)\phi_1d^3
+2\left[10\phi_1^2-2(2C_1+1)\phi_1+C_1\right]d^2\notag\\
&\hspace*{2.cm}+2(2C_1-8\phi_1+1)d+4\Big]\phi_0^{d-4}
\Bigg\},\label{E2-k-d}\\
&E_{\rm{sym},4}^{\rm{kin},d}(\rho)=\frac{1}{24d^3(d-4)(d+2)}\frac{k_{\rm{F}}^2}{2M}\Bigg\{
dC_0\phi_0^{d-4}\Bigg[
-(d-4)(d-5)(d-6)(d-7)d^4\phi_1^4\notag\\
&\hspace*{2.cm}+4d^3(d-4)(d-5)(d-6)(dC_1+d+2)\phi_1^3\notag\\
&\hspace*{2.cm}-12d^2(d-4)(d-5)(d+2)(dC_1+1)\phi_1^2
-24d(d-4)(d+2)\left(dC_1-\frac{d}{3}+\frac{2}{3}\right)\phi_1\notag\\
&\hspace*{2.cm}+8(d-2)(d+2)\left(dC_1-\frac{d}{2}+\frac{1}{2}\right)\Bigg]\notag\\
&\hspace*{1.cm}
+(d-2)\Bigg[
C_0d^4(d-4)^2(d-3)(d-5)\phi_1^4\phi_0^{d-2}
+4C_0d^3(d-4)^2(d-3)(dC_1+d+2)\phi_1^3\phi_0^{d-2}\notag\\
&\hspace*{2.cm}
+12C_0d^2(d-4)(d-3)(d+2)(dC_1+)\phi_1^2\phi_0^{d-2}\notag\\
&\hspace*{2.cm}+24C_0d(d-4)(d+2)\left(dC_1-\frac{d}{3}+\frac{2}{3}\right)\phi_1\phi_0^{d-2}\notag\\
&\hspace*{2.cm}-8dC_0C_1\left(d^2\phi_0^{d-2}-2d\phi_0^{d-2}-8\phi_0^{d-2}+8\right)
\notag\\
&\hspace*{2.cm}+(d-1)\left[4d^2\left(C_0\phi_0^{d-2}+1\right)-8d\left(C_0\phi_0^{d-2}+3\right)-32C_0\left(\phi_0^{d-2}-1\right)+3\right]\Bigg]\Bigg\}.\label{E4-k-d}
\end{align}
If one takes the high momentum cutoff to $\phi_0=1$ and $\phi_1=0$, all the kinetic symmetry energies reduce to the FFG predictions e.g., $
E_{\rm{sym},d}^{\rm{kin},\rm{FFG}}(\rho)={k_{\rm{F}}^2}/{2dM}$, see (\ref{4qEsymFF_1}).
Moreover, the $E_{\rm{sym},4}^{\rm{kin},d}(\rho)$ in dimensions $d=1$ and $d=2$ are zero consistent with the FFG prediction given by the expression (\ref{4qEsym4FF_1}).

\begin{figure}[h!]
\renewcommand*\figurename{\footnotesize Fig.}
\centering
 \includegraphics[height=3.1cm]{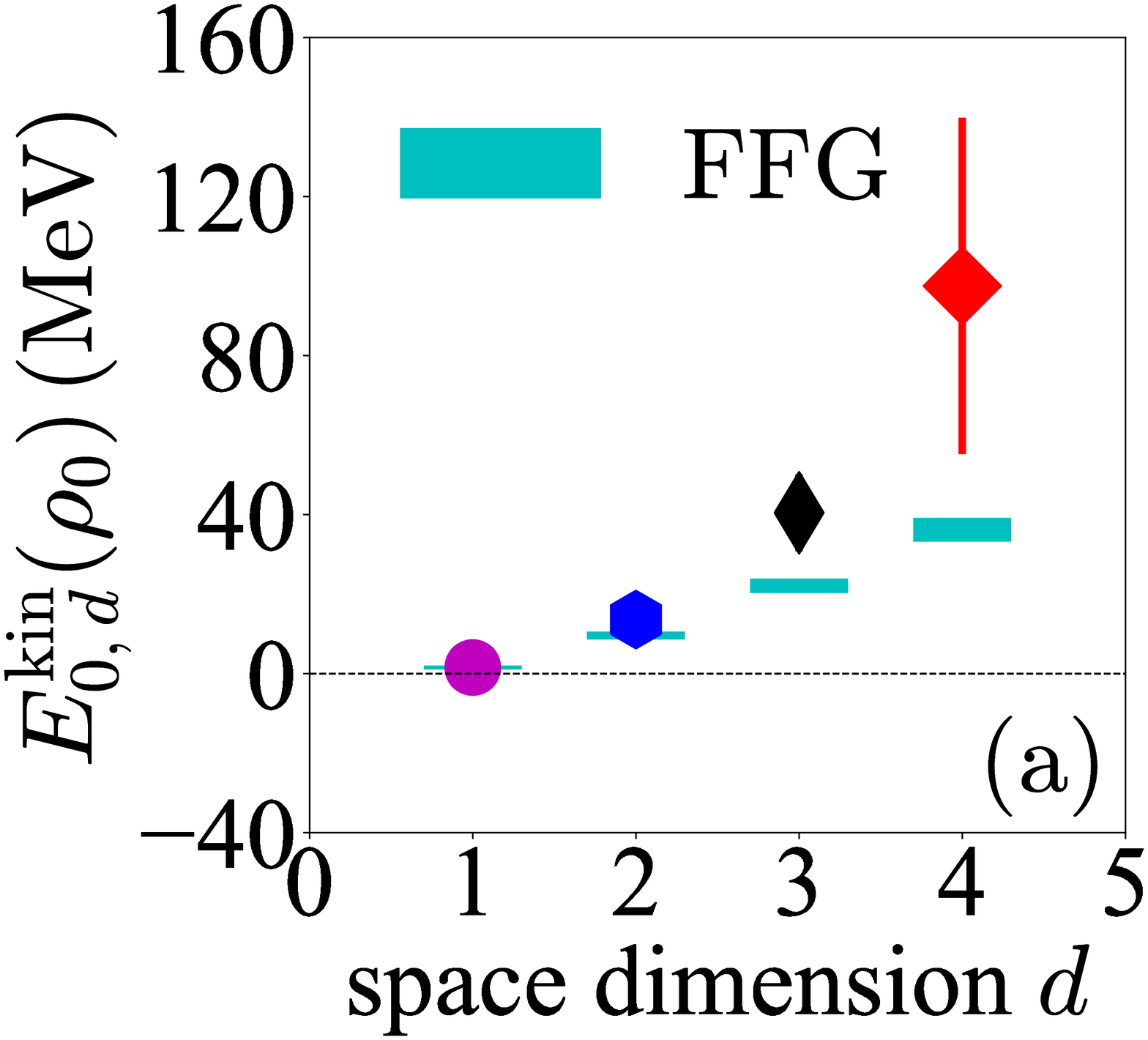}\quad
 \includegraphics[height=3.1cm]{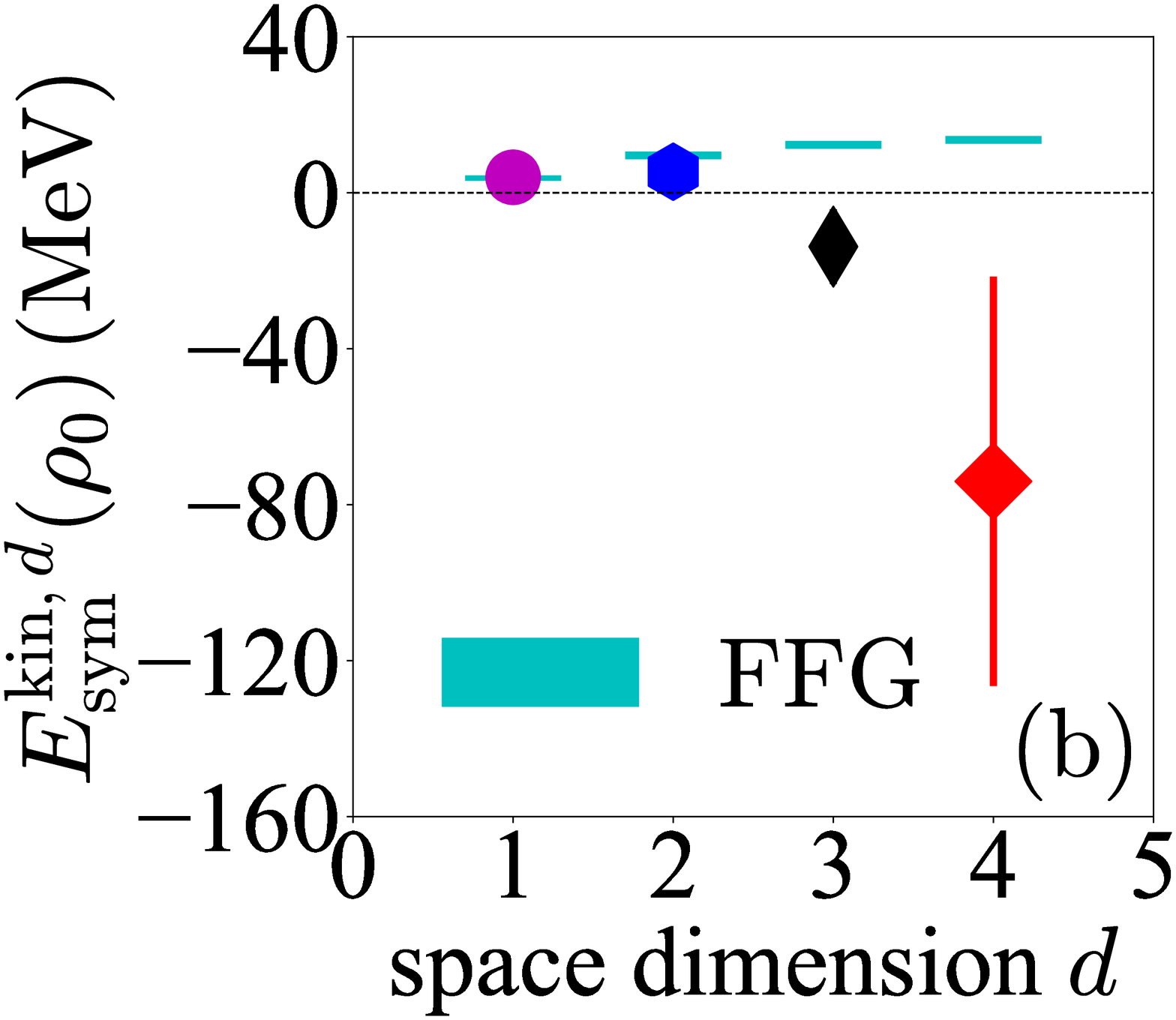}\quad
 \includegraphics[height=3.1cm]{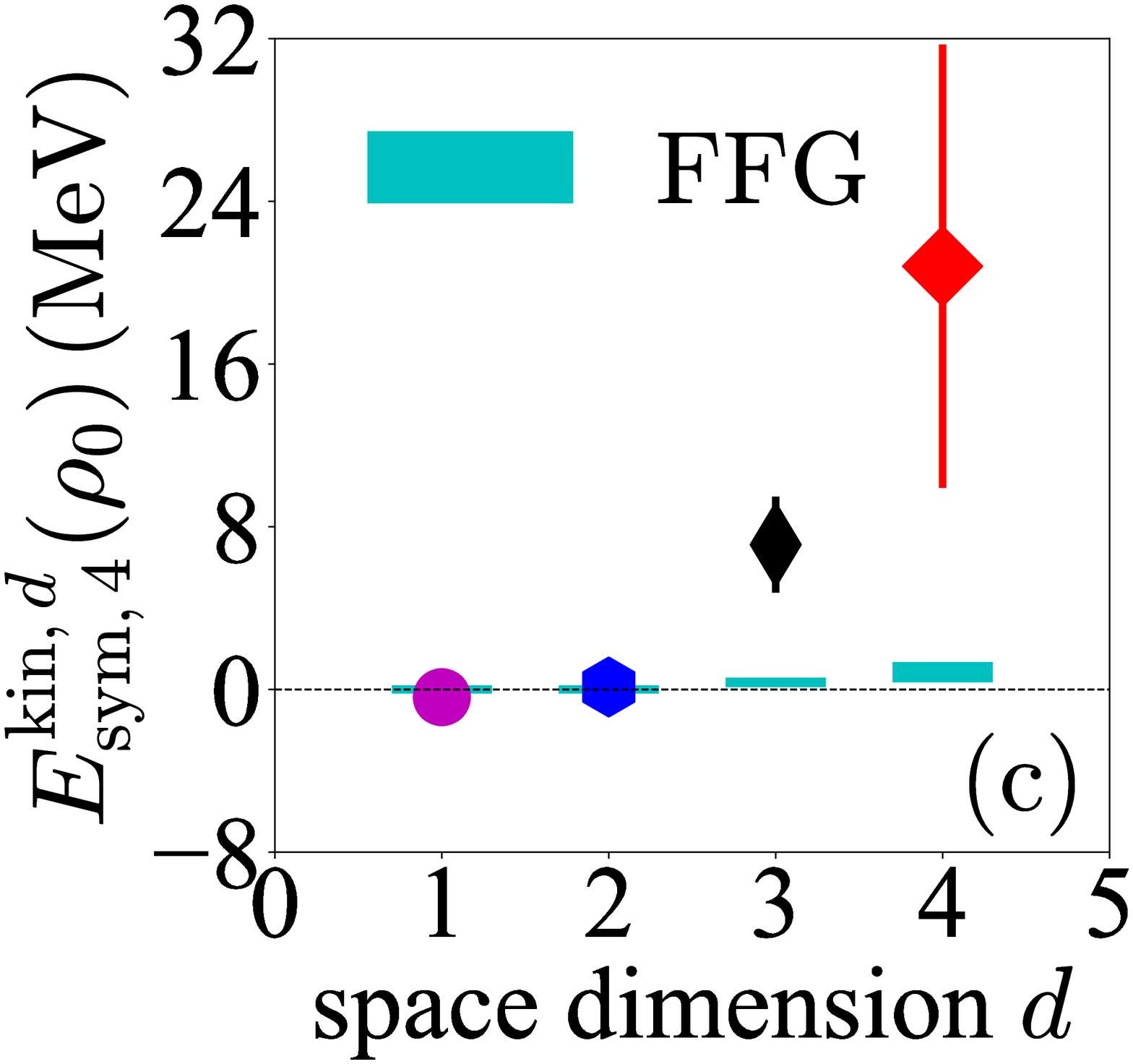}\quad
 \includegraphics[height=3.1cm]{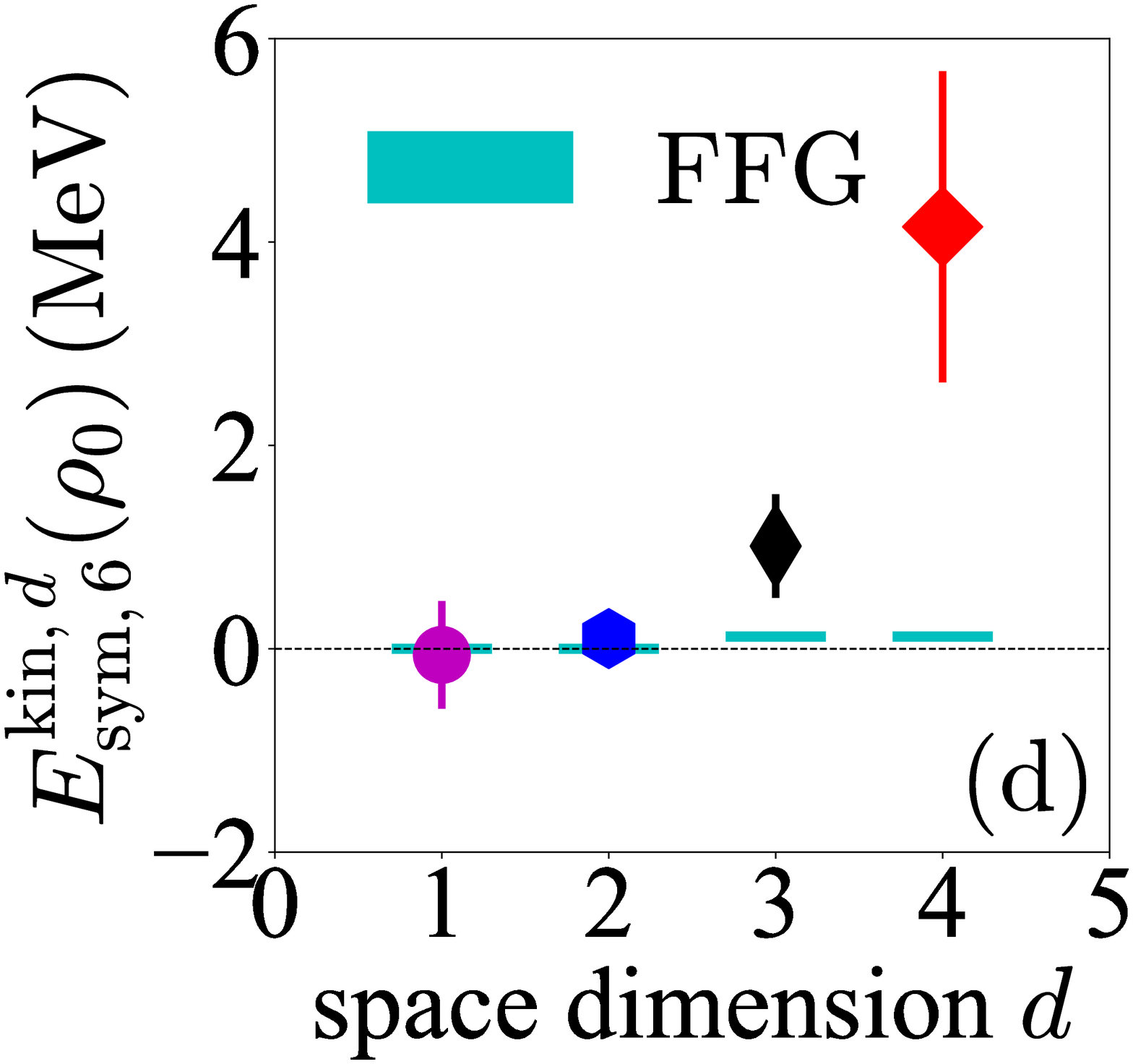}
\caption{\footnotesize Kinetic EOS of ANM in different dimensions.}
\label{fig_kinetic-E-d}
\end{figure}

To have more quantitative evaluations of the kinetic symmetry energies in $d$D spaces, we now adopt $C_0\approx0.161\pm0.015$, $C_1\approx-0.25\pm0.07$, $\phi_0\approx2.38\pm0.56$, $\phi_1\approx-0.56\pm0.10$\,\cite{CaiBJ2015PRC} as constrained in the conventional 3D situation and $\rho_0\approx0.16\pm0.02\,\rm{fm}^{-3}$. Then using the $k_{\rm{F}}$ given by (\ref{def_kF_d}) where $\rho_d=\rho^{d/3}$ and $\rho$ is the conventional 3D density, we find that the fourth-order symmetry energies in $d$D are $E_{\rm{sym},4}^{\rm{kin},\rm{(1D)}}(\rho_0)\approx-0.38\pm0.14\,\rm{MeV}$, $E_{\rm{sym},4}^{\rm{kin},\rm{(2D)}}(\rho_0)\approx0.12\pm0.44\,\rm{MeV}$, $E_{\rm{sym},4}^{\rm{kin},\rm{(3D)}}(\rho_0)\approx7.12\pm2.36\,\rm{MeV}$\,\cite{CaiBJ2015PRC}, and $E_{\rm{sym},4}^{\rm{kin},\rm{(4D)}}(\rho_0)\approx20.81\pm10.91\,\rm{MeV}$, respectively.
Similarly, one can also obtain the values of the kinetic EOS of SNM and the quadratic kinetic symmetry energy in $d$D spaces.
Quantitatively, for the former we have $E_{0,\rm{(1D)}}^{\rm{kin}}(\rho_0)\approx1.55\pm0.14\,\rm{MeV}$, $E_{0,\rm{(2D)}}^{\rm{kin}}(\rho_0)\approx13.61\pm1.79\,\rm{MeV}$, $E_{0,\rm{(3D)}}^{\rm{kin}}(\rho_0)\approx40.47\pm9.48\,\rm{MeV}$\,\cite{CaiBJ2015PRC}, and $E_{0,\rm{(4D)}}^{\rm{kin}}(\rho_0)\approx97.53\pm42.30\,\rm{MeV}$, respectively. While for the quadratic kinetic symmetry energy, we have $E_{\rm{sym}}^{\rm{kin},\rm{(1D)}}(\rho_0)\approx3.98\pm0.41\,\rm{MeV}$, $E_{\rm{sym}}^{\rm{kin},\rm{(2D)}}(\rho_0)\approx5.41\pm1.45\,\rm{MeV}$, $E_{\rm{sym}}^{\rm{kin},\rm{(3D)}}(\rho_0)\approx-13.80\pm9.63\,\rm{MeV}$\,\cite{CaiBJ2015PRC}, and $E_{\rm{sym}}^{\rm{kin},\rm{(4D)}}(\rho_0)\approx-74.04\pm52.50\,\rm{MeV}$, respectively.
In addition, the sixth-order kinetic symmetry energy could also be given. Due to its generally complicated nature of the long analytical expression, here we do not give its explicit form. Numerically, they are found to be 
$E_{\rm{sym},6}^{\rm{kin},\rm{(1D)}}(\rho_0)\approx-0.06\pm0.05\,\rm{MeV}$, $E_{\rm{sym},6}^{\rm{kin},\rm{(2D)}}(\rho_0)\approx0.10\pm0.14\,\rm{MeV}$, $E_{\rm{sym},6}^{\rm{kin},\rm{(3D)}}(\rho_0)\approx1.01\pm0.51\,\rm{MeV}$, and $E_{\rm{sym},6}^{\rm{kin},\rm{(4D)}}(\rho_0)\approx4.51\pm2.25\,\rm{MeV}$, respectively.

\begin{figure}[h!]
\renewcommand*\figurename{\footnotesize Fig.}
\centering
 \includegraphics[height=6.8cm]{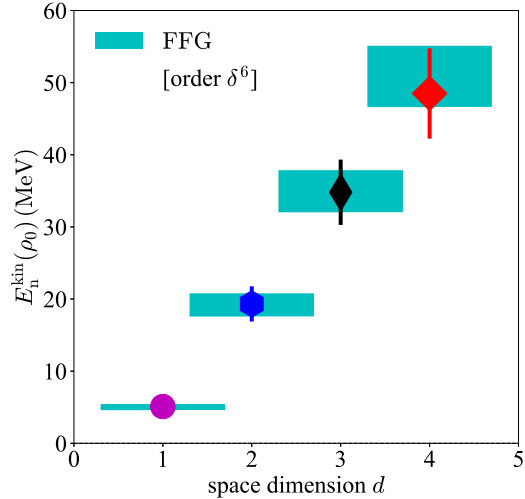}
\caption{\footnotesize Kinetic EOS of PNM to order $\delta^6$ in the FFG/HMT model.}
\label{fig_Enkin-d}
\end{figure}

For comparisons, we notice that the FFG model predicts $E_0^{\rm{kin}}(\rho)=dk_{\rm{F}}^2/(d+2)$, $E_{\rm{sym}}^{\rm{kin}}(\rho)=k_{\rm{F}}/2dM$, $E_{\rm{sym,4}}^{\rm{kin}}(\rho)=(2/d-1)(2/d-2)k_{\rm{F}}^2/24dM$, and 
$E_{\rm{sym},6}^{\rm{kin},d}(\rho)=(3d-2)(d-1)(d-2)(2d-1)k_{\rm{F}}^2/180d^5M$, respectively. Numerically, we then have $E_{0,d}^{\rm{kin}}(\rho_0)\lesssim36.20\,\rm{MeV}$, $E_{\rm{sym},2}^{\rm{kin},d}(\rho_0)\lesssim13.57\,\rm{MeV}$, $E_{\rm{sym},4}^{\rm{kin},d}(\rho_0)\lesssim0.85\,\rm{MeV}$, and $E_{\rm{sym},6}^{\rm{kin},d}(\rho_0)\lesssim0.12\,\rm{MeV}$, respectively, for $1\leq d\leq 4$ based on the FFG model predictions. Fig.\,\ref{fig_kinetic-E-d} shows the components of the kinetic EOS considering the SRC-induced HMT in comparison with the FFG model predictions .
It is seen that for $d=1$ and $d=2$, all components of the kinetic EOS with the SRC/HMT are very close to their FFG counterparts. In addition, the fourth-order and the sixth-order symmetry energies from the FFG model in 1D or 2D are exactly zero. Even when the SRC-induced HMT is considered in 1D and 2D, the predicted kinetic symmetry energies are still very close to the FFG model predictions. Interestingly, however, at higher dimensions the SRC/HMT effects on all the kinetic energies studied become more apparent. Fig.\,\ref{fig_Enkin-d} shows the kinetic EOSs of PNM to order $\delta^6$ in the FFG and HMT models. It is seen that the two models predict very similarly for $d\leq4$ that the kinetic energy in PNM increases with the dimension $d$. The negligible difference between the kinetic energies in PNM calculated with the two models is expected. This is because the HMT model parameters were based on the experimental finding that the HMT fraction is about 20\% in SNM while it is less than about 2\% in PNM \cite{CaiBJ2015PRC}. 

\begin{figure}[h!]
\renewcommand*\figurename{\footnotesize Fig.}
\centering
 \includegraphics[height=7.3cm]{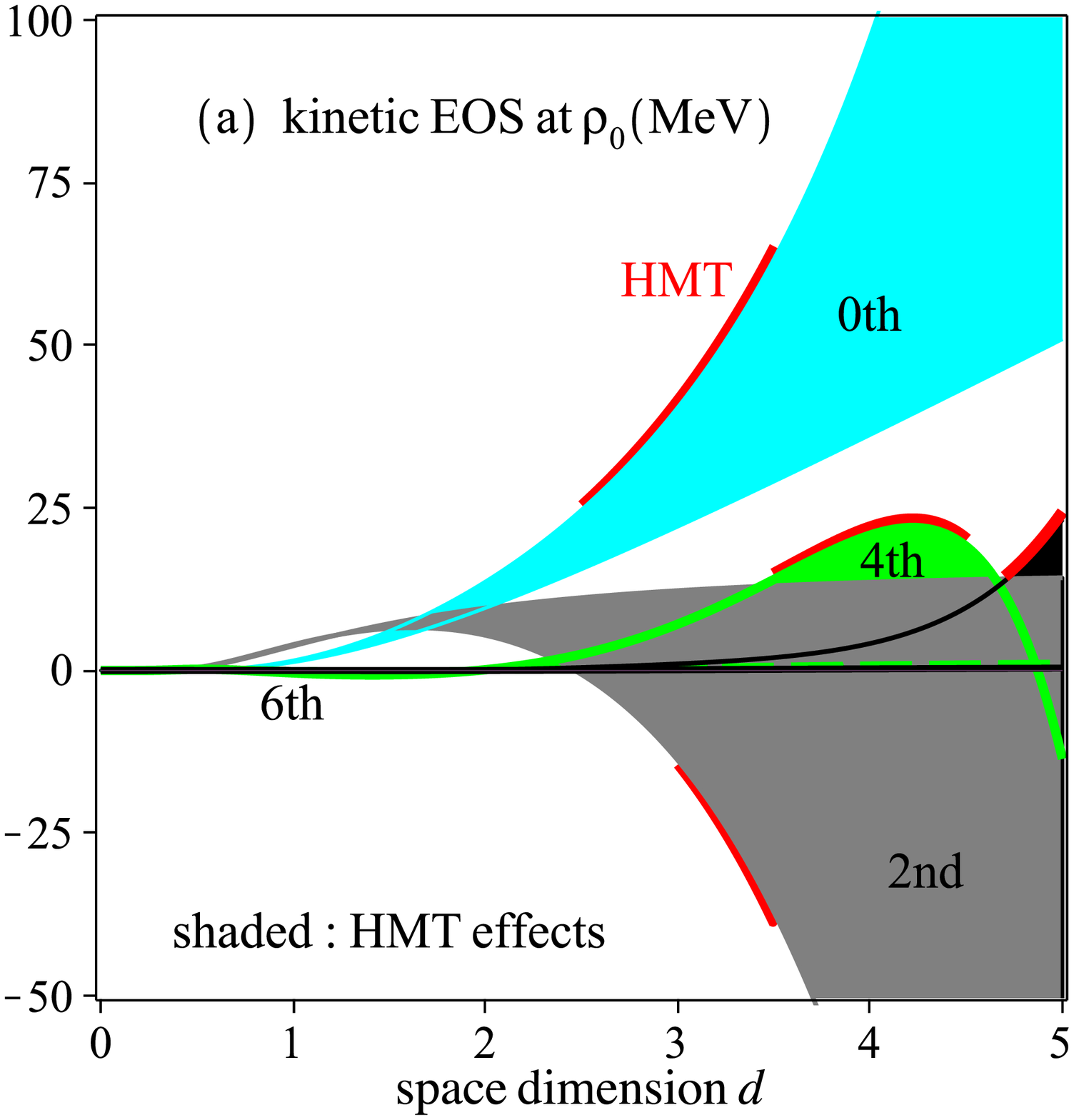}\qquad
 \includegraphics[height=7.3cm]{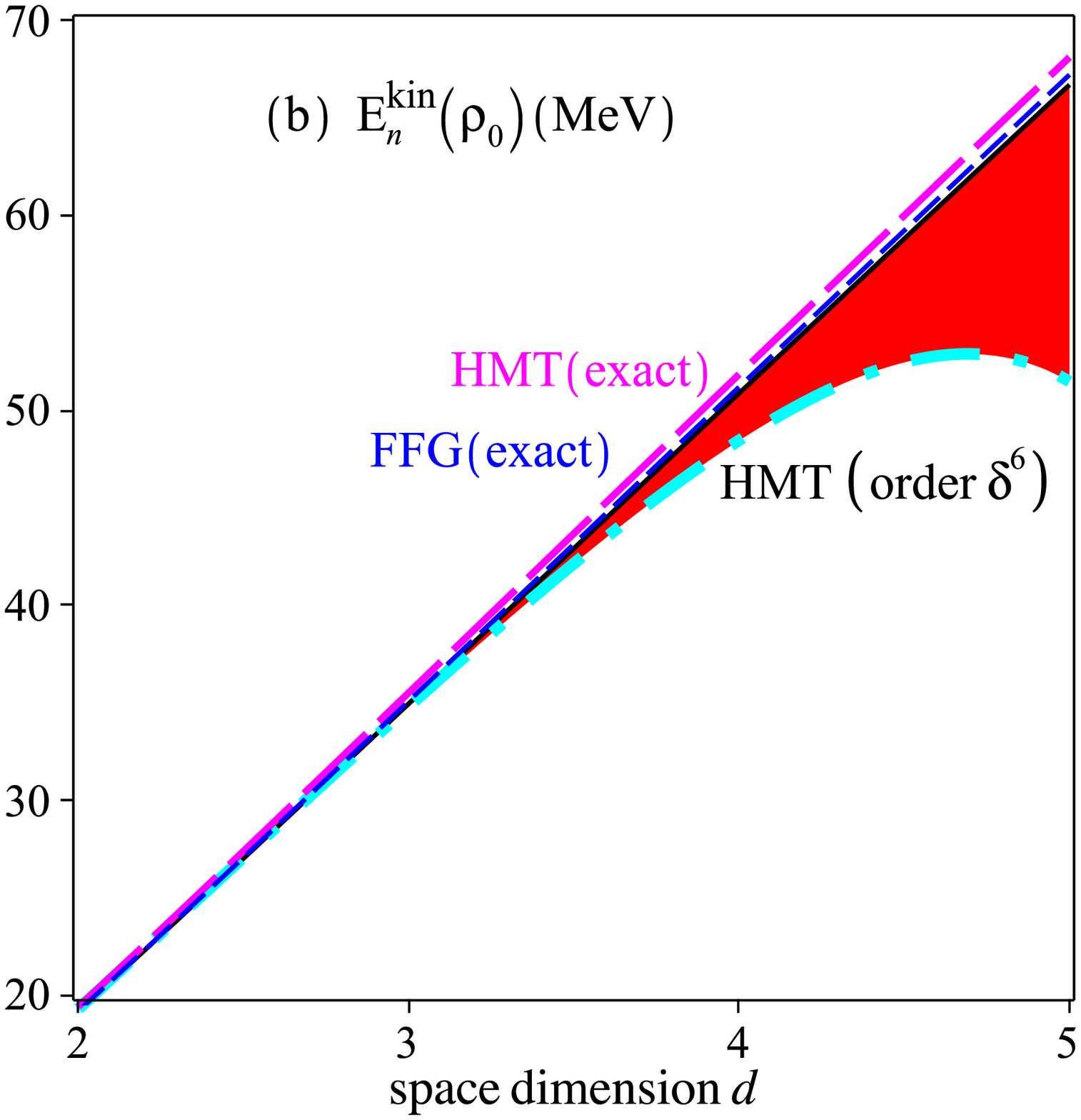}
\caption{\footnotesize Kinetic EOS of ANM as a function of $d$ at saturation density $\rho_0$.
The upper boundary of the red area in the right panel corresponds to the FFG prediction (to order $\delta^6$), while the blue dash line is for the exact kinetic EOS of PNM from the FFG model.}
\label{fig_Ekin024-d}
\end{figure}

If one treats the dimension $d$ as a continuous variable, then we can plot different-order terms of the kinetic EOS as a continuous function of $d$, see the left panel of Fig.\,\ref{fig_Ekin024-d}, where the red segment on each shaded area corresponds to the kinetic EOS including the SRC-induced HMT. The kinetic fourth-order symmetry energy and the sixth-order symmetry energy are comparatively smaller than the kinetic EOS of SNM and the quadratic kinetic symmetry energy. The maximum value of the $E_{\rm{sym},4}^{\rm{kin},d}(\rho_0)$ occurs at about $d\approx4$, see also the third panel of Fig.\,\ref{fig_kinetic-E-d}.
The kinetic EOS of SNM and the kinetic symmetry energy as well as the fourth- (sixth-) order symmetry energy are all analytical functions of $d$ (as $d\to0$). In addition, at $d=2$ and/or $d=4$, both $E_{\rm{sym}}^{\rm{kin},d}(\rho_0)$ and $E_{\rm{sym},4}^{\rm{kin},d}(\rho_0)$ are continuous (with respect to $d$), showing that the apparent singular points of the general expressions of $E_{\rm{sym}}^{\rm{kin},d}(\rho)$ and $E_{\rm{sym},4}^{\rm{kin},d}(\rho)$ at $d=2$ and/or $d=4$ are unreal as we discussed earlier. Moreover, for $d\lesssim2$, consistent with the results shown in Fig.\,\ref{fig_kinetic-E-d}
all the kinetic EOSs shown in Fig.\,\ref{fig_Ekin024-d} deviate only slightly from their FFG correspondences (i.e., ideal EOS without interaction). This may indicate that the $d=2$ is perhaps a critical dimension. This is also consistent with the relative momentum fluctuation $\Upsilon_k$ investigated in the last subsection, i.e., the prediction on $\Upsilon_k$ from the HMT and the FFG models becomes very close as $d$ decreases, see the right panel of Fig.\,\ref{fig_fluc-d}.

In the right panel of Fig.\,\ref{fig_Ekin024-d}, we show the corresponding results for the kinetic EOS of PNM.
Here the blue dash line and the magenta long dash line are exact kinetic EOS of PNM for the FFG and the HMT models, respectively. The red region corresponds to the approximation to order $\delta^6$ (with the upper (lower) boundary corresponding to the FFG (HMT) model). It is seen that the kinetic EOS of PNM is less affected by the SRC-induced HMT at least for $d\lesssim3$, see the blue dash line (exact FFG prediction) and the magenta long dash line (exact HMT prediction). As we mentioned earlier, this is because there are few HMT nucleons in PNM as indicated by experiments for the conventional 3D case.

It is instructive and educational to examine the PNM results for several limiting cases. As a reference, in 3D we notice that the $E_{\rm{n}}^{\rm{kin}}(\rho_0)=2^{1/3}\cdot 3k_{\rm{F}}^2/10M\approx35.1\,\rm{MeV}$ for the exact FFG model. One can show that as $d\to0$,
\begin{align}
E_{\rm{n}}^{\rm{kin}}(\rho)
\to&\frac{k_{\rm{F}}^2}{2M}\frac{d}{d+2}\cdot2^{1+2/d}\left[1+C_{\rm{n}}\left(1-\frac{1}{\phi_{\rm{n}}^2}\right)\right]\notag\\
\approx&\frac{k_{\rm{F}}^2}{2M}\frac{d}{d+2}\left(2+\frac{4\ln2}{d}\right)
\left[1+C_{\rm{n}}\left(1-\frac{1}{\phi_{\rm{n}}^2}\right)\right]\notag\\
\approx&\frac{k_{\rm{F}}^2\ln 2}{M}\left[1+C_{\rm{n}}\left(1-\frac{1}{\phi_{\rm{n}}^2}\right)\right]\to0,
\end{align}
since $k_{\rm{F}}\approx\rho_d2^{1-2/d}\sqrt{\pi}\to0$ as $d\to0$ ($\rho_d=\rho^{d/3}$ is a constant as $d\to0$), where $C_{\rm{n}}=C_0(1+C_1)$ and $\phi_{\rm{n}}=\phi_0(1+\phi_1)$.
On the other hand, the kinetic EOS of PNM to order $\delta^6$ from the HMT model (lower boundary of the red region indicated by cyan solid line) is somewhat smaller than the exact HMT prediction (magenta long dash line) especially for $d\gtrsim3$. For example, $E_{\rm{n}}^{\rm{kin}}(\rho_0)\approx51.8\,\rm{MeV}$ adopting the exact formula while it is about $48.5\,\rm{MeV}$ for the EOS up to $\delta^6$ in 4D. It indicates that even higher order kinetic symmetry energies beyond order 6 are still sizable (in the presence of the HMT) in spaces with higher dimensions.  

Finally, similar to the kinetic EOS of SNM in dimension 1, there would be an extra contribution from the HMT to the kinetic symmetry energy if $\phi_0\to\infty$ is taken. This term is $k_{\rm{F}}^2/2M\cdot[8C_0(1+C_1)/3]$, see the expression (\ref{E2-d1}), which is greater than zero since $1+C_1>0$. However, the corresponding extra term contributing to the fourth-order kinetic symmetry energy is $4C_0C_1k_{\rm{F}}^2/9M$, see the expression (\ref{E4-d1}). It is generally negative if $C_1<0$. Combining the three extra terms associated with the kinetic EOS of SNM, the quadratic kinetic symmetry energy and the kinetic fourth-order symmetry energy, we have
\begin{equation}
\Delta E^{\rm{kin}}_{\rm{n},(\rm{1D})}(\rho)=({16k_{\rm{F}}^2}/{9M})\cdot C_{\rm{n}},~~\rm{as}~~\phi_0\to\infty.
\end{equation}
This then gives us the EOS of PNM in 1D as
$
E_{\rm{n}}(\rho)=2k_{\rm{F}}^2/3M+({16k_{\rm{F}}^2}/{9M})\cdot C_{\rm{n}}$, since the sixth-order and even higher order symmetry energies in dimension 1 approach zero as $\phi_0\to\infty$.
Numerically one find in 1D that $E_{\rm{n}}(\rho_0)\approx5.03\,\rm{MeV}$ and $\Delta E^{\rm{kin}}_{\rm{n},(\rm{1D})}(\rho_0)\approx1.62\,\rm{MeV}$ adopting the conventional 3D saturation density as $\rho_0\approx0.16\,\rm{fm}^{-3}$, i.e., the density in 1D is about $0.16^{1/3}\,\rm{fm}^{-1}\approx0.54\,\rm{fm}^{-1}$.

\setcounter{equation}{0}
\section{The $E_{\rm{sym}}(\rho)$ in $\rm{2D}$ with perturbative contribution from $\epsilon^2$-order}\label{SEC_Esym2D-O2}

In this section, we perturb the symmetry energy in 3D to second order of $\epsilon=d-d_{\rm{f}}=d-3$ and study its preliminary features. In particular, we have by generalizing (\ref{ddef_Esym}),
\begin{align}
E_{\rm{sym}}(\rho)=&\frac{\overline{k}_{\rm{F}}^2}{6M}+\left.\frac{\overline{k}_{\rm{F}}}{6}
\frac{\partial U_0}{\partial
|\v{k}|}\right|_{|\v{k}|=\overline{k}_{\rm{F}}}+\frac{1}{2}U_{\rm{sym}}(\rho,\overline{k}_{\rm{F}})\notag\\
&+\epsilon\left[\frac{\overline{k}_{\rm{F}}^2}{6M}\left(2\sigma-\frac{1}{3}\right)
+\frac{\overline{k}_{\rm{F}}}{6}\left({\displaystyle\sigma\overline{k}_{\rm{F}}\frac{\partial^2
U_0}{\partial
|\v{k}|^2} }+\left(\sigma-\frac{1}{3}\right)\frac{\partial
U_0}{\partial
|\v{k}|}\right)
+\frac{\sigma}{2}\frac{\partial U_{\rm{sym}}}{\partial
|\v{k}|}\cdot\overline{k}_{\rm{F}}\right]_{|\v{k}|=\overline{k}_{\rm{F}}}\notag\\
&+\epsilon^2
\left[
\frac{\sigma^2\overline{k}_{\rm{F}}^2}{12M}\left(
\overline{k}_{\rm{F}}M\frac{\partial^3U_0}{\partial|\v{k}|^3}+2M\frac{\partial^2U_0}{\partial|\v{k}|^2}+3M\frac{\partial^2U_{\rm{sym}}}{\partial|\v{k}|^2}+2\right)
-\frac{\sigma\overline{k}_{\rm{F}}^2}{18M}\left(M
\frac{\partial^2U_0}{\partial|\v{k}|^2}+\frac{M}{\overline{k}_{\rm{F}}}\frac{\partial U_0}{\partial|\v{k}|}+2\right)
\right.\notag\\
&\hspace*{2cm}
+\left.\frac{\sigma'\overline{k}_{\rm{F}}^2}{6M}\left(
M\frac{\partial^2U_0}{\partial|\v{k}|^2}+\frac{M}{\overline{k}_{\rm{F}}}\frac{\partial U_0}{\partial |\v{k}|}+\frac{3M}{\overline{k}_{\rm{F}}}
\frac{\partial U_{\rm{sym}}}{\partial |\v{k}|}+2\right)
+\frac{\overline{k}_{\rm{F}}^2}{54M}\left(\frac{M}{\overline{k}_{\rm{F}}}\frac{\partial U_0}{\partial|\v{k}|}+1\right)
\right]_{|\v{k}|=\overline{k}_{\rm{F}}},
\label{ddef_Esym-o2}
\end{align}
where $\sigma'\approx-0.0307$ characterizes the second-order correction of the Fermi momentum, i.e., via $k_{\rm{F}}\approx\overline{k}_{\rm{F}}(1+\sigma\epsilon+\sigma'\epsilon^2)$.
Tab.\,\ref{Tab_2DEsym-o2} gives the contributions from different models at order $\epsilon^2$ to the symmetry energy.
In the ImMDI model, if the $y$ parameter is taken to be about $-115\,\rm{MeV}$\,\cite{XuJ2015PRC}, then all the contributions to $E_{\rm{sym}}(\rho_0)$ at order $\epsilon^2$ are negative.
Another feature is that for each model, the second-order contribution can not change the qualitative conclusion from the linear-order. For example, in the Skyrme model\,\cite{WangR2018} although the second-order contribution is negative the overall result is still positive determined by the linear term.
In this sense, the second-order term is really a ``perturbation'' (small correction) although $\epsilon=-1$ is applied for the 2D symmetry energy.
One can similarly define the effective form of the symmetry energy in 2D as in (\ref{ddef_Esym_eff}).

\renewcommand*\tablename{\footnotesize Tab.}
\begin{table}[h!]
\centering
\begin{tabular}{c|c|c|c|c}\hline
&$\epsilon$-contribution &$\epsilon^2$-contribution&sign of $\epsilon^2$-term&reference\\
\hline\hline
Skyrme&$+2.2\epsilon\,\rm{MeV}$&$-0.82\epsilon^2\,\rm{MeV}$&negative&Ref.\,\cite{WangR2018}\\\hline
MDI&$-1.1\epsilon\,\rm{MeV}$&$-0.92\epsilon^2\,\rm{MeV}$&negative&Ref.\,\cite{Das2003PRC}\\\hline
ImMDI&$4.4\epsilon\,\rm{MeV}+0.14y\epsilon$&$5.3\epsilon^2\,\rm{MeV}+0.05y\epsilon^2$&undetermined&Ref.\,\cite{XuJ2015PRC}\\\hline
toy model&$+3.1\epsilon\,\rm{MeV}$&$-0.48\epsilon^2\,\rm{MeV}$&negative&$/$ \\
\hline
\end{tabular}\caption{Second order of $\epsilon$ correction to the symmetry energy from different models.}\label{Tab_2DEsym-o2}
\end{table}

If one neglects the momentum dependence of the $U_0$ to second order and third order, i.e., $\partial^2U_0/\partial |\v{k}|^2\approx0$ and $\partial^3U_0/\partial|\v{k}|^3\approx0$, as well as the momentum dependence of the $U_{\rm{sym}}$ to second order, i.e., $\partial^2 U_{\rm{sym}}/\partial |\v{k}|^2\approx0$, then the perturbed symmetry energy to $\epsilon^2$-order could be approximated as,
\begin{align}
E_{\rm{sym}}(\rho)=&\frac{\overline{k}_{\rm{F}}^2}{6M}+\left.\frac{\overline{k}_{\rm{F}}}{6}
\frac{\partial U_0}{\partial
|\v{k}|}\right|_{|\v{k}|=\overline{k}_{\rm{F}}}+\frac{1}{2}U_{\rm{sym}}(\rho,\overline{k}_{\rm{F}})\notag\\
&+\epsilon\left[\frac{\overline{k}_{\rm{F}}^2}{6M}\left(2\sigma-\frac{1}{3}\right)
+\frac{\overline{k}_{\rm{F}}}{6}\left(\sigma-\frac{1}{3}\right)\frac{\partial
U_0}{\partial
|\v{k}|}
+\frac{\sigma}{2}\frac{\partial U_{\rm{sym}}}{\partial
|\v{k}|}\cdot\overline{k}_{\rm{F}}\right]_{|\v{k}|=\overline{k}_{\rm{F}}}\notag\\
&+\epsilon^2
\left[
\frac{\sigma^2\overline{k}_{\rm{F}}^2}{6M}
-\frac{\sigma\overline{k}_{\rm{F}}^2}{18M}\left(\frac{M}{\overline{k}_{\rm{F}}}\frac{\partial U_0}{\partial|\v{k}|}+2\right)
\right.\notag\\
&\hspace*{1.cm}
+\left.\frac{\sigma'\overline{k}_{\rm{F}}^2}{6M}\left(\frac{M}{\overline{k}_{\rm{F}}}\frac{\partial U_0}{\partial |\v{k}|}+\frac{3M}{\overline{k}_{\rm{F}}}
\frac{\partial U_{\rm{sym}}}{\partial |\v{k}|}+2\right)
+\frac{\overline{k}_{\rm{F}}^2}{54M}\left(\frac{M}{\overline{k}_{\rm{F}}}\frac{\partial U_0}{\partial|\v{k}|}+1\right)
\right]_{|\v{k}|=\overline{k}_{\rm{F}}}.
\label{ddef_Esym-o2-1}
\end{align}
Then the $\epsilon^2$-order contribution to the symmetry energy become $-0.46\epsilon^2,-0.36\epsilon^2,8.2\epsilon^2\,\rm{MeV}-0.003y\epsilon^2$, and $-0.65\epsilon^2$, respectively, for the four models listed in Tab.\ref{Tab_2DEsym-o2}.
In addition, the correction to the kinetic symmetry energy in 2D to order $\epsilon^2$ is given as $
[{\overline{k}_{\rm{F}}^2}/{54M}]
\cdot[(3\sigma-1)^2+18\sigma']\approx
-0.65\,\rm{MeV}$,
and consequently $E_{\rm{sym}}^{\rm{kin}}(\rho_0)\approx9.84\,\rm{MeV}$.
Compared with the result of the linear-order calculation of about 10.49\,MeV (see the estimate (\ref{ia-2})), the second-order correction makes the final result much closer to the exact value of about 9.60\,MeV.
Even higher order corrections (i.e., those beyond $\epsilon^2$) are expected to be small for $|\epsilon|\lesssim1$.

\section{Summary and discussions of some unresolved issues for future studies}\label{SEC_SUM}

Freeing up the spatial dimension $d$ for neutron-rich matter, we derived the general expression of its EOS in terms of the nucleon  isoscalar and isovector potentials based on the generalized HVH theorem. We find that the SNM EOS is more bounded at a higher 3D-equivalent saturation density while the symmetry energy becomes smaller in spaces with lower dimensions compared to the conventional 3D case. While we speculated that there are objects and/or sub-systems of nucleons that may be considered as living in spaces with reduced dimensions, implications of our findings for nuclear physics and astrophysics remain to be studied. In the following we summarize our main results and point out a few unresolved issues deserving further studies.
\begin{enumerate}
\item[(a)] Analytical expressions of the nucleon specific energy $E_0(\rho)$, pressure $P_0(\rho)$, incompressibility coefficient $K_0(\rho)$ and skewness coefficient $J_0(\rho)$ of SNM are derived in a general $d$D space. The corresponding expressions for the quadratic symmetry energy $E_{\rm{sym}}(\rho)$, its slope parameter $L(\rho)$ and curvature coefficient $K_{\rm{sym}}(\rho)$ as well as the fourth-order symmetry energy $E_{\rm{sym,4}}(\rho)$ are also given.

\item[(b)] The general features of the EOS of ANM in 2D are analyzed in some details. In particular, we found that the EOS of ANM (including the symmetry energy and the EOS of SNM) could be obtained from an effective expansion based on $\sqrt{\rho}$.  Moreover, there are no kinetic terms appearing in the general expressions for $K_0(\rho), J_0(\rho)$ and $E_{\rm{sym,4}}(\rho)$. Based on this finding, we have shown that the quartic term $E_{\rm{sym,4}}(\rho)$ is generally far smaller than the quadratic symmetry energy $E_{\rm{sym}}(\rho)$ in the sense that $E_{\rm{sym,4}}(\rho_0)/E_{\rm{sym}}(\rho_0)\lesssim1\%$, validating the conventional parabolic approximation for the EOS of ANM in 2D. 
Furthermore, the fourth-order symmetry energy in other dimensions (from 1D to higher dimensions) is also shown to be small due to the specific structure of $E_{\rm{sym,4}}(\rho)$ where the dimension $d$ plays an essential role (see the expression (\ref{4qEsym4FF_1})). Thus, from the viewpoint in a $d$D space in which the nucleons live, the effectiveness of the parabolic approximation of its EOS is consequently natural and the conventional 3D case is not unique.

\item[(c)] The EOSs in $d$D spaces are derived from the $\epsilon$-expansion (using an assumed small perturbative dimension $\epsilon=d-d_{\rm{f}}$ with $d_{\rm{f}}$ being a reference dimension) based on the EOS in 3D. 
In particular, we investigated the kinetic EOS both for the FFG model (section \ref{SEC_EXP} with $d_{\rm{f}}=3$ to linear order of $\epsilon$) and for the HMT model (section \ref{SEC_HMT} with general $d_{\rm{f}}$). We found that starting from the conventional 3D kinetic EOS, one can reasonably approximate the corresponding kinetic EOS of SNM and the kinetic symmetry energy in 2D, see (\ref{ia-1}) and (\ref{ia-2}). In addition, the $\epsilon$-expansion based on $d_{\rm{f}}\approx1\sim3$ even with the SRC-induced HMT included is effective, see results of Fig.\,\ref{fig_pert-ep}.
Moreover, the tendency of the EOS in a perturbed dimension is analyzed using the relevant analytical expressions.
Specifically, we found that the nucleon specific energy and pressure in SNM will be reduced (enhanced) if the perturbative dimension $\epsilon$ is negative (positive) compared with the 3D case, e.g., the EOS of SNM in 2D will be correspondingly softened compared with the 3D EOS of SNM, see the formula (\ref{ddef_E0}). The correction to the symmetry energy from the $\epsilon^2$ contribution does not qualitatively change the prediction from the linear approximation, as shown in section\,\ref{SEC_Esym2D-O2}.
In a nutshell, the 3D EOS of ANM in fact encapsulates the very relevant/useful information on the EOS in $d_{\rm{f}}$D with $d_{\rm{f}}$ being near 3, and the $\epsilon$-expansion provides a useful technique to explore the low-$d$ EOS of ANM from its 3D counterpart.

\item[(d)] A toy model for the single-nucleon potential is constructed by enforcing the corresponding EOS of ANM to fulfill all the available empirical constraints, including the one on the pressure in SNM from studying nuclear collective flow data in relativistic heavy-ion collision\,\cite{Pawel2002Science}, the constraint on the single-nucleon potential in SNM from optical model analyses of nucleon-nucleus scattering data\,\cite{Hama1990}, the scalar nucleon effective mass around the saturation density\,\cite{LiBA2018PPNP},  the causality constraint at densities $\lesssim6\rho_0$, the isobaric analog state constraint on the symmetry energy near the saturation density\,\cite{Pawel2014} as well as the constraint on the EOS of PNM from chiral effective field theories\,\cite{Tews2013}. The constructed model (which owns certain momentum dependence) is then applied to study the EOS of ANM in $d$D spaces. Specifically, we found that the EOS of SNM is enhanced (reduced) if $d$ is upwardly (downward) perturbed with respect to the 3D case, verifying the general conclusion found in section \ref{SEC_EXP}.
This means that the many-nucleon system tends to be more bounded in spaces with reduced dimensions, while it may become completely unbounded in spaces with $\gtrsim4$. The latter phenomenon in fact reflects some deep principles from the viewpoint of the high-dimensional geometry regarding the $d$-dimensional sphere.
In addition, the symmetry energy in 2D is found to be reduced compared to its correspondence in 3D.

\item[(e)] The kinetic EOS of ANM in the presence of SRC-induced HMT in the single-nucleon momentum distribution function in $d$D is studied. Besides the effectiveness of the $\epsilon$-expansion of the kinetic EOS of SNM, we also found that the first two moments of the nucleon momentum distribution in lower dimensions become smaller in both the FFG and HMT models, while the relative nucleon momentum fluctuation significantly increases as $d$ decreases below 2, see Fig.\,\ref{fig_fluc-d}.
More interestingly, the prediction on the relative momentum fluctuation $\Upsilon_k$ from the HMT and the FFG model becomes closer as $d$ decreases, indicating that the many-nucleon system in low dimensions behaves like a free system in terms of their kinetic EOSs. The same phenomenon is also found through the calculations on the isospin-expansion of the kinetic EOS of ANM, i.e., the predictions on them from the FFG and the HMT model become very similar in low dimensions (e.g., in 1D or 2D), as shown in the left panel of Fig.\,\ref{fig_Ekin024-d}. The nearly free property of the kinetic EOS of ANM in low dimensions may provide a useful tool to explore the EOS in 3D from the corresponding counterpart in 2D or 1D, or vice versa.
Furthermore, the predictions on the kinetic EOS of PNM by combining the $E_0^{\rm{kin}}(\rho)$, $E_{\rm{sym}}^{\rm{kin}}(\rho)$, $E_{\rm{sym,4}}^{\rm{kin}}(\rho)$ and $E_{\rm{sym,6}}^{\rm{kin}}(\rho)$ from the HMT and the FFG models are very close for $d\lesssim3$, verifying the conventional picture in 3D that there exist very few high-momentum neutrons in PNM.
However, as the dimension $d$ increases beyond 3, the above predictions from the HMT and the FFG models tend to deviate, see the right panel of Fig.\,\ref{fig_Ekin024-d}, indicating that the EOS of PNM in high dimensions may behave very differently from the one in $d\lesssim3$.
\end{enumerate}

There is currently no systematic study on the EOS of ANM in a space of general dimension $d$, partially due to the experimental limitation, theoretical inability or lack of interest. While our work presented here is exploratory in nature, we found the exploration very interesting and feel confident that the results summarized above are useful. Certainly, much more work is needed to further 
investigate many of the issues we have touched on in the present work. There are also many other interesting issues to be studied. To stimulate further studies along this line we list below a few possible such issues:

\begin{enumerate}
\item[(a)] We studied in this work the kinetic EOS (both with and without the HMT) in some details in the general dimension $d$, the next important step is to effectively/reasonably incorporate the nucleon-nucleon interactions/potentials in a self-consistent way,  either at the phenomenological or at the microscopic levels. For example, one needs to consider how to determine the model parameters in $d$D, and the basic principles and/or symmetries to be considered, etc.
Before detailed schemes (fitting schemes and/or the principles) are available, the $\epsilon$-expansion provides a practical starting point, e.g., one can use (\ref{ddef_Esym_eff}) to investigate how the 3D knowledge on the symmetry energy could induce relevant information in dimensions $\epsilon+d_{\rm{f}}$.

\item[(b)] Considering the HMT issue, how can one obtain the EOS of ANM based on the HVH theorem in general $d$D, i.e., the close relation between the partial derivative of the energy density with respect to the density and the single energy at $k_{\rm{F}}$, since when the momentum distribution has depletion as well as the HMT, one does not have $\partial\varepsilon/\partial\rho=e(k_{\rm{F}})$, where $e$ is the single-nucleon energy as the sum of the kinetic part and the potential part, and $\varepsilon$ is the energy density of the system, i.e., $e(k_{\rm{F}})$ is generally not the chemical potential of the nucleons (which however holds for the FFG model). This is in fact a fundamental problem in (nuclear) many-body theories\,\cite{AGD}.

\item[(c)] As discussed in the last section, considering that the many-nucleon system in low dimensions is close to its FFG counterpart, can one rely on it (e.g., via the higher-order $\epsilon$-expansions) to extract useful information for the corresponding EOS in 3D?
A related issue is on the correlation between the symmetry energy and its slope parameter. It is known that the latter (at $\rho_0$) in 3D is constrained to be about $L\approx fS$ with $f$ being about 2-3\,\cite{LiBA2021Universe} and $S\equiv E_{\rm{sym}}(\rho_0)$, which is in fact quite close to its FFG prediction $L=3S$.
If the low-dimensional system is close to the free model, can one establish the corresponding correlation (even with the effective potential included) to be near $L\approx 3S$ in 2D or $L\approx 6S$ in 1D, or more generally $L\approx 6S/d$?

\item[(d)]In this work, we study the EOS of ANM in a space of general dimension $d$ totally adopting the non-relativistic calculations. Since the relativistic kinematics, the nucleon-nucleon interactions, and/or the dimensionality together may effectively affect the EOS of ANM\,\cite{CaiBJ2022}, it would be extremely interesting to combine the relativistic kinematics and the fruitful nucleon-nucleon interactions (we have already found that the momentum dependence of the single-nucleon potential could effectively influence certain problems) in the general dimension $d$ and to explore how they influence each other coherently.
See the following overall scheme (see (\ref{sjk_1}) for the definition of $\theta_{\rm{f}}$ and $e_J$ is the general relativistic single-nucleon energy\,\cite{Serot1986}) on its inner relations.
\[
\boxed{
\begin{CD}
\begin{array}{c}
e_J(\rho,\delta,|\v{k}|)~\mbox{in~$d$D}\\
\left(\hspace*{-0.15cm}
\begin{array}{c}
\rm{NN~interactions}\\
\rm{dimensionality}
\end{array}
\hspace*{-0.15cm}
\right)
\end{array}@>{\delta\approx0}>{\rm{HVH/others}}>
\begin{array}{l}
\mbox{Relativistic}\\
\rm{characteristics:}\\
E_{\rm{sym}}(\rho),L(\rho),\cdots 
\end{array}@>{|\v{k}|,k_{\rm{F}}\ll M}>{\epsilon=d-d_{\rm{f}}}>
\left\{\hspace*{-0.15cm}
\begin{array}{l}
\rm{NR~formulas}\\
\mbox{$d_{\rm{f}}$D~quantities}
\end{array}\right.
\\
@VV{|\v{k}|,k_{\rm{F}}\ll M}V @V{\theta_{\rm{f}}\approx0}VV  \\
\begin{array}{c}
|\v{k}|^2/2M+U_J(\rho,\delta,|\v{k}|)\\
(\rm{nucleon~potential})
\end{array}@. 
\begin{array}{l}
\rm{Characteristics:}\\
P(\rho),K_0(\rho),J_0(\rho),\cdots
\end{array}
\\
@VVV @V {|\v{k}|,k_{\rm{F}}\ll M}V{\epsilon=d-d_{\rm{f}}}V  \\
\rm{NR~formulas}@>\delta\approx0,\theta_{\rm{f}}\approx0> \epsilon=d-d_{\rm{f}}\approx0>
\begin{array}{l}
\mbox{NR~formulas~in $d_{\rm{f}}$D:}\\
P(\rho),K_0(\rho),J_0(\rho),\cdots
\end{array}\\
\end{CD}}
\]

\item[(e)] The role played by the momentum-dependences of the single-nucleon potentials especially that of the symmetry potential $U_{\rm{sym}}$ on the symmetry energy is revealed. Currently, we can not determine whether an enhancement or a reduction may be induced to the $E_{\rm{sym}}(\rho)$ if $d$ is perturbed, see the formula (\ref{ddef_Esym}) and Tab.\,\ref{Tab_2DEsym}, the discussions after Eq.\,(\ref{edf}) and those after Eq.\,(\ref{djk-2}).
To be deterministic, more detailed information about the momentum dependence of the symmetry potential $U_{\rm{sym}}$ and  more accurate calculations (e.g., to higher orders of $\epsilon$) are needed.

\item[(f)] The last but not least, it is really important to find applications where the low-dimensional EOS of ANM could be used. As we have speculated in the {\sc Introduction}, the crust of a neutron star may be treated as a quasi-2D system while its non-rotating core may be consider as a 1D system, and certain sub-systems of particles in heavy-ion reactions may be also treated as a low-dimensional manifold. Finding novel signatures of these problems adopting the general $d$D EOS of ANM is a big challenge and also an exciting task. For example, the radial equation relating the crust EOS of neutron star matter with observables should also be derived starting from a suitable 2D setting\,\cite{Zarro2009,Rahaman2014} (e.g., via the 2D Christoffel coefficients). Thus, the conventional Tolman-Oppenheimer-Volkoff equation\,\cite{Misner1973} and its input EOS may both need to be modified to account for the changes from 3D to $d$D spaces.

\end{enumerate}

\section*{Acknowledgment}
This work is supported in part by the U.S. Department of Energy, Office of Science,
under Award No. DE-SC0013702, the CUSTIPEN (China-
U.S. Theory Institute for Physics with Exotic Nuclei) under
US Department of Energy Grant No. DE-SC0009971.
The authors would like to thank Zhen Zhang for helpful discussions on the possible low-dimensional nuclear systems created in experiments or in astrophysics.


{\small
\setlength{\bibsep}{4.5pt plus 1.ex}
\bibliography{mybibfile1}
}

\end{document}